%% file: main.tex
\documentclass[acmtog,screen,authorversion,nonacm,balance=true]{acmart}

\AtBeginDocument{%
  }

\usepackage{booktabs} 

\copyrightyear{2024}
\acmYear{2024}
\setcopyright{rightsretained}

\setcitestyle{nosort,square} 

\title{\textsf{RNA: Relightable Neural Assets}}

\author{Krishna Mullia} 
    \email{mulliala@adobe.com}
    \affiliation{%
        \institution{Adobe Research}%
        \city{San Francisco}
        \state{CA}
        \country{USA}
    }
    
\author{Fujun Luan}
    \email{fluan@adobe.com}
    \affiliation{%
        \institution{Adobe Research}%
        \city{San Jose}
        \state{CA}
        \country{USA}
    }

\author{Xin Sun}
    \email{xinsun@adobe.com}
    \affiliation{%
        \institution{Adobe Research}%
        \city{San Jose}
        \state{CA}
        \country{USA}
    }
    
\author{Milo\v{s} Ha\v{s}an}
    \email{mihasan@adobe.com}
    \affiliation{%
        \institution{Adobe Research}%
        \city{San Jose}
        \state{CA}
        \country{USA}
    }

\renewcommand\footnotetextcopyrightpermission[1]{} 

\usepackage{mathtools,stmaryrd,nicefrac,bbm}
\usepackage[many]{tcolorbox}

\usepackage[linesnumbered,ruled,vlined]{algorithm2e} 

\SetAlFnt{\small}
\SetAlCapFnt{\small}
\SetAlCapNameFnt{\small}
\SetAlCapHSkip{0pt}
\let\oldnl\nl
\newcommand{\nonl}{\renewcommand{\nl}{\let\nl\oldnl}}

\newcommand{\ignorethis } [1] {}

\usepackage{tikz}
\usetikzlibrary{calc}
\usepackage{pgfplots}

\usepackage{enumitem}
\setlist[itemize]{parsep=0pt,partopsep=0pt,leftmargin=*,itemsep=5pt}
\setlist[enumerate]{parsep=0pt,partopsep=0pt,leftmargin=*,itemsep=5pt}

\input{chapters/packages}

\begin{document}

\input{chapters/symbols}
\input{chapters/macros}

\input{chapters/teaser}

\input{chapters/abstract}

    

\maketitle

\input{chapters/introduction}
\input{chapters/relatedwork}

\input{chapters/model}

\input{chapters/training}
\input{chapters/render}
\input{chapters/results}

\input{chapters/conclusion}

\input{chapters/acknowledgements}

\bibliographystyle{ACM-Reference-Format}
\bibliography{references}

\end{document}

%% file: chapters/packages.tex
\usepackage[T1]{fontenc}
\usepackage[utf8]{inputenc}
\usepackage{color}
\usepackage{graphicx} 
\usepackage{ifthen}
\usepackage{float}
\usepackage[tableposition=above]{caption}
\usepackage{alltt}
\usepackage{mathenv}
\usepackage{amsmath}

\usepackage{amsthm}
\usepackage{newlfont} 
\usepackage{floatflt}
\usepackage{wrapfig}
\usepackage{fixltx2e}
\usepackage{multirow}
\usepackage{CJKutf8} 
\usepackage[ruled]{algorithm2e} 
\usepackage{bm}
\usepackage{diagbox}
\usepackage{array}
\usepackage{graphicx}
\usepackage{caption}
\usepackage{subfig}
\usepackage{makecell}

\usepackage[percent]{overpic}
\usepackage{contour}


%% file: chapters/symbols.tex

\newcommand{\eg}{e.g.\@}
\newcommand{\etal}{et al.\@}
\newcommand{\etc}{etc.\@\xspace}
\newcommand{\ie}{i.e.\@\xspace}
\newcommand{\wrt}{w.r.t.\@}
\newcommand{\vs}{vs.\@\xspace}

\newcommand{\eqnref}[1]{Eq.~\ref{#1}}
\newcommand{\secref}[1]{Sec.~\ref{#1}}
\newcommand{\figref}[1]{Fig.~\ref{#1}}
\newcommand{\tabref}[1]{Tab.~\ref{#1}}
\newcommand{\dtilde}[1]{\tilde{\tilde{#1}}}
\newcommand{\ttilde}[1]{\tilde{\tilde{\tilde{#1}}}}

\renewcommand{\deg}[0]{^{\circ}}
\makeatletter
\newcommand{\rmnum}[1]{\romannumeral #1}
\newcommand{\Rmnum}[1]{\expandafter\@slowromancap\romannumeral #1@}
\makeatother

\newcommand{\adaptSteps}{18}
\newcommand{\adaptEpisodes}{600}
\newcommand{\adaptTime}{30}

%% file: chapters/macros.tex

\newcommand{\ignore}[1]{}

\newcommand{\bo}{\boldsymbol{\omega}}
\newcommand{\bx}{\mathbf{x}}
\newcommand{\bn}{\mathbf{n}}
\newcommand{\props}{\boldsymbol{\xi}}
\newcommand{\asset}{\mathcal{T}}
\newcommand{\bzeta}{\boldsymbol{\zeta}}
\newcommand{\bd}{\mathbf{d}}

\newcommand{\propsA}{\tilde{\boldsymbol{\xi}}}
\newcommand{\propsB}{\hat{\boldsymbol{\xi}}}
\newcommand{\propsX}{\tilde{\bx}}
\newcommand{\vis}{\mathbf{V}}

%% file: chapters/teaser.tex
\begin{teaserfigure}
    \centering
    \includegraphics[width=1\textwidth]{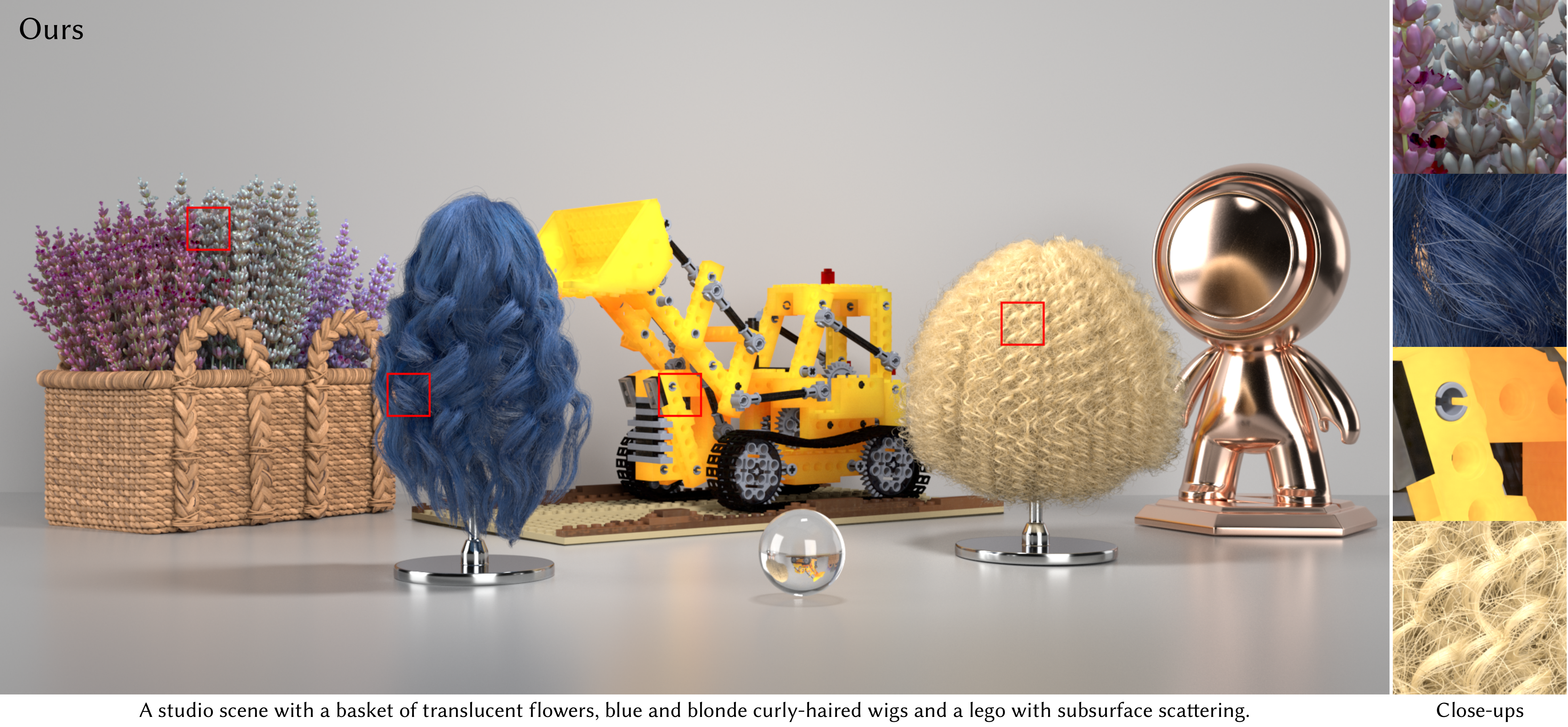}
    \caption{
        We propose a compact relightable neural 3D asset representation for geometries with complex shading, such as the fiber BCSDF model~\cite{chiang2016} for blue and blonde curly-haired wigs that exhibit glossy highlights and strong multiple fiber scattering, and the translucent Burley-Christensen shader~\cite{christensen2015approximate} combined with shading graphs in the Lego and flowers. Our results are realistic, even at extreme zoom levels, and closely match path-tracing references (shown later). Our representation combines explicit geometry (mesh or fibers) with a neural feature grid and an MLP decoder. This allows for view variation and full relightability, while reducing rendering costs and implementation complexity compared to the original asset representation. The assets can be integrated into a full renderer for path tracing with fast shading under arbitrary lighting conditions, correctly handling light transport within and between assets. On the right we show the close-ups of these assets, demonstrating high-fidelity shading such as complex multiple scattering effects and global illumination.
    }
    \label{fig:teaser}
\end{teaserfigure}

%% file: chapters/abstract.tex

\begin{abstract}
\label{abstract}

High-fidelity 3D assets with materials composed of fibers (including hair), complex layered material shaders, or fine scattering geometry are critical in high-end realistic rendering applications. Rendering such models is computationally expensive due to heavy shaders and long scattering paths. Moreover, implementing the shading and scattering models is non-trivial and has to be done not only in the 3D content authoring software (which is necessarily complex), but also in all downstream rendering solutions. For example, web and mobile viewers for complex 3D assets are desirable, but frequently cannot support the full shading complexity allowed by the authoring application.
Our goal is to design a neural representation for 3D assets with complex shading that supports full relightability and full integration into existing renderers. We provide an end-to-end shading solution at the first intersection of a ray with the underlying geometry. All shading and scattering is precomputed and included in the neural asset; no multiple scattering paths need to be traced, and no complex shading models need to be implemented to render our assets, beyond a single neural architecture.
We combine an MLP decoder with a feature grid. Shading consists of querying a feature vector, followed by an MLP evaluation producing the final reflectance value. Our method provides high-fidelity shading, close to the ground-truth Monte Carlo estimate even at close-up views. We believe our neural assets could be used in practical renderers, providing significant speed-ups and simplifying renderer implementations.

\end{abstract}

%% file: chapters/introduction.tex

\section{Introduction}
\label{sec:introduction}

High-fidelity 3D assets with materials using complex layered material shaders (subsurface scattering, coatings, weathered surfaces), or composed of fibers (including hair, fur, or detailed fabrics) are critical in high-end realistic rendering applications. Rendering such models is computationally expensive due to heavy shaders and long scattering paths. Moreover, all downstream renderers need to implement the exact same shading and scattering models as the source authoring system to correctly support the 3D asset. This is non-trivial: for example, web and mobile viewers for 3D assets are unlikely to support the full set of features of advanced 3D content authoring software.

Recent progress in neural rendering suggests converting the 3D asset to a suitable neural representation; however, no existing method provides a sufficient answer.  Earlier NeRF representations \cite{mildenhall2020nerf,muller2022instant,chen2022tensorf} focus on view synthesis only; these methods typically bake the original scene lighting into the neural asset, and cannot relight (that is, respond to the illumination of a new scene), which is critical for a high-fidelity asset representation. The spatial resolution of models based on volume rendering is limited: thin primitives like fibers cannot be fully resolved, and close-up views are blurry. However, in our use case the ground-truth geometry is available, and can be used as is, letting us focus on the challenging problem of representing the high-dimensional reflectance accurately.

More recent neural capture methods support relighting \cite{bi2020,NeuralPBIR,jin2023tensoir}, but this typically works by fitting analytic reflectance models to the observed views, which necessarily degrades complex material appearance. There are some exceptions; the recent work of \citet{NRHints} does not have this analytic BRDF limitation, and produces fully neural relightable assets of very high quality from real captures, but still has limitations when representing high-complexity digital 3D assets and does not focus on integrating the results into full-featured renderers. Neural materials \cite{kuznetsov2021neumip,kuznetsov2022rendering} are naturally relightable, but represent standalone materials rather than full assets. We would like to represent the entire asset with its texturing and material assignments, rather than just a flat tileable material patch. Adapting the ideas from the above methods to our asset representation setting is possible, but requires new approaches.

The key contribution of this paper is a neural representation for 3D assets with complex materials that supports high accuracy (even at strong zoom levels), full relightability and correct integration in Monte Carlo path tracers. We keep an explicit geometry, since the cost and implementation complexity of casting primary and shadow rays is reasonable and not a core challenge. Our neural model handles all shading and scattering; no multiple-scattering paths need to be traced, and no complex shaders need to be implemented in the deployment rendering system.
Our neural architecture combines an MLP (multi-layer perceptron) decoder with a feature grid, which can be defined using the triplane formulation \cite{EG3D}. A shading operation consists of querying the feature vector from the grid at the shading position, followed by passing the feature vector, combined with local information about the geometry intersection, view and light directions, which then produces the final shading color. 

Our method provides high fidelity shading, close to the ground truth Monte Carlo estimate even at strong close-up views. Moreover, our assets can be integrated into a full path tracer, interacting correctly with any other scene elements (objects and lights). The data generation consists of rendering $400$ camera views of the asset under different random light directions per pixel, which is tractable even for expensive hair/fur assets. The training on a single NVIDIA A100 ($40GB$) GPU takes about $90$ minutes, with an asset size of about $29 MB$ ($30$ minutes and $24 MB$ for small model).

In summary, our contributions are:
\begin{itemize}
    \item A neural 3D representation for assets with complex shading as a combination of explicit geometry, neural feature grid and MLP, allowing for full variation in view and lighting. The representation achieves higher accuracy and rendering performance than previous relightable neural representations.
    \item An efficient data generation and training pipeline specialized for this representation.
    \item A full integration of the neural representation into a production renderer. The target renderer can display the asset correctly within a scene consisting of other objects, with full global light transport, at high performance, and with no need to implement the complex material models encoded in the asset.
\end{itemize}
In the following sections, we cover the background, theory, precomputation, training and rendering of our relightable neural assets, demonstrating high-fidelity renderings, including videos in the supplementary materials. 

%% file: chapters/relatedwork.tex

\section{Related Work}
\label{sec:relatedwork}

\input{chapters/pipeline}

Ours is a neural relightable rendering method, so we cover neural rendering broadly, and relightable neural capture specifically. Since we target both surface shading and fiber shading, we cover classical techniques in these areas. Additionally, we discuss precomputed radiance transfer (PRT).

\paragraph{Neural rendering in graphics}
\label{sec:relatedwork:neural}
Recently, deep learning has demonstrated success in a wide range of disciplines, including the field of computer graphics. Neural Radiance Fields (NeRF)~\cite{mildenhall2020nerf} and follow-up neural scene representations~\cite{chan2022efficient, chen2022tensorf, yu2021plenoxels, muller2022instant} enable photorealistic novel view synthesis on complex real-world scenes. However, these neural fields typically do not support relighting, since the lighting and reflectance are baked in the radiance field. Earlier work trained multilayer perceptrons (MLPs) for fast global illumination rendering~\cite{ren2013global, ren2015image}.

Several neural graphics papers targeted specific rendering effects. \citet{kallweit2017deep} enables fast cloud rendering with radiance-predicting neural networks. \citet{chu2017data} applied CNNs on efficient fluid simulation. \citet{vicini2019learned} learns a shape-adaptive BSSRDF model that better approximates subsurface scattering. \citet{zhu2021neural} presented a neural complex luminaire representation that supports the compression, evaluation and importance sampling of the lightfield based on a simplified geometric proxy.

Our work is related to neural materials~\cite{Rainer2019Neural,kuznetsov2021neumip, kuznetsov2022rendering} in its neural architecture and relighting ability, but unlike learning the complex material appearance on planar or curved surfaces, our model is more focused on learning full assets, combining material with geometry. 
Our approach is similar to the neural architecture of the recent NeuMIP work~\shortcite{kuznetsov2021neumip, kuznetsov2022rendering}, which can theoretically fit any material (e.g. a complex synthetic micro-geometry or measured BTF data) with MLPs.

\paragraph{Relightable neural capture.} Recent advancements in inverse rendering have significantly leveraged neural representations. Various methods, including NeRFactor~\cite{zhang2021nerfactor}, Neural Reflectance Fields~\cite{bi2020}, NeRD~\cite{boss2021nerd}, and TensoIR~\cite{jin2023tensoir}, utilize an implicit neural density field for inverse volume rendering, among others~\cite{srinivasan2021nerv, boss2021neural, kuang2022neroic, yao2022neilf, wu2023nefii}. Out of these methods, we compare with \citet{bi2020}, because of its high quality (partly due to the dark room capture setup) and its ability to specially handle hair/fur shading.

Alternatively, representing shape via a Signed Distance Field (SDF) is explored in methods such as PhySG~\cite{zhang2021physg}, IRON~\cite{zhang2022iron}, and more~\cite{zhang2022modeling, mao2023neus, bangaru2022differentiable}. Hybrid methods combining neural and explicit representations with physics-based differentiable rendering are also emerging~\cite{cai2022physics, sun2023neural, luan2021unified, munkberg2022extracting, hasselgren2022shape}. However, these methods predominantly depend on analytic BRDF models, limiting their ability to represent and relight complex materials and visual appearances that do not easily fit these models, such as hair, fur, translucency, or skin shading graphs. Deferred Neural Lighting~\cite{gao2020deferred} inputs proxy geometry and rendered AOVs, employing a 2D CNN decoder for final rendering details. However, it does not support full path tracing integration, as evaluating path throughput for arbitrary positions and ray directions and interleaving path-tracing with CNN passes is non-trivial. NeMF~\cite{zhang2023nemf} uses a microflake-like phase function instead of a surface BRDF model, but still relies on fitting an analytic parametric material model, inheriting similar limitations.

NRHints~\cite{NRHints} is a recent method that overcomes the analytical BRDF constraint, enabling high-fidelity relighting of neural assets from real captured data. However, NRHints involves expensive training and rendering, and did not study full light transport applied to the resulting assets. Representing digital 3D assets with complex geometry and/or multiple scattering is possible with NRHints but less accurate than with our method. In contrast, our method requires significantly less training time, supports full relightability, and achieves full path-tracing integration into existing renderers at real-time frame rates. Of course, our method is designed for representing digital assets while NRHints focuses on capturing real objects; however, NRHints is still the closest alternative method to ours, so we provide extensive comparisons to it.

\paragraph{Complex surface and subsurface shading} 

Translucency, specifically subsurface scattering (SSS), plays an important role for skin and many other materials. Various shading techniques study efficient simulation of SSS effects~\cite{habel2013photon, donner2006spectral, jensen2001practical, donner2008layered}. Recently, Monte Carlo random walks~\cite{wrenninge2017path} have been preferred for translucency in high-end production. We use the Burley-Christensen method~\cite{christensen2015approximate} but our approach makes no assumptions on the method used to compute the effect in training data. Layered mixture models~\cite{guo2018position, belcour2018efficient, guo2016rendering} are capable of faithfully representing complex surface material layering and coating. Lastly, increasing efforts have been made in developing shading languages and material definition tools such as Nvidia MDL, MaterialX, and Adobe Substance. The complexity of materials achievable in a system that allows combining all of these tools using large shading graphs can become intractable for downstream deployment of the assets. A neural representation like ours helps to limit this complexity to the precomputation stage, and produces assets that are much easier to render downstream.

Hair, fur and fabrics are crucial components in representing realistic appearance in our world. \citet{marschner2003light} proposed a realistic hair reflectance model by representing hair fibers as rough dielectric cylinders with reflectance, absorption and transmission, which was later extended by \citet{d2011energy} and \citet{khungurn2017azimuthal}.  \citet{yan2015physically} presented an animal fur model that represents the fiber geometry with a double cylinder. \citet{chiang2016} presented a practical fiber shading model for hair and fur (or other fiber-based materials) that is efficient for production path tracing. We use this model to precompute our fiber-based assets, though any similar model could be used. This model is \emph{near-field}, meaning it defines a full BSDF for points on the fiber, allowing reflectance to vary across the width of the fiber; an important effect that we fully capture.  Dual scattering~\shortcite{zinke2008dual}
accounts for the multiple fiber scattering effects in human hair with fast global and local approximations, but requires user intervention in parameter tweaking and only applies to hair.

\paragraph{Precomputed radiance transfer (PRT)} 
Efficient scene relighting through PRT has been explored in computer graphics research for two decades~\cite{Sloan:2002:PRT,Ramamoorthi:2009:PBR}, often using the spherical harmonic basis, but also using other bases such as wavelets \cite{Ng:2003:Wavelet,Ng:2004:TripleProduct}, polynomials~\cite{Ben-Artzi:2008:BRDF}, spherical Gaussians~\cite{Tsai:2006:Gaussian,Xu:2013:AnisotropicGaussian} or neural basis functions~\cite{Xu:2022:NeuralBasisFunctions}.
More recently, Neural PRT~\cite{Rainer:2022:NPRT} aims for more compact storage and higher quality shading, however, it is designed for screen-space relighting in a deferred neural shader at scene scale and individual assets cannot be integrated in path tracers to interact with other assets.
It is important to note that the motivation behind PRT methods is fundamentally different from our work. Scenes rendered using PRT are constrained in that they cannot easily incorporate assets that have not undergone the same PRT representation. 
PRT models do not allow querying the lighting in an arbitrary direction. This lowers the lighting frequencies that can be represented, but even more importantly, the lighting is hard to obtain in a specific basis when shading a given scene point during path tracing. This limitation hinders the compatibility of PRT methods with production Monte Carlo path tracing.

In contrast, our relightable neural assets provide an end-to-end shading solution, which can be seamlessly integrated into existing renderers for Monte Carlo path tracing along with any other classical scene elements for global illumination.

%% file: chapters/pipeline.tex
\begin{figure*}[ht]
    \centering
    \includegraphics[width=\textwidth]{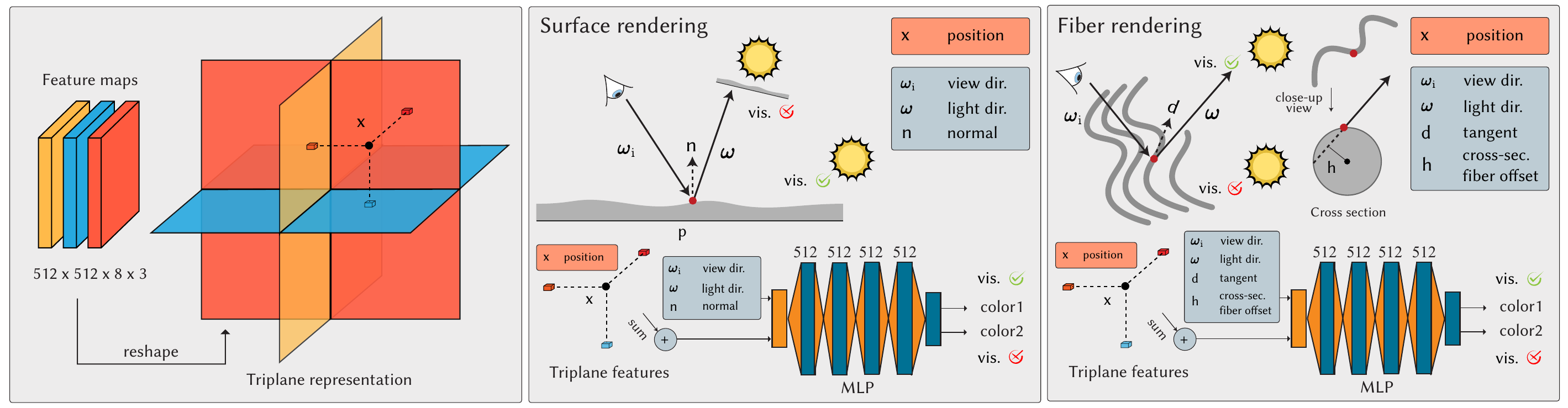}
    \caption{\textbf{Overview of the pipeline}. On the left, we illustrate our triplane representation, consisting of XY, XZ and YZ planes each with 8 feature channels and a default resolution of $512 \times 512$. The feature vectors queried from the triplane representation are summed and passed into an MLP, along with additional properties. We show two configuration variants of relightable neural asset pipelines designed for surface rendering (middle) and fiber rendering (right), respectively. The main difference is that surfaces use a normal input, while fibers use tangent and cross-section offset.
    Both variants output two colors, one of which will be picked according to the visibility at render time.
    } 
    \label{fig:pipeline}
\end{figure*}

%% file: chapters/model.tex

\section{Neural Shading Model}
\label{sec:model}

Our goal is to take a complex asset and precompute its light transport (including its materials and multiple light interactions with itself) in isolation and compress it using a neural model. Such an asset is fully relightable with respect to the underlying physical light transport and can be inserted into other scenes by quickly evaluating the neural model, for different positions, camera and light directions. No multiple scattering paths need to be traced, and no complex shading models need to be implemented to render our assets, beyond a single neural architecture. 

\subsection{Relightable Asset Definition}
\label{sec:transport}

To achieve this, we treat the asset as a scene by itself, lit by some incoming light distribution. The outgoing radiance $L_o$ at the shading position $\bx_o$ with the viewing direction $\bo_o$ is an integral of the light transport from all positions with all incoming lighting directions
\begin{align}
\label{eqn:transport}
L_o (\bx_o, \bo_o) = \int_{\bx_i \in \mathcal{A}, \bo_i \in \bo} L_i (\bx_i, \bo_i) {T(\bx_o, \bo_o, \bx_i, \bo_i) \mathrm{d} \bx_i \mathrm{d} \bo_i},
\end{align}
where $L_i$ is the incoming radiance at position $\bx_i$ with lighting direction $\bo_i$,
whose domains are $\mathcal{A}$ and $\boldsymbol{\Omega}$ respectively.
$T$ is the light transport function from $\{\bx_i, \bo_i\}$ to $\{\bx_o, \bo_o\}$, globally depending on the geometries and materials of the entire scene, and including light paths of all lengths.

Representing the above transport would still require 8-dimensional data: each outgoing pair $(\bx_o, \bo_o)$ is an integral over all incoming pairs $(\bx_i, \bo_i)$.  We make the further approximation of assuming distant directional light. More precisely, we propose a neural asset $\asset$, for which the outgoing radiance is computed as: 
\begin{align}
\label{eqn:neuraltransport}
L_o (\bx, \bo_o) = \int_{\bo_i \in \boldsymbol{\Omega}} L_i (\bo_i) {\asset(\bx, \bo_o, \bo_i) \mathrm{d} \bo_i},
\end{align}
where the shading position $\bx_o$ is denoted as $\bx$ for brevity, and $\asset$ is defined by integrating out the dependence on $\bx_i$, that is,
\begin{align}
    \asset(\bx, \bo_o, \bo_i) = \int_{\bx_i \in \mathcal{A}} T(\bx_o, \bo_o, \bx_i, \bo_i) \mathrm{d} \bx_i. 
\end{align}
Equivalently, $\asset(\bx, \bo_o, \bo_i)$ is the radiance leaving $\bx$ into direction $\bo_o$ when lit by a unit-irradiance directional light from direction $\bo_i$, and including all self-occlusions and inter-reflections. This makes our asset definition similar to a bidirectional texture function (BTF), but defined on an arbitrary geometry instead of a plane. In other words, our assets are akin to 3D BTFs defined over explicit geometry. We show that despite the distant directional light assumption in this definition, our assets can be used with any illumination including near-field lighting, much like BTFs.

Our goal is to represent the asset as a combination of the geometry itself and a neural module capable of evaluating $\asset$ for a given shading point $\bx$ and given lighting and viewing directions. The similarity of this problem to neural BTF compression also suggests that the design of neural techniques for compressing BTFs could be adapted to our asset compression problem, as we will see shortly.

\subsection{Neural Shading Architecture}
\label{sec:model-neural}

The neural shading architecture is composed of two steps, as shown in Fig.~\ref{fig:pipeline}.
At a shading point $\bx$, we query a feature vector from the triplane feature grid, $\bzeta(\bx)$; this feature vector is simply the sum of the features bilinearly interpolated from the three orthographically projected planes (XY, YZ and XZ), similar to \citet{EG3D}.

The shading point $\bx$ has other properties, such as the normal (for surface assets), tangent and position along fiber width (for fiber assets). We combine these properties with other available information, such as the camera direction $\bo_o$, light direction $\bo_i$.  We concatenate all of these properties into a property vector $\propsX = \propsX(\bx, \bo_i, \bo_o)$, which can be thought of as an extended shading point, with all easily available information from the rendering process added to it. More detail on the property vector $\propsX$ is given below in subsection \ref{sec:model-properties}. We concatenate $\bzeta(\bx)$ and $\propsX$ into a final input vector $\props$.  

In the second step, the transport $\asset(\bx, \bo_i, \bo_o)$ is evaluated by an MLP (multilayer perceptron) decoder taking the concatenation of the feature vector $\bzeta(\propsX)$ and properties $\propsX$ as input and returning \emph{two} RGB colors, one of which will be picked according to the visibility at render time. This design choice is to facilitate the full integration into production renderers --- where BSDF evaluation is often done before a shadow ray is traced for visibility. We denote the full input of the MLP as $\props$. 
The Monte Carlo simulation of global transport is not required any more for runtime evaluation of $\asset$, since for any point on the geometry and any incoming and outgoing directions, it can be quickly evaluated through a combination of querying the triplane feature grid and evaluating the MLP.

The neural asset is thus the combination of the geometry, the feature grid $\bzeta$ and the MLP weights. Once the feature grid and MLP are jointly trained, the scene can be efficiently re-rendered with arbitrary light and camera directions.
Importantly, the lighting does not necessarily have to be directional in the final scene where the asset is used, because the neural asset $\asset$ is parameterized with the differential of lighting direction $\bo_i$, similar to other shading models employed in Monte Carlo path tracing.
The lighting direction $\bo_o$ points to a sampled location on any light source and can be different every time $\asset$ is evaluated,
hence the asset can be used with any lighting that is sampled in a manner similar to traditional Monte Carlo path tracers.

In this sense, our asset is similar to a BTF \cite{dana1999btf}, acquired or synthesized under distant directional lighting but used in a final renderer with any lighting (such as area/directional/point emitters and IBLs). Recent neural material representations like NeuMIP \cite{kuznetsov2021neumip} are essentially compressed BTFs and make the same assumptions; they are also similar in combining feature grid lookups with MLP decoders, though their MLPs take somewhat different inputs and represent different phenomena on planar surfaces.

In summary, explicit geometry combined with the trained triplane grid $\bzeta$ and the MLP is a relightable asset that can be evaluated for new camera angles within new illumination conditions.


\subsection{Surface Properties}
\label{sec:model-properties}

The choice of surface properties $\propsX$ and $\props$ differs between surface-based and fiber based assets. We append the surface normal $\bn(\bx)$ for surfaces, and direction (tangent) $\bd(\bx)$ for fibers, to the property vector. This is not strictly required, and the model will learn without it, but it improves the fitting accuracy. For fibers, we additionally supply $h(\bx)$, the offset across the fiber width from the fiber axis, normalized to $[-1,1]$, as introduced by Marschner et al. \shortcite{marschner2003light}. This is critical for the ability of the model to learn spatial variation in lighting across (typically tiny) fiber width, which is a feature of near-field fiber shading models \cite{chiang2016}.

\begin{figure}[b]
\includegraphics[width=1.0\linewidth]{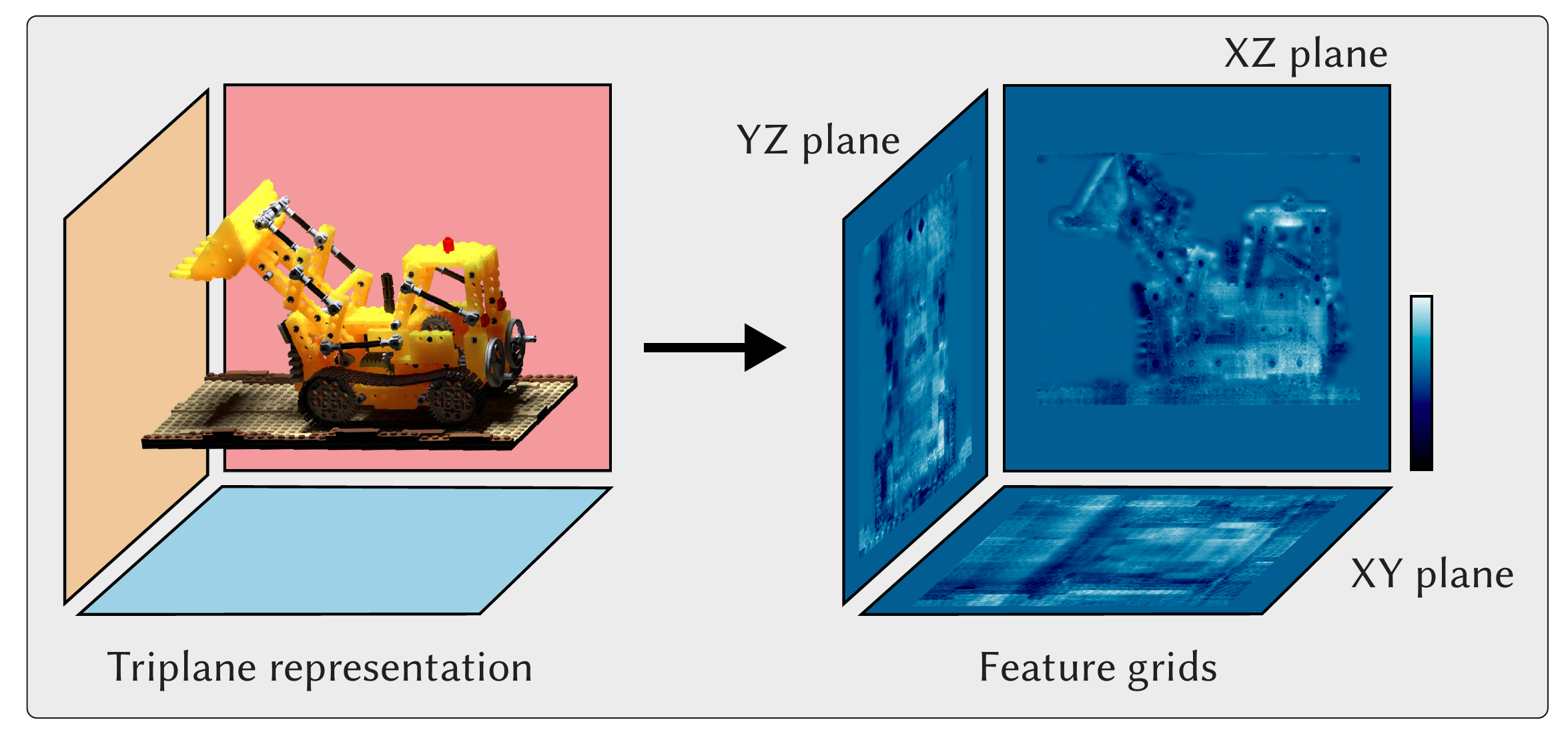}
\caption{\textbf{Visualization of the triplane representation.} We show the triplane representation for the subsurface lego asset, visualizing the first channel of XY, YZ, and XZ planes, respectively.
 } 
\label{fig:triplane_vis}
\end{figure}

\subsubsection*{Triplane representation}
\label{sec:model-properties-positions}

If complete and high-quality UV coordinates are available, our method can use a 2D feature texture, but this is not always feasible. It is challenging to build non-overlapping, low-distortion and compact texture coordinates, especially for meshes with complex topology and fiber assemblies. A more general solution is to use a triplane representation \cite{EG3D}.
The world position at $\bx$ is a $3D$ vector $\{x, y, z\}$, which is converted to three $2D$ vectors of $\{x, y\}$, $\{y, z\}$ and $\{z, x\}$. The neural feature grid $\bzeta$ is composed of three $2D$ tables for each of them, $\bzeta_{xy}$, $\bzeta_{yz}$ and $\bzeta_{zx}$, as illustrated in Figure~\ref{fig:triplane_vis}. Each query outputs an 8-channel vector. By summing the three output vectors, we obtain the final 8-channel feature vector $\bzeta(\bx)$.

\subsubsection*{Lighting Visibility for Self-Shadowing}
\label{sec:model-properties-visibility}

Direct shadowing is an important effect, easily handled by classical rendering methods, but difficult to learn for a neural network due to its discontinuous nature. Therefore, we also consider the binary visibility of the light direction $\vis(\bx, \bo_i)$, specifying whether the light ray self-intersects with the asset. We use this visibility value as a hint to the MLP, but unlike \citet{NRHints} we do not use it as an input, but rather have the MLP produce two outputs (assuming a visibility value of true or false) and pick the right value at render time. This solution provides more practical render integration, as detailed later in Sec. \ref{sec:render}. The visibility hint significant improves rendering quality with noise reduction around the shadowing edges.

%% file: chapters/training.tex

\section{Data Generation and Training}
\label{sec:training}
In this section, we describe our data generation pipeline and training details in Sec.~\ref{sec:training:data} and Sec.~\ref{sec:training:training}, respectively.

\subsection{Data generation}
\label{sec:training:data}

The training dataset is generated using the Python bindings of Blender 3.5, with a customized CPU version of the \emph{Cycles} path tracer. Surface scenes are modeled as meshes and can use arbitrary shaders allowed by Blender, including custom shading graphs. For scenes composed of fibers, the fibers are modeled as curved cylinders and their material properties are defined using the model of \citet{chiang2016}, known as \emph{Principled Hair BSDF} in Blender.

\begin{figure}[tb!]
    \centering
    \includegraphics[width=1.0\linewidth]{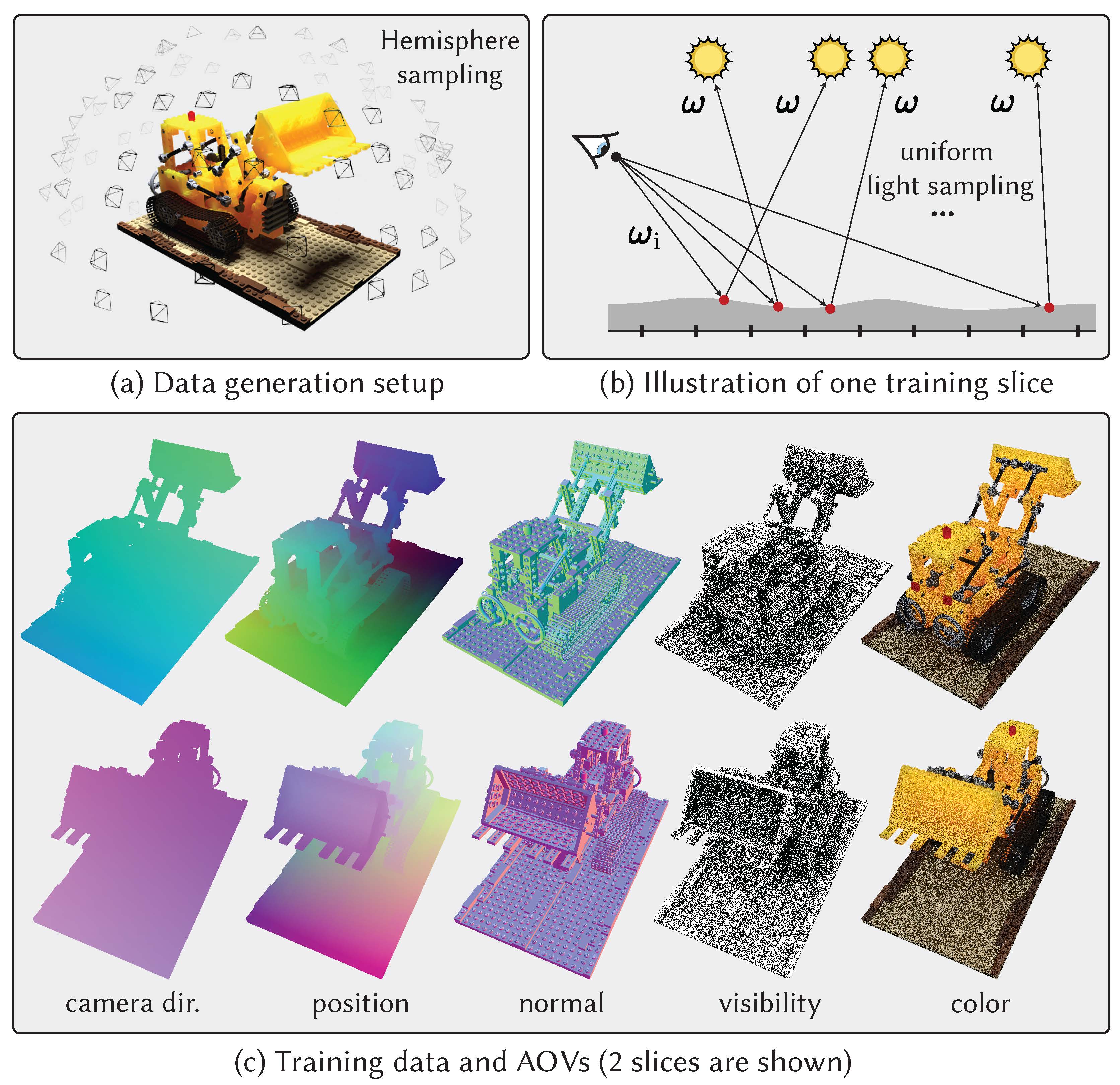}
    \caption{\textbf{Data generation of our pipeline.} (a) We sample camera views around the asset (upper hemisphere) and for each training slice (b) the light directions are randomly sampled for each pixel's ray hit shading point. In (c), we visualize two slices of the training data and AOVs obtained from Blender Cycles path tracer.  }
    \label{fig:datagen}
\end{figure}

Figure \ref{fig:datagen} shows our data generation setup along with a visualization of the data generated. To render each image, a perspective camera is randomly placed on a sphere with a user-defined radius centered at the asset, with the field of view set such that the asset is taking most of the view.
During data generation, the primary rays within a pixel are all traced through its center, to compute the outgoing radiance of a single visible point $\bx$ on the geometry, rather than an average of a footprint. In other words, we are interested in point sampling the appearance spatially, rather than pre-integrating it. Since the primary rays always intersect the same geometry at every pixel, the ``alpha'' channel of the RGBA rendered result is binary and indicates the pixels that contain valid intersections; these pixels become valid data points for training.

Instead of choosing a single light direction for each image or slice of the training data, we randomize the light direction for each pixel, improving the coverage of the space of view/light pairs available in the dataset. In practice, our path tracer allows lighting to be overridden such that each pixel is rendered with a different light direction $\bo_i$. 
As a result, each rendered pixel with a valid intersection provides a separate data point with distinct camera and light directions, which becomes a training sample.
Note that even though we use a fixed camera per rendered image in the dataset, it is possible to customize the data generator to produce a random view direction per pixel for shading purposes, just like the light direction. However, we did not find this necessary, and opted for simplicity in keeping the original view directions. Even in this case, the camera directions are slightly different for each pixel since we use a perspective camera, but not random.

In addition to the output radiance, we also utilize a number of arbitrary output variables (AOVs) in the shading graph and output them using Blender's compositing nodes for the values required by $\propsX$ and $\props$, such as the position $\bx$, viewing direction $\bo_o$, lighting direction $\bo_i$ and optionally the lighting visibility $\vis$. For an asset composed of fibers, we also output the fiber direction $\bd$ and the offset across the fiber $h$. 

We apply a direct light radiance clamping of value 20.0 and an indirect light clamping of value 10.0. This scales down very bright samples and prevents issues when combining very low roughness and purely directional lighting. Alternatively, we support turning the directional lights into lights with a small angular radius, which also has the same effect of restricting the dynamic range. However, we found that clamping is simpler and works just as well, while making it easier to get accurate visibility hints. Note that previous methods \cite{NRHints,bi2020} use low dynamic range images, thus essentially clamping at 1.0.

We render $400$ cameras with a resolution of $1024^2$ for a given asset.
For some assets, we restrict the lighting directions and viewing directions to the top hemisphere, as it may not be useful to learn bottom hemisphere appearance for certain assets. A typical dataset for a surface asset is about $17.2GB$, while fiber assets increase to about $18GB$.
We render $128$ to $4096$ samples per pixel, depending on the complexity of the scattering paths for an asset.
Although the rendered data has some remaining noise, it is sufficient as our training regularizes away the noise. 
We render the images on a cloud instance with $96$ CPU cores. The data for our assets (with complex effects like subsurface scattering and hair with long scattering paths) is generated within $4$ to $6$ hours. We also render out an additional $40$-view validation dataset with fixed light direction per view, enabling us to judge the fitting quality on realistic configurations numerically and visually.

\begin{figure}[b]
    \centering
    \includegraphics[width=0.49\linewidth]{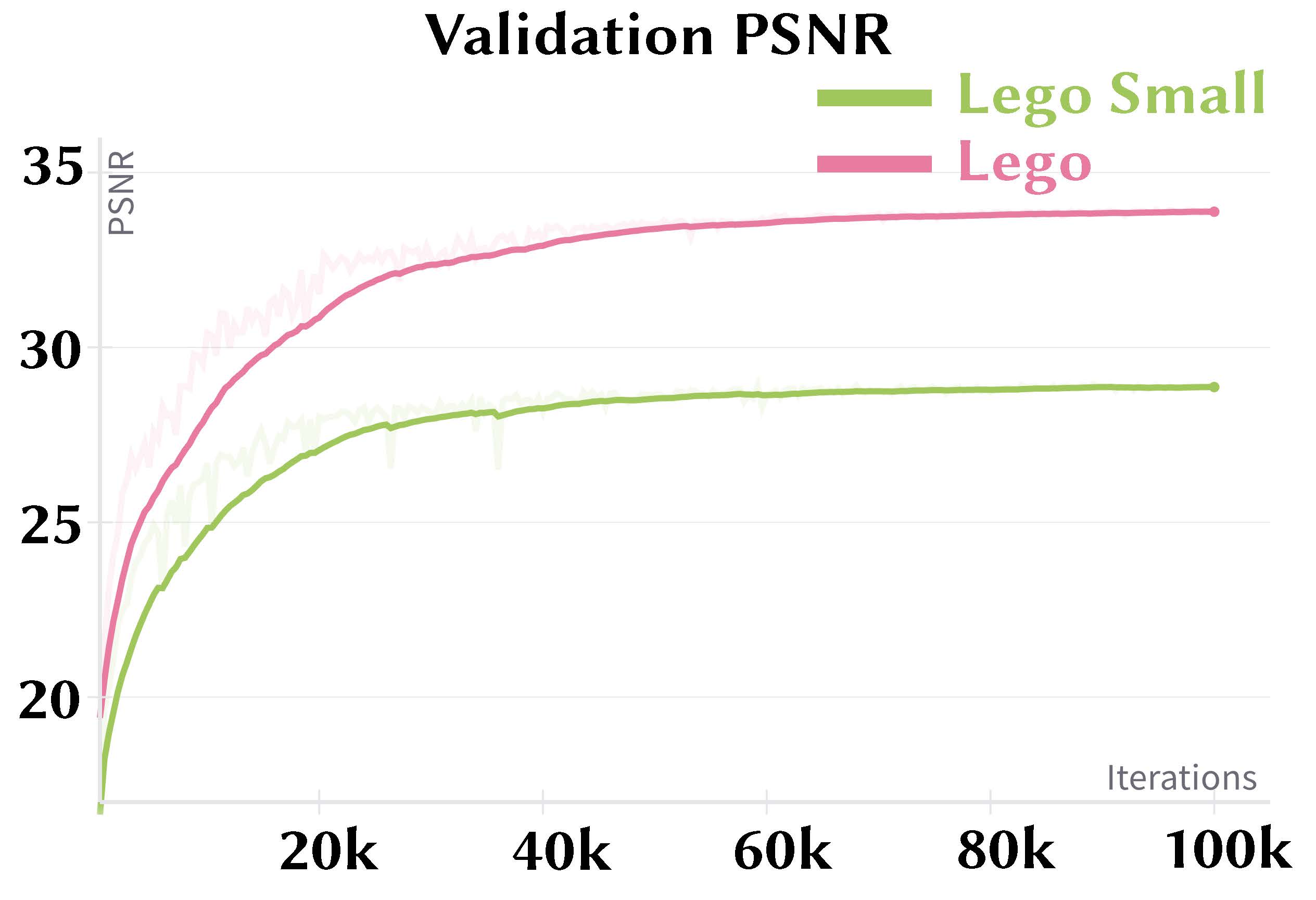}
    \includegraphics[width=0.49\linewidth]{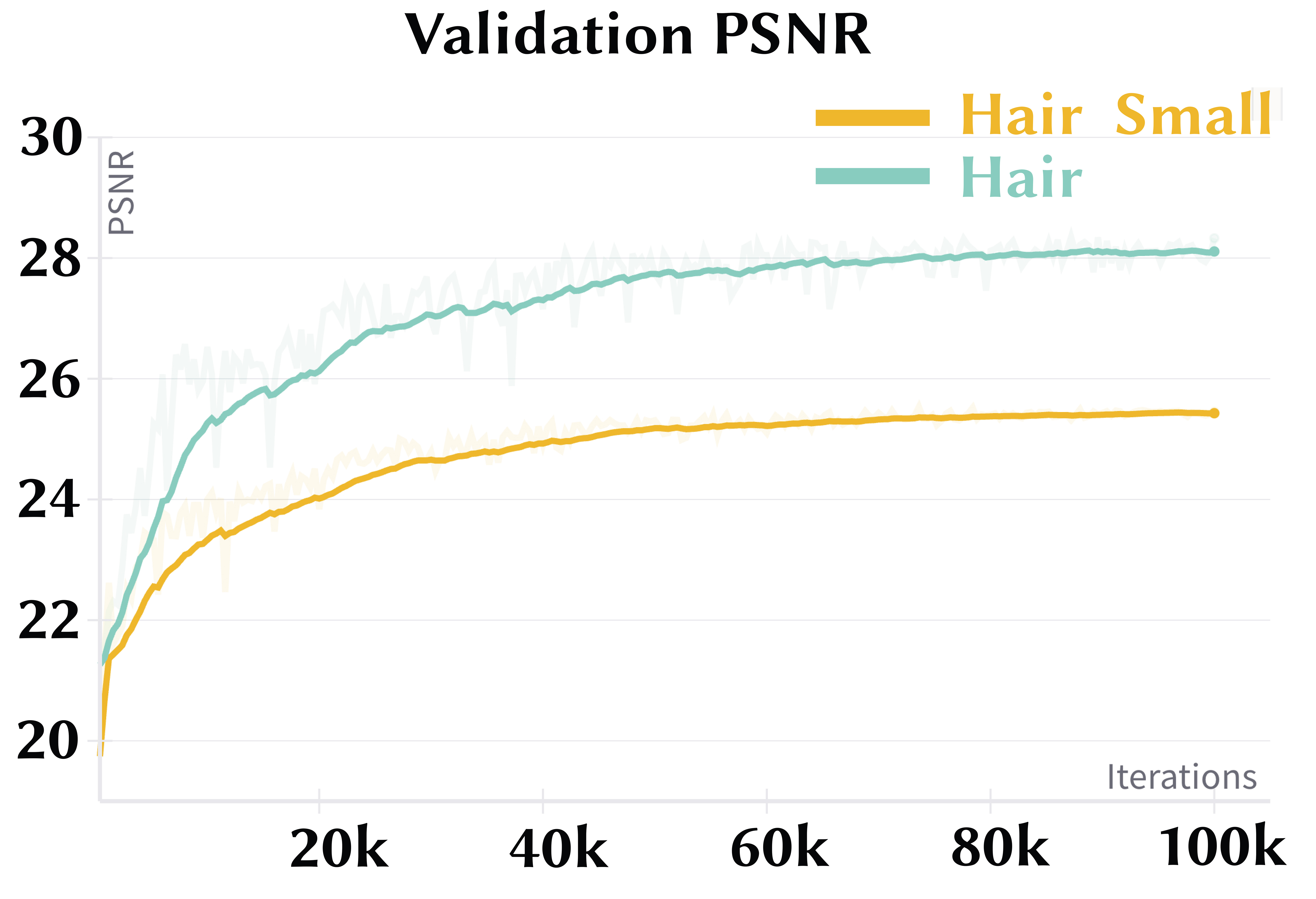}
    \caption{\textbf{Training convergence plots on translucent lego and blue hair assets.} We show the convergence plots for two versions of our model on a surface asset (lego) and a fiber asset (blue hair). The \emph{Lego Small} and \emph{Hair Small} are trained with a small variant of our neural model with reduced MLP size for real-time rendering in path tracer, at the cost of compromising some accuracy.}
    \label{fig:training_convergence}
\end{figure}

\input{chapters/eclair_lego}

\subsection{Training}
\label{sec:training:training}

For each sample point $\bx$, the input properties tuple $\props$ is used to predict two RGB transport values $\asset$ in the forward pass of our neural architecture.
We use the L2 error with $\log(x+1)$ applied to prediction and ground truth, and backpropagate to update the grid $\bzeta$ and the MLP decoder weights. The network outputs two RGB values (with and without visibility) and we apply the loss only to the output matching the visibility value for the given data point, ignoring the other output.

During training, one batch consists of one image slice from our dataset; each batch contains a different number of valid data samples since every camera can have a different number of valid geometry intersections. As described in \ref{sec:training:data}, the ''alpha'' channel of the radiance RGBA data is binary; our dataloader only loads data in each image slice with ''alpha'' of 1.0, pixels with 0.0 alpha are discarded since they represent ''empty space'' with no valid intersections. One training epoch includes the full $400$ batches of data.
Our training is implemented using the PyTorch Lightning framework.
We use a single Nvidia A100 GPU (40 GB memory) and our entire dataset fits into GPU memory during training, making the training very efficient.

Our training runs for $100,000$ iterations ($250$ epochs) using an Adam optimizer~\cite{kingma2014adam}, along with a \emph{StepLR} learning rate scheduler using an initial learning rate of $1*10^{-3}$, reducing it by half every $50$ epochs. With these settings, an average training run takes around $90$ to $120$ minutes, and $30$ to $40$ minutes for our small model; so it is much faster than data generation. In contrast, NRHints~\cite{NRHints} and Neural Reflectance Fields~\cite{bi2020} would take roughly a day to train their models on multiple GPUs (e.g. 4).

Note that during the initial iterations of training, we apply a blurring kernel on the three feature grids; starting at an initial footprint of $4$ pixels and gradually decaying it down to $1$ pixel by $20,000$ iterations ($50$ epochs); the remaining training keeps this minimum blur of $1$ pixel. This is inspired by NeuMIP~\cite{kuznetsov2021neumip} and shows noticeably better results due to spatial low-frequency data sharing, before optimization moves on high frequency details.

During training, we monitor the average PSNR values on the 40 image slices from the validation data set as shown for the Lego and Blue Hair assets in Figure~\ref{fig:training_convergence}. The plots show that the models already converge at around $80,000$ iterations ($200$ epochs), but we notice that letting them train another $20,000$ iterations ($50$ epochs) at the lowest learning rate helps resolve some more high frequency detail.

\input{chapters/eclair_hair}

We use PyTorch Lightning's checkpointing system to save out the three best models in terms of PSNR values on the validation dataset as well as the last checkpoint. The trained model includes $7.4$ million parameters ($6.3M$ for small model) between the feature grids and MLP weights, and the total uncompressed size is about $29 MB$ ($25 MB$ for small model).

%% file: chapters/eclair_lego.tex
\begin{figure*}
    \centering
    \setlength\tabcolsep{3.0pt}
    \resizebox{1.0\textwidth}{!}{
    \begin{tabular}{c|c|c|c|c|c}
         Blender Cycles  &
         Ours &
         Blender Cycles &
         Ours (small model) &
         Ours (small model) &
         Blender Cycles \\
         (4096 spp, 247 s) &
         (16 spp, 34 s) &
         (equal-time, 34 s) &
         (16 spp, 1.7 s on CPU) &
         (16 spp, 0.9 s on GPU) &
         (equal-time, 1 s) 
            \\
         \includegraphics[trim={100 100 100 100},clip,width=0.1666\textwidth]{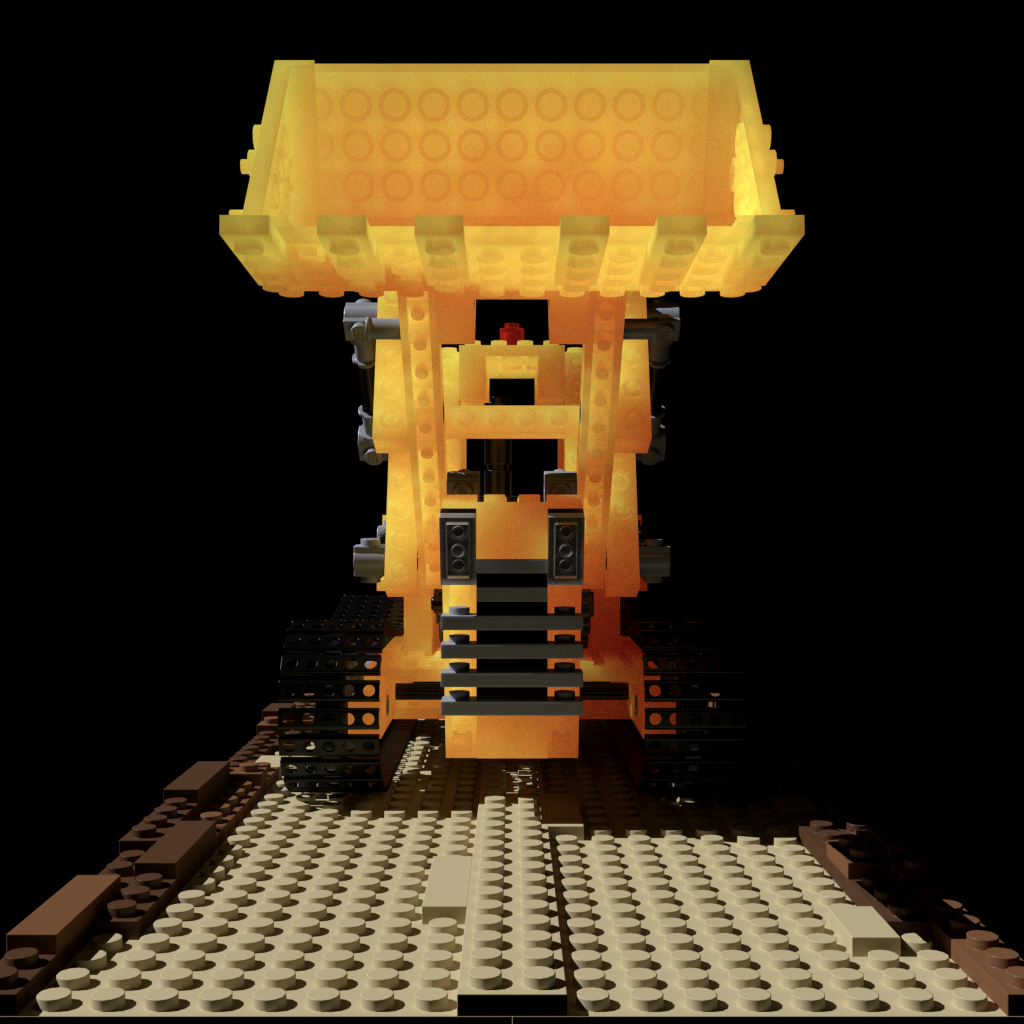} & 
         \includegraphics[trim={100 100 100 100},clip,width=0.1666\textwidth]{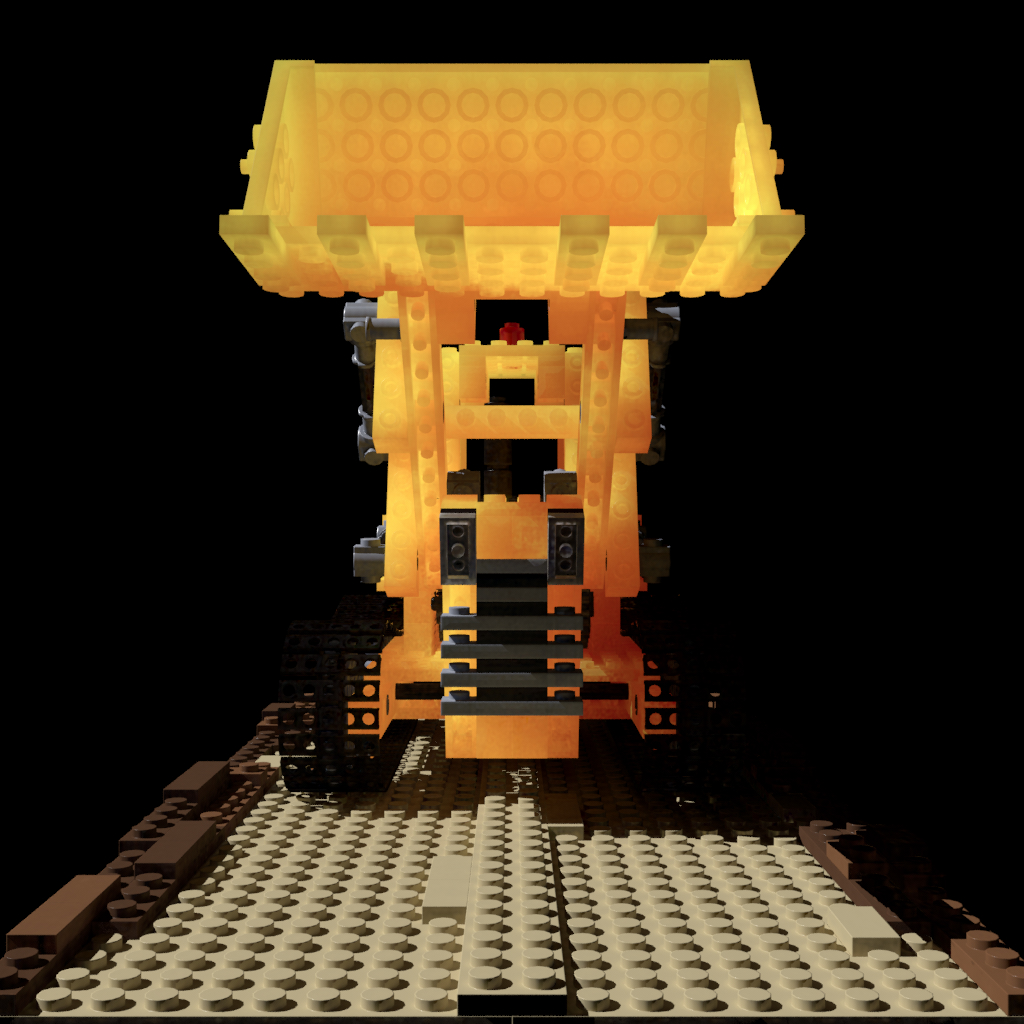} & 
         \includegraphics[trim={100 100 100 100},clip,width=0.1666\textwidth]{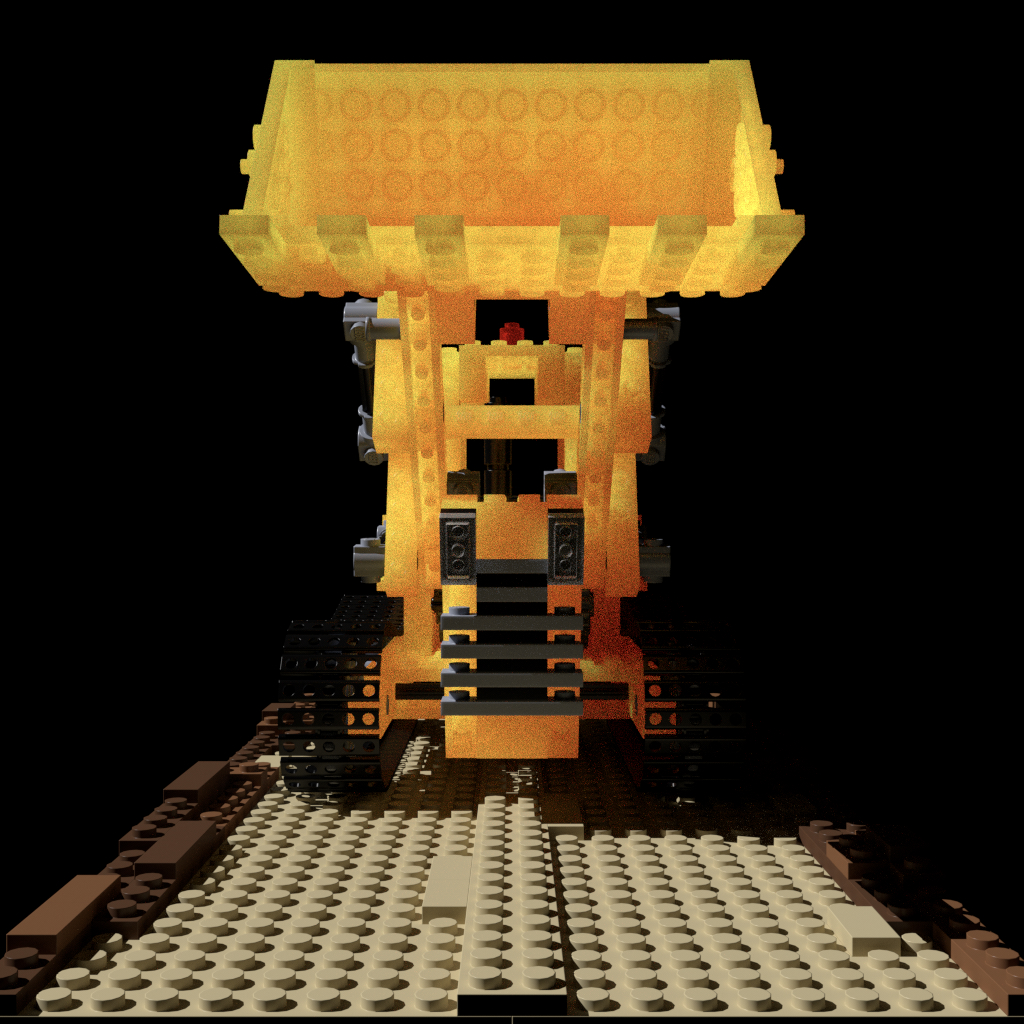} & 
         \includegraphics[trim={100 100 100 100},clip,width=0.1666\textwidth]{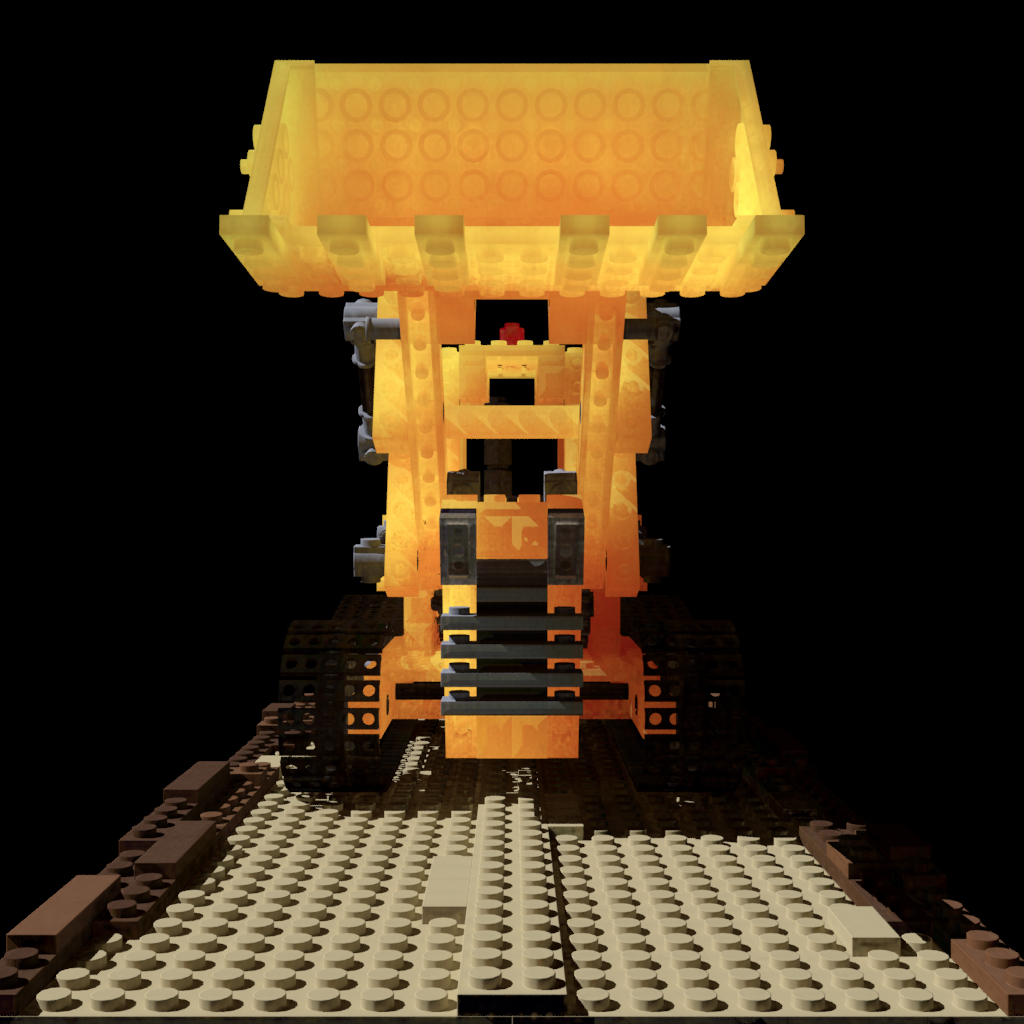} & 
         \includegraphics[trim={100 100 100 100},clip,width=0.1666\textwidth]{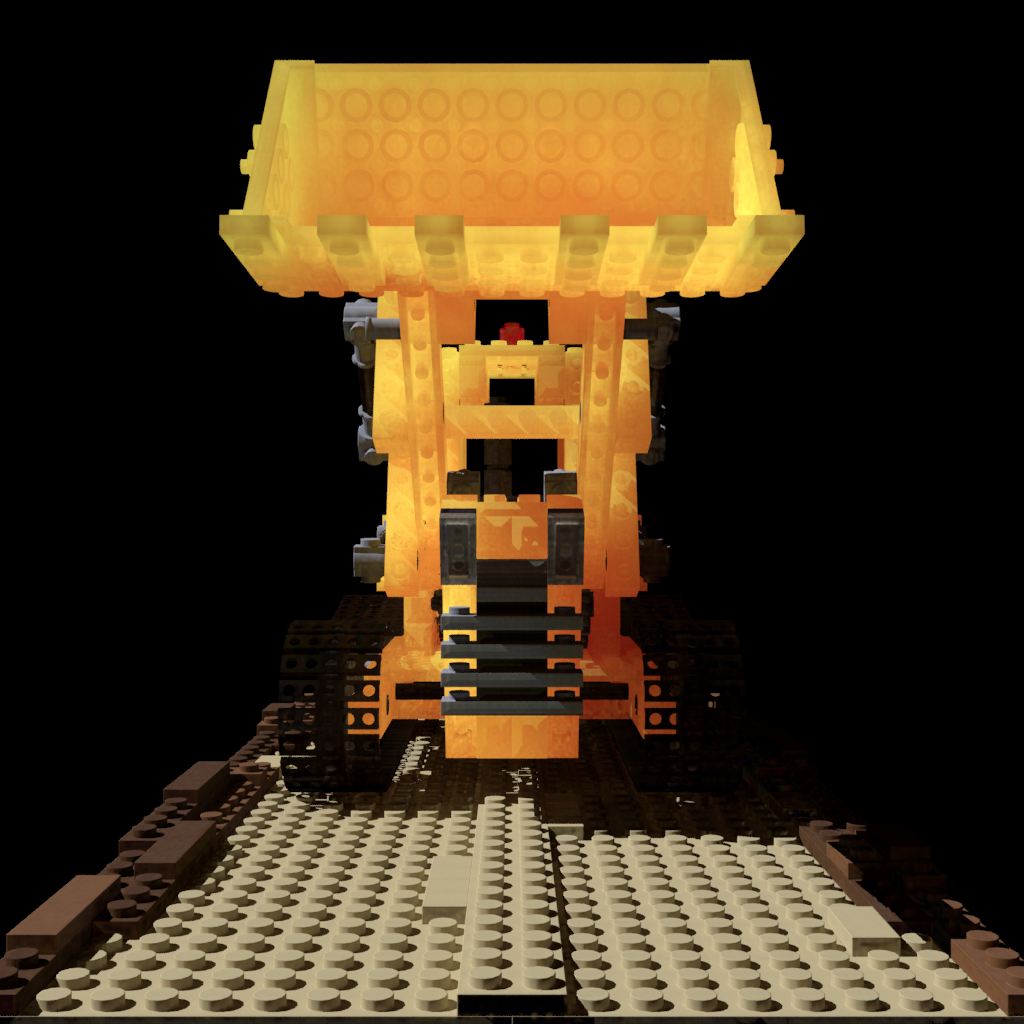} &
         \includegraphics[trim={100 100 100 100},clip,width=0.1666\textwidth]{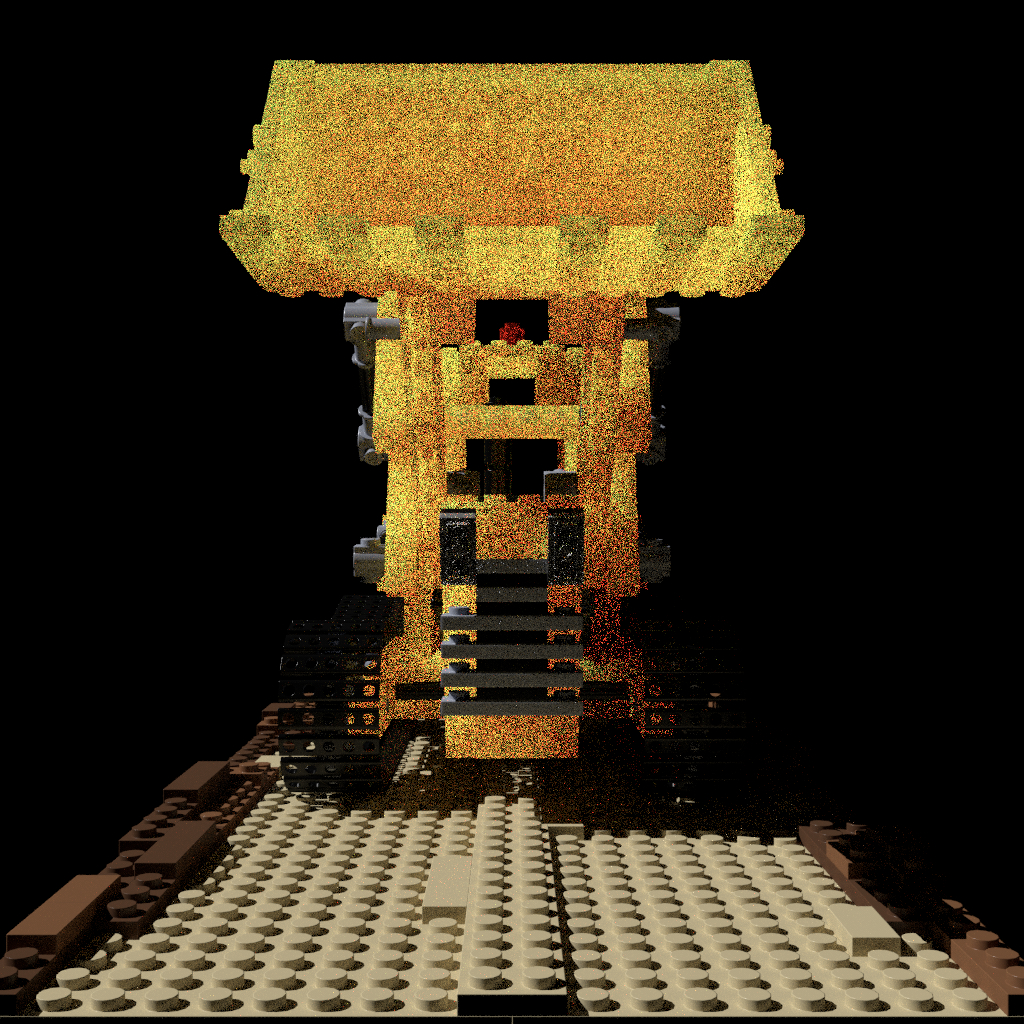}
          \\
         \includegraphics[trim={400 400 400 400},clip,width=0.1666\textwidth]{figs/eclair_perf/lego/blender_cpu_4kspp_247s.jpg} & 
         \includegraphics[trim={400 400 400 400},clip,width=0.1666\textwidth]{figs/eclair_perf/lego/eclair_heavy_cpu_16spp_34s_2150ms.jpg} & 
         \includegraphics[trim={400 400 400 400},clip,width=0.1666\textwidth]{figs/eclair_perf/lego/lego_blender_34s.jpg} & 
         \includegraphics[trim={400 400 400 400},clip,width=0.1666\textwidth]{figs/eclair_perf/lego/eclair_rt_16spp_1.7s_105ms.jpg} & 
         \includegraphics[trim={400 400 400 400},clip,width=0.1666\textwidth]{figs/eclair_perf/lego/eclair_rt_gpu_16spp_0.9s_46ms.jpg} &
         \includegraphics[trim={400 400 400 400},clip,width=0.1666\textwidth]{figs/eclair_perf/lego/lego_blender_1s.jpg}
    \end{tabular}
    }
    \caption{{\bf Path tracer integration on surface-based asset.} We integrate surface models into a production path tracer on both CPU and GPU rendering. Our model significantly simplifies the shader implementation and improves the rendering performance. In contrast, blender path tracing exhibits severe Monte Carlo noise at equal rendering time budget.}
    \label{fig:eclair_perf_surface}
\end{figure*}

%% file: chapters/eclair_hair.tex
\begin{figure*}
    \centering
    \setlength\tabcolsep{3.0pt}
    \resizebox{1.0\textwidth}{!}{
    \begin{tabular}{c|c|c|c|c}
         Blender Cycles &
         Ours &
         Blender Cycles &
         Ours (small model) &
         Blender Cycles \\
         (4096 spp, 1 hour) &
         (16 spp, 74.4 s) &
         (equal-time, 74.6 s) &
         (16 spp, 17.7 s on CPU) &
         (equal-time, 17.7 s) 
            \\
         \includegraphics[width=0.1999\textwidth]{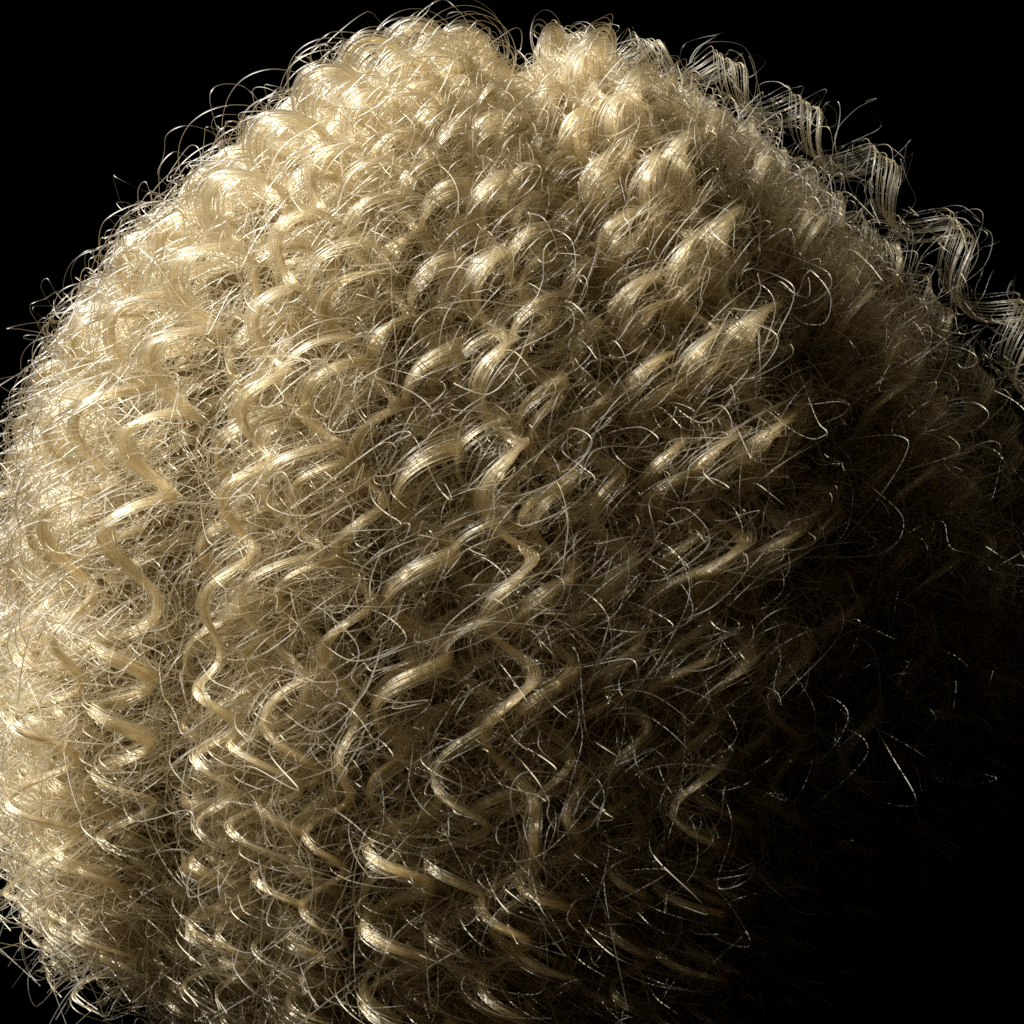} & 
         \includegraphics[width=0.1999\textwidth]{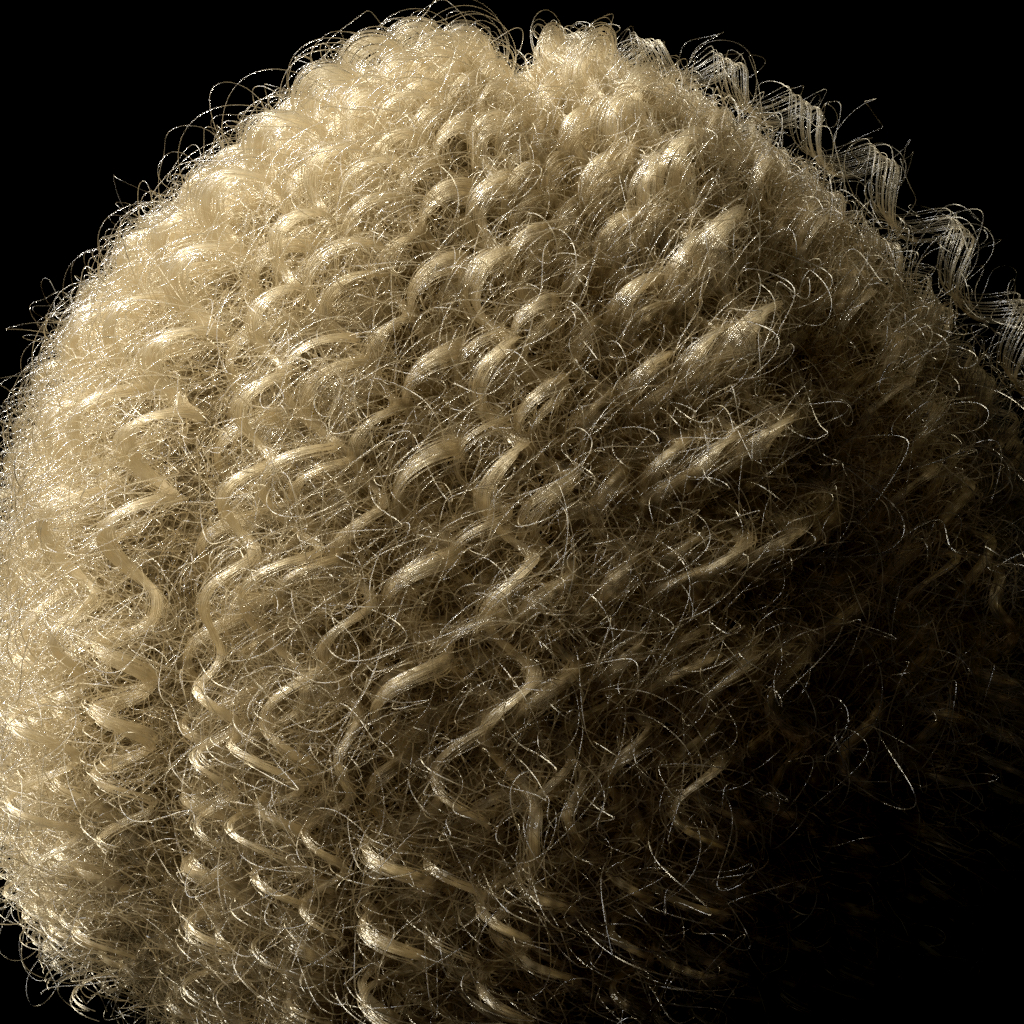} & 
         \includegraphics[width=0.1999\textwidth]{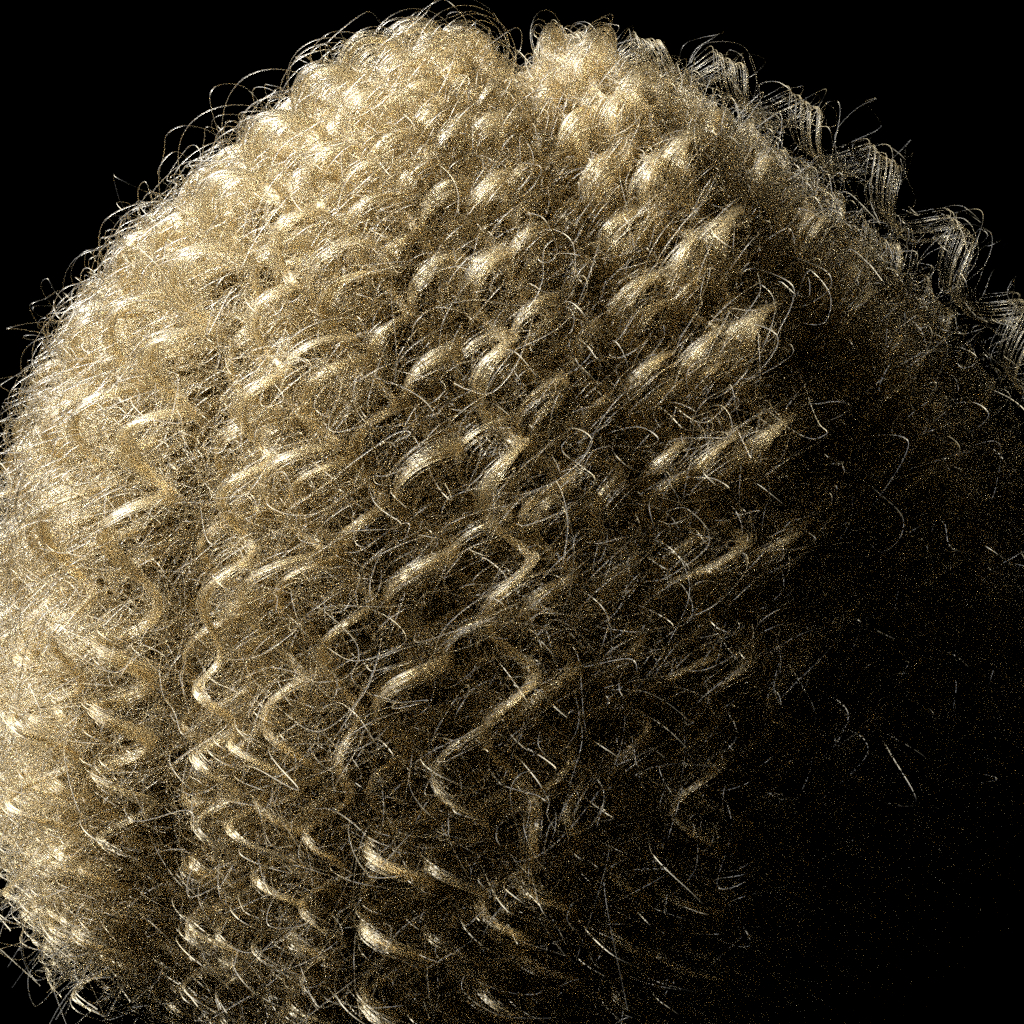} & 
         \includegraphics[width=0.1999\textwidth]{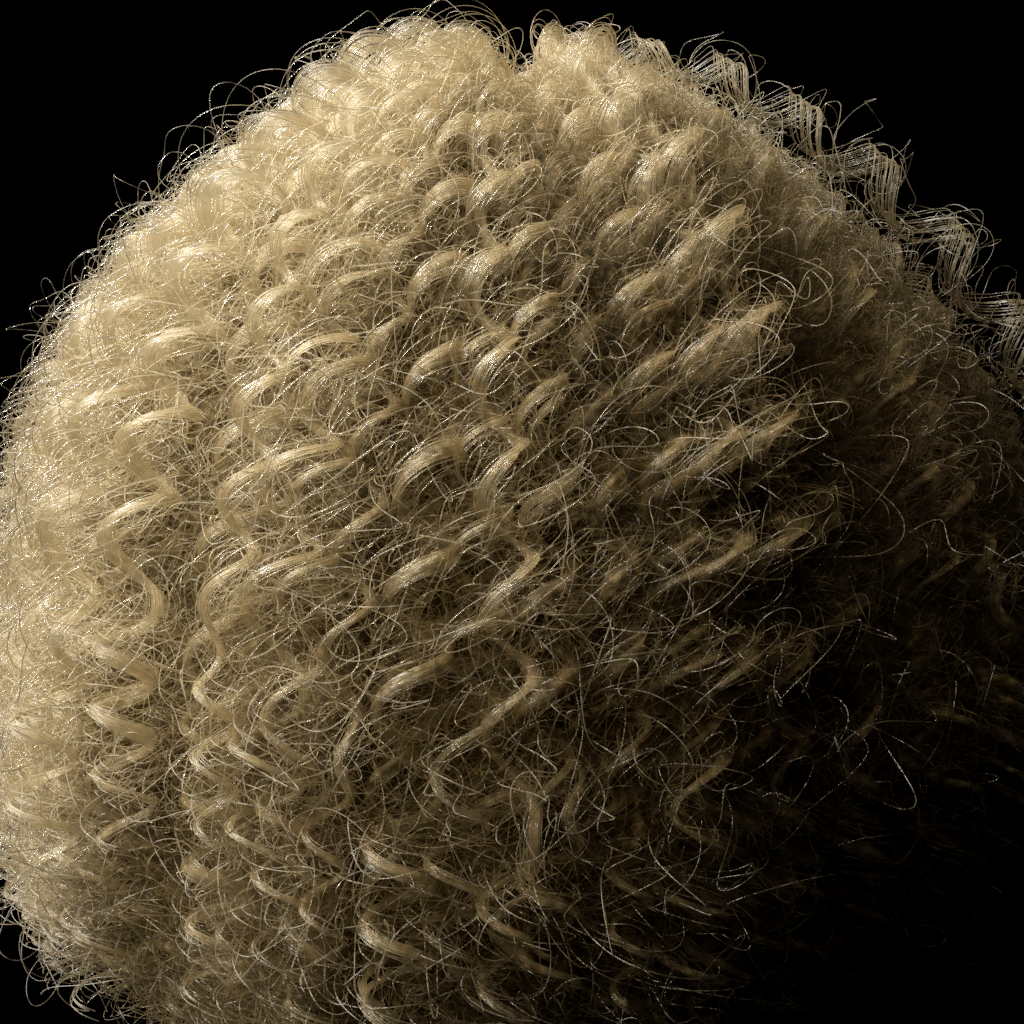} & 
         \includegraphics[width=0.1999\textwidth]{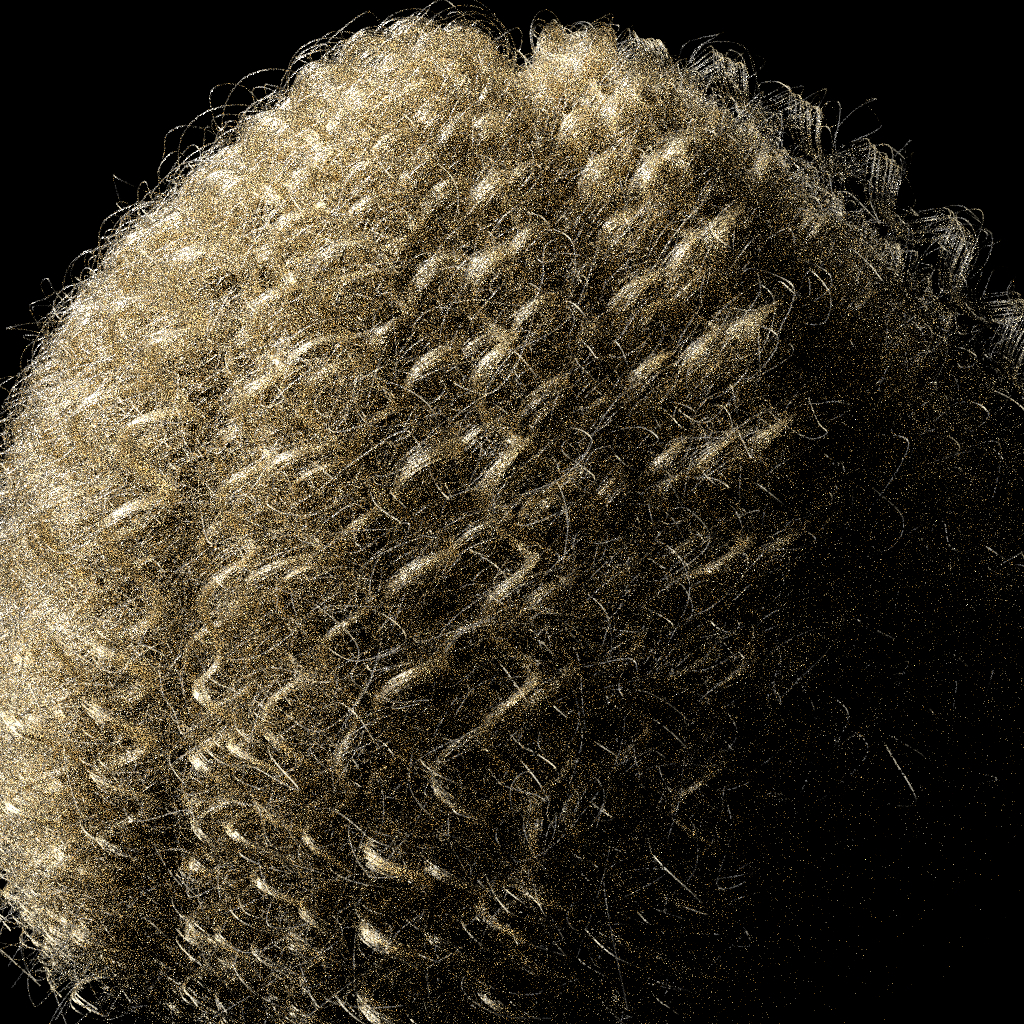} 
          \\
         \includegraphics[trim={0 0 600 600},clip,width=0.1999\textwidth]{figs/eclair_perf/hair/hair_blonde_blender_cpu_4kspp_59m30s.jpg} & 
         \includegraphics[trim={0 0 600 600},clip,width=0.1999\textwidth]{figs/eclair_perf/hair/hair_blonde_eclair_cpu_16spp_74.4s_4650ms.jpg} & 
         \includegraphics[trim={0 0 600 600},clip,width=0.1999\textwidth]{figs/eclair_perf/hair/hair_blonde_blender_74.6s.jpg} & 
         \includegraphics[trim={0 0 600 600},clip,width=0.1999\textwidth]{figs/eclair_perf/hair/hair_blonde_ec_cpu_rt_17.7s_1105ms.jpg} & 
         \includegraphics[trim={0 0 600 600},clip,width=0.1999\textwidth]{figs/eclair_perf/hair/hair_blonde_blender_17.7s.jpg}
    \end{tabular}
    }
    \caption{{\bf Path tracer integration on fiber-based asset.} We integrate hair models into a production path tracer on CPU rendering. Our model significantly simplifies the shader implementation and improves the rendering performance. In contrast, blender path tracing exhibits severe Monte Carlo noise at equal rendering time budget. 
    }
    \label{fig:eclair_perf_fiber}
\end{figure*}

%% file: chapters/render.tex

\section{Rendering}
\label{sec:render}

The relightable neural asset can be integrated within a full-featured production path tracer. We demonstrate an integration of our asset in a path tracer that allows for environment (IBL) lighting, local area lighting, other geometries in the scene, and global illumination (Figure \ref{fig:teaser}, \ref{fig:eclair_perf_surface}, \ref{fig:eclair_perf_fiber}, \ref{fig:relighting} and \ref{fig:relighting2}). 

If a ray intersects the geometry, the shading on the neural asset is computed according to \eqnref{eqn:neuraltransport}.
The original materials are not needed; the neural asset is fully defined by its geometry, feature triplane and MLP weights. Anti-aliasing is applied as usual in the renderer, as our assets encode point-wise rather than pixel-aggregated transport. A rasterizer integration may be possible as well, but we have not studied it yet.

Rendering with a point or directional light simply requires a feature grid query and an MLP evaluation for each hit point. For other illuminations, Monte Carlo light sampling can be used.
Indirect rays can be traced from our asset as well, currently based on uniform sampling. Neural importance sampling could be applied in the future, by adapting some of the methods discussed by \citet{NeuSample}. We combine the light samples and indirect samples using multiple importance sampling (MIS).

The implementation needs to take care to recreate the same conditions that the asset has been trained in. All vectors need to be converted to training space (normally equivalent to object space) since this is what the neural model will assume.

Below, we discuss some subtleties of integrating our neural assets into a production path tracer that may not be immediately obvious. Figure~\ref{fig:eclair_perf_surface} and \ref{fig:eclair_perf_fiber} show the correctness of our path tracer integration by comparing to ground truth renders generated using Blender Cycles.

\subsubsection*{Visibility hint application}

In production renderers, BSDF evaluation is frequently done before a shadow ray is traced for the shading point. For example, in frameworks like OptiX and Vulkan raytracing, BSDF evaluation is normally done in a \emph{closest hit shader}, while visibility checking requires a subsequent \emph{ray generation shader} to construct the appropriate shadow rays, and the ordering of these shaders would be very tedious to change. Our model essentially replaces the BSDF evaluation in the shading pipeline, so the visibility being unknown at this point is a practical obstacle. To resolve this efficiently, we designed our MLP to output both values (shadowed and unshadowed); we simply hold on to the two values on the ray payload and pick one of these values later on based on the results of the visibility checking.

If the visibility hint is used as an additional MLP input, such as in the assets produced by \citet{NRHints}, we can still use this approach, though at the cost of double shading cost: we can evaluate the MLP twice for both cases, put the values on the ray payload and continue as above.

\subsubsection*{Correct light transport handling}

The neural representation $\asset$ models all transport within the asset; multi-bounce Monte Carlo simulation is only required for the transport between different assets. A shadow ray or a secondary indirect ray should not be occluded by the asset itself, but still intersects the geometries of other assets to allow for inter-asset global transport in the scene.

In practice, for both rays (shadow and indirect), self-intersection is tracked by comparing the instance ID (unique per asset) on the origin of the ray to the ID on the hit point. If a hit is a self-intersection, we mark the visibility hint as zero on the ray payload and continue the ray until it hits the light, or any other asset, or exits the scene. Note that the visibility hint needs to be applied accordingly on both the shadow ray and the indirect ray.

This accounts for correct global illumination, ignoring multiple contributions from the neural asset, which have already been handled in precomputation. Shadow rays need to use closest-hit (rather than any-hit) queries to implement this logic correctly. Multiple importance sampling (MIS) can be used to weight the direct and indirect ray as usual.

%% file: chapters/results.tex

\section{Results}
\label{sec:result}

Our novel relightable neural asset model demonstrates versatility and precision across a variety of materials and lighting conditions. We showcase our method on four surface-based and three fiber-based assets, showing comparisons to a path-traced reference and baselines from previous work \cite{NRHints,bi2020}. The supplementary video shows animated relighting results, as well as an integration into an interactive path tracer. 

The path-traced reference images are rendered with the Blender Cycles renderer, while ours are rendered by evaluating the fitted neural model in PyTorch. This is separate from the full path-tracer integration shown in Figure \ref{fig:teaser} and involves a deferred shading pipeline that uses the Cycles renderer to generate the deep buffers for inputs to our neural model and a forward pass in PyTorch for model evaluation. The buffers are $4 \times 4$ super-sampled for anti-aliasing. This PyTorch rendering pipeline serves to validate the correctness of the fitted neural models.

\paragraph*{High quality and small models.}
In the following results, we showcase two model sizes using our method: the \emph{high-quality} and \emph{small} models. The only difference between the two models is the size of the MLP, the high-quality model having $512$ neurons in each hidden layer, while the small model having only $64$. The larger MLP size allows for more network capacity with high-fidelity results close to ground truth, while the smaller MLP allows for a high-performance version trading off some quality; even running in real-time on the GPU for most assets.

\subsection{Rendering Performance}

All our renderings are done on a Windows 11 workstation with an AMD Threadripper 3990x 64 core CPU (128 threads), 128GB of RAM and an Nvidia Quadro RTX A6000 GPU. As mentioned earlier, the results generated with the PyTorch deferred shading pipeline are only intended for correctness validation, so we focus on the rendering performance of our path tracer integration.

Figure~\ref{fig:eclair_perf_surface} and \ref{fig:eclair_perf_fiber} show the performance gained by leveraging our neural asset representation to encode complex light transport on the surface of assets, especially with lots of scattering effects in assets like the Lego and Blonde Hair shown. Our path tracer integration in addition to the CPU-based backend, includes a Vulkan-based GPU backend support for surface-based assets; so we are able to verify real-time ray tracing performance on these assets. Both assets are rendered at a resolution of $1024x1024$ for the performance metrics shown.

The Lego, rendered with a single directional light in Blender Cycles on the CPU to $4096spp$ takes $247s$. It is to be noted that even at $4096spp$, there is still some residual Monte-Carlo noise remaining, but we use this as reference as our training data for this model was rendered to the same sample count. Rendering with our model, using the same scene setup in our CPU-based path tracer to $16spp$, the general number of samples needed for anti-aliasing, takes $34s$, which is over a $7x$ performance gain. The performance gain is even more apparent with our small model, which can render the Lego asset to $16spp$ instantly (with the loss of some quality in scattering); with the CPU-backend taking an average of $105ms$ per sample, and on the GPU $46ms$, including ray tracing time. We also show the noisy renders that can be obtained on path-tracing the original non-neural asset to an equal time in Blender Cycles.

The Blonde Hair asset tells a similar story, but with even larger performance gains: up to $60x$ on the high-quality model, and $200x$ with the small model because one of the major benefits of our model is that it cuts down on ray tracing time by not having to trace any indirect rays inside the asset for multiple scattering. The loss of some definition in the highlights is visible in the small model, however, the overall appearance is preserved. This asset is particularly heavy containing $41,164$ hair strands, with $8.2$ million vertices; hence, the CPU path tracer needs about $1.1s$ per sample. Our GPU backend does not support curves, so we do not have an estimate of speedup for fiber-based assets on the GPU. 

\subsection{Comparison to NRHints on surface assets}

Figure \ref{fig:surface_nrhints} shows a comparison of the path-traced reference (left) to our results (middle, both high-quality and small models) and NRHints \cite{NRHints} (right).

We show renderings across four distinct surface-based assets: Subsurface Lego, Jug and Dice, Ten24 Head and Flowers. These assets have been specifically selected to underscore our model's ability to handle intricate light transport scenarios, including phenomena such as translucent subsurface scattering, complex self-occlusion, and multiple interreflections. One of the assets is from the NRHints paper (Jug and Dice, top row). The Subsurface Lego features strong translucent appearance, modified from the original NeRF Lego scene~\cite{mildenhall2021nerf}. The Ten24 Head presents a complex dermis skin shader and layered BSSRDFs and BRDFs. The Flowers scene shows a combination of translucency and complex intricate geometry.

Our model faithfully replicates the path-traced reference, confirming its potential to operate effectively under unfamiliar light/view conditions. Our relightable neural asset model successfully captures the soft translucent appearance as well as the high-frequency details of these surface assets.

The PSNR values achieved by both methods are specified within each image. Overall, even our real-time model performs better than \citet{NRHints} despite its smaller MLP size, and our high-quality model is consistently better. Specifically, both on zoomed-out views, and when zooming in on details, our method achieves better PSNR values. NRHints has worse depiction of geometric details (Lego and Flowers) and high frequency texture/shading details (jug/dice and head). This is despite NRHints using more training images (1,000) than our results (400), and requiring highlight hints (while our method only uses shadow/visibility hints). The improvement is due to several factors: our ground-truth geometry, triplane feature grid, and per-pixel randomized light directions in training data.

\input{chapters/compare_nrhints}

\input{chapters/compare_sai}

\subsection{Comparison to NRHints on fiber assets}

In Figure \ref{fig:fiber_nrhints}, we show a comparison of our method to path traced reference and NRHints \cite{NRHints} on fiber-based assets (hair and fur), where the effects of detailed geometry and multiple-scattered light paths are even more significant than for surface assets. This is reflected in the corresponding PSNR values, but (in our opinion) it is even more true perceptually.

Our asset representation is sufficiently versatile to replicate widely varying fiber effects. For instance, we showcase two hair models, Blonde Hair and Brown Hair, with dramatically different material settings. The Blonde Hair exhibits a much lighter color due to pronounced multiple scattering, while the Brown Hair is characterized by dominant low-order scattering, accompanied by a distinct secondary (transmit-reflect-transmit) highlight, an important effect detailed by \citet{marschner2003light} that our model reproduces without issues.

Even though \cite{NRHints} learns plausible shading cues and color (for example, the pink highlights in the blonde hair), the results are blurry due to fine geometric details not being reproducible. On the other hand, our results reproduce even the fine structured ``glinty'' appearance of the fibers. Our PSNR values are consistently better than NRHints, even for our \textit{small} model, which uses a much smaller neural network than NRHints.

Our method not only uses the ground truth fiber geometry, but also accurately models radiance variation across fiber width; we believe ours is the first neural 3D representation that captures fiber shading at this level of accuracy.

\subsection{Comparison to Neural Reflectance Fields}

In Figure \ref{fig:comparison_sai}, we show a comparison with Neural Reflectance Fields~\cite{bi2020}, which fits views with collocated point lighting. We chose one surface and one fiber based scene, and compare the methods with both collocated and non-collocated novel point lighting. In all cases, our method (both high-quality and small models) clearly outperforms \citet{bi2020}, as their method approximates the shading by estimating the parameters of a simple surface BRDF model (or Kajiya-Kay hair model for the fiber asset). Even with collocated lighting (first and third row), which matches its lighting assumptions, \citet{bi2020} is not able to recover the same amount of detail and contrast in the shading as our method.

\subsection{Comparison to geometry-aware variants of NRHints}

In Figure \ref{fig:nrhints_geometry}, we show a comparison of our method against two variations of NRHints that benefit from ground-truth geometry, to find out how much better NRHints could perform if information from ground-truth geometry was available (as is normally the case for synthetic assets).

The first variant takes our pipeline with surface-based primary intersections on ground-truth geometry. The inputs to the network remain the same; however, we replace our triplane and MLP architecture with that of NRHints, using a large MLPs with frequency encoding on the inputs. Not only does our method produce results with higher visual quality, it also has faster run-time and faster training time due to our MLPs being smaller and with fewer optimizable parameters, showing the benefit of our triplane and MLP architecture.

The second variant uses the NRHints pipeline and volume rendering formulation, but introduces signals derived from ground-truth geometry to aid the training process, in the form of depth and alpha-mask losses. The depth and alpha masks are rendered alongside the multi-view training RGB images and included as additional inputs to the NRHints network. Though there is some benefit derived from these signals, our method still outperforms this variant in terms of visual quality and speed.

\input{chapters/relighting}

\input{chapters/relighting2}

\subsection{Showcase of our relighting results}
In addition, we present our model's capability to relight surface and fiber assets under varying lighting conditions. As illustrated in Figure~\ref{fig:relighting}, we render a translucent Lego and a basket of flowers, both with subsurface scattering, under four distinct IBL illuminations, each with a shadow-casting ground plane. The selected IBL environments—\emph{Lake}, \emph{Office Hallway}, \emph{City Plaza}, and \emph{Above the Clouds}—encompass a diverse array of lighting scenarios, both indoor and outdoor. Our relightable neural assets consistently demonstrate the visual transformation of a single asset under varied illuminations but also emphasize the rich detail and photorealism that can be achieved, highlighting the intricate subtleties of light interaction with complex illumination and materials. 

In Figure~\ref{fig:relighting2}, we further demonstrate our model's relighting capability on fiber assets. Four distinct hair models are each illuminated by an outdoor IBL (\emph{Sunset}) and an indoor IBL (\emph{Studio Light}). Our model is capable of accommodating novel perspectives while accurately responding to illumination changes with a complex interplay of highlights, textures, self-shadowing and multiple fiber scattering. We believe no neural rendering methods for full 3D assets have shown comparably powerful relighting ability under arbitrary views.

\input{chapters/nearfield}

\input{chapters/ablations}

\subsection{Discussion and limitations}

In summary, these results affirm our model's robustness in managing diverse lighting conditions and viewpoints, as well as its precision in modeling an array of surface types and complex light transport phenomena. We believe our approach has broad applicability, combined with its high performance, positions is as potentially broadly applicable in practice.

Several limitations exist and we would like to address them in future work.
One limitation of our current path-tracer integration is that we use simple uniform sampling as an approximation, similar to NeuMIP \cite{kuznetsov2021neumip}. A more advanced importance sampling module could be trained, similar to \citet{NeuSample} or \citet{NvidiaNeuralMaterials}. These approaches would require a new sampling neural network, but could reuse the same feature grid from our method.

Furthermore, while our method supports specular objects (e.g. the skin material is fairly specular), very low roughness (close to mirror or glass) materials will show blurring in our fits. Our assets are currently static; animated assets would be an interesting future challenge, and could be supported through mapping shading queries into a canonical pose and training over different poses ins addition to camera and light directions. Furthermore, while primary ray-tracing of the explicit geometry is cheaper than full light transport, it may still be expensive for some applications; this could be addressed by an approximate proxy geometry.

%% file: chapters/compare_nrhints.tex
\begin{figure*}
    \centering
    \includegraphics[width=0.9\textwidth]{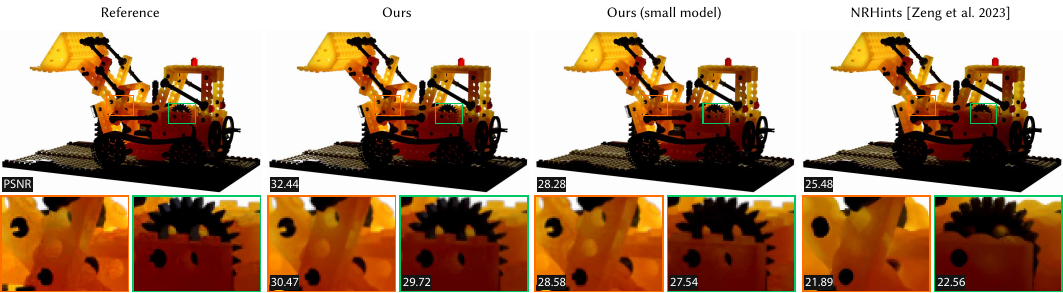}
    \includegraphics[trim={0 0 0 8},clip,width=0.9\textwidth]{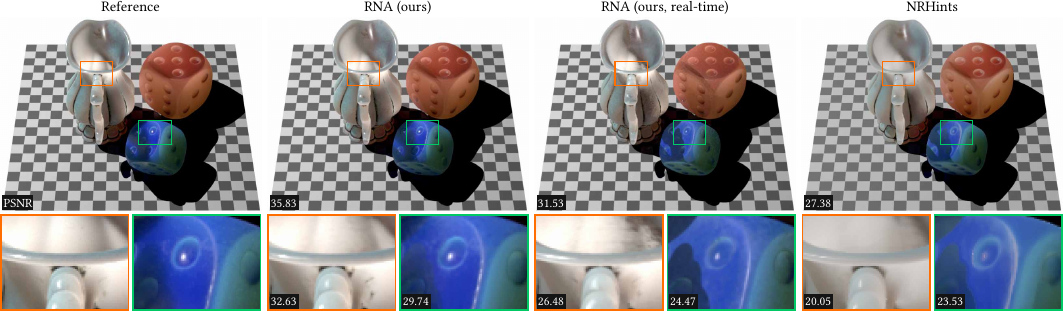}
    \includegraphics[trim={0 0 0 8},clip,width=0.9\textwidth]{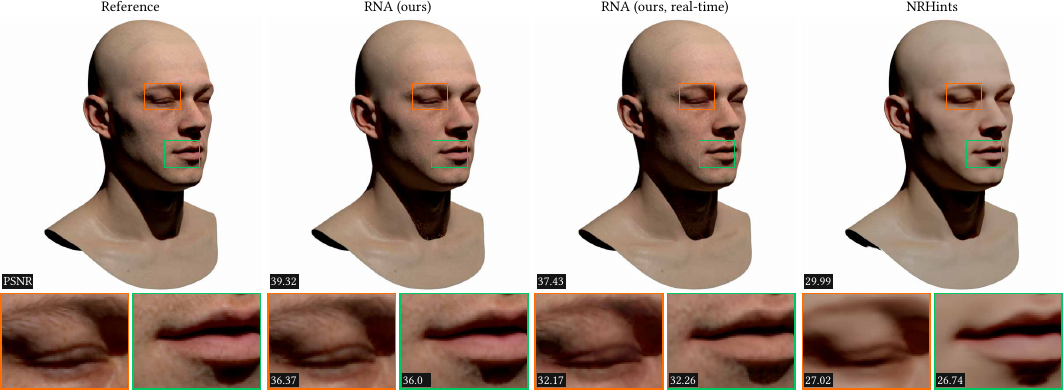}
    \includegraphics[trim={0 0 0 8},clip,width=0.9\textwidth]{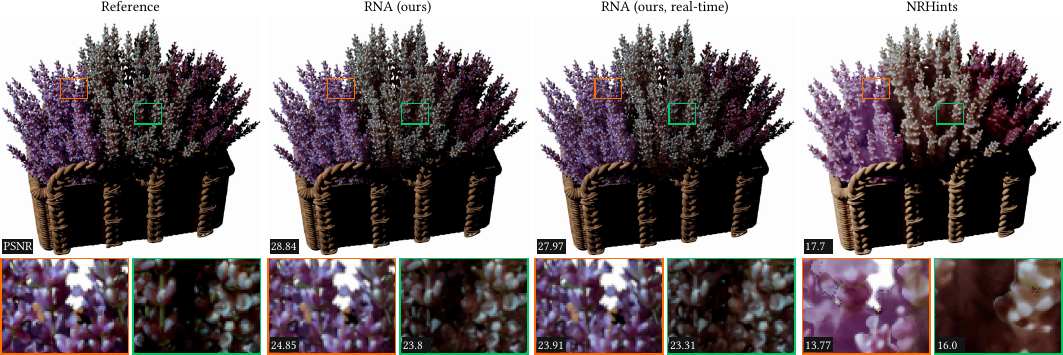}
    \caption{{\bf Comparisons with NRHints~\cite{NRHints} on surface-based assets.} We show a comparison with four mesh-based assets under point light illumination. All of them show effects of long light paths due to subsurface scattering applied to varying degrees. One of the assets is from the NRHints paper (jug and dice, second row). Overall, even our real-time model performs as well or better than \citet{NRHints}, and our high-quality model is consistently better. NRHints has worse depiction of geometric details (lego and flowers) and high frequency texture/shading details (jug/dice and head).}
    \label{fig:surface_nrhints}
\end{figure*}

\begin{figure*}
    \centering
    \includegraphics[width=1\textwidth]{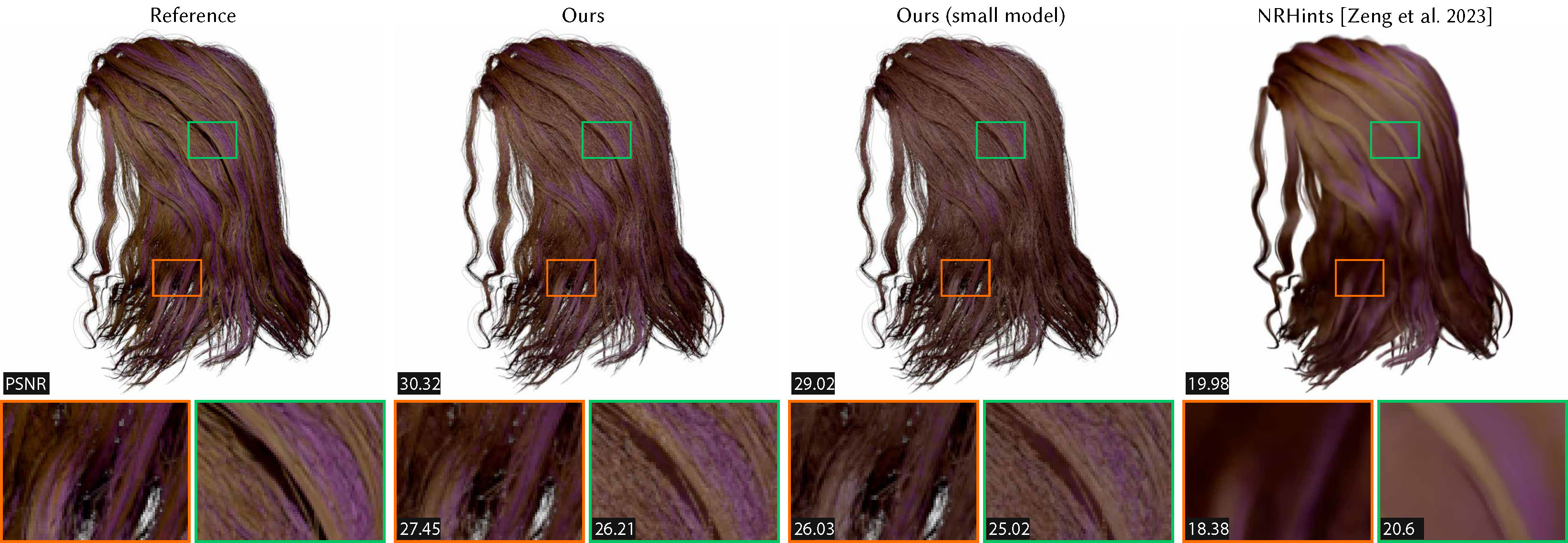}
    \includegraphics[trim={0 0 0 8},clip,width=1\textwidth]{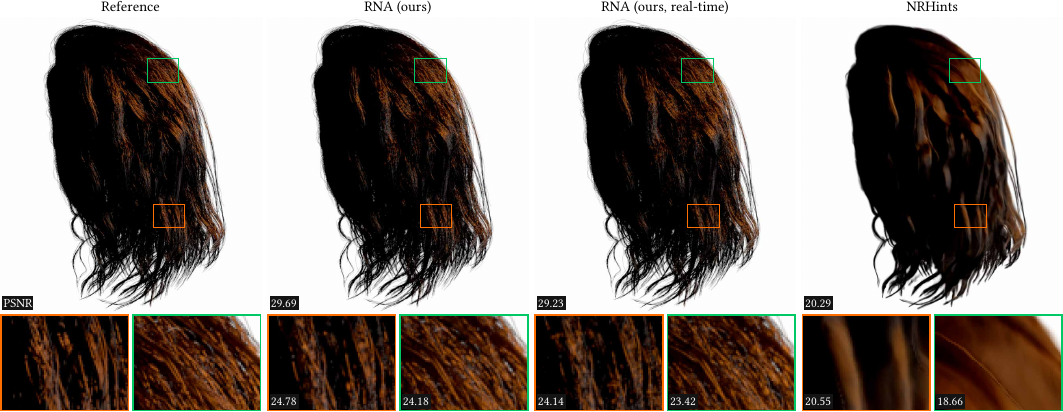}
    \includegraphics[trim={0 0 0 8},clip,width=1\textwidth]{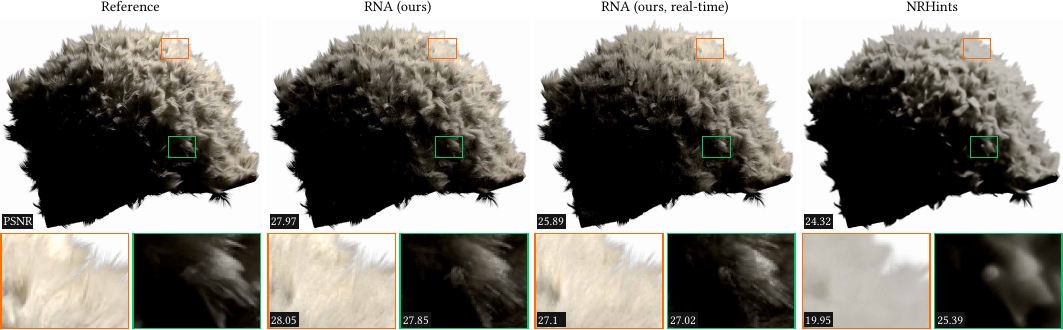}
    \caption{{\bf Comparisons with NRHints~\cite{NRHints} on fiber-based assets.} Fiber-based assets show even more clearly the advantages of our method. Even though \cite{NRHints} learns plausible shading cues and color (for example, the pink highlights in the blonde hair), the results are blurry due to fine geometric details not being reproducible. On the other hand, our results reproduce even the fine structured ``glinty'' appearance of the fibers, since our method not only uses the ground truth fiber geometry, but also accurately models radiance variation across fiber width. Moreover, the comparison results are rendered under point light illumination to align with the capabilities of the NRHints architecture, which is currently designed to support point lights.
    }
    \label{fig:fiber_nrhints}
\end{figure*}

\begin{figure*}
\centering
\includegraphics[width=1.0\textwidth]{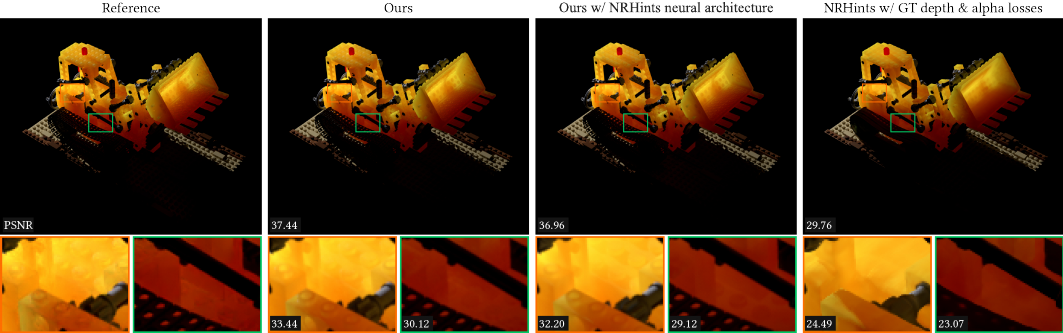}
\includegraphics[width=1.0\textwidth, clip, trim={0cm 0cm 0cm 0.3cm}]{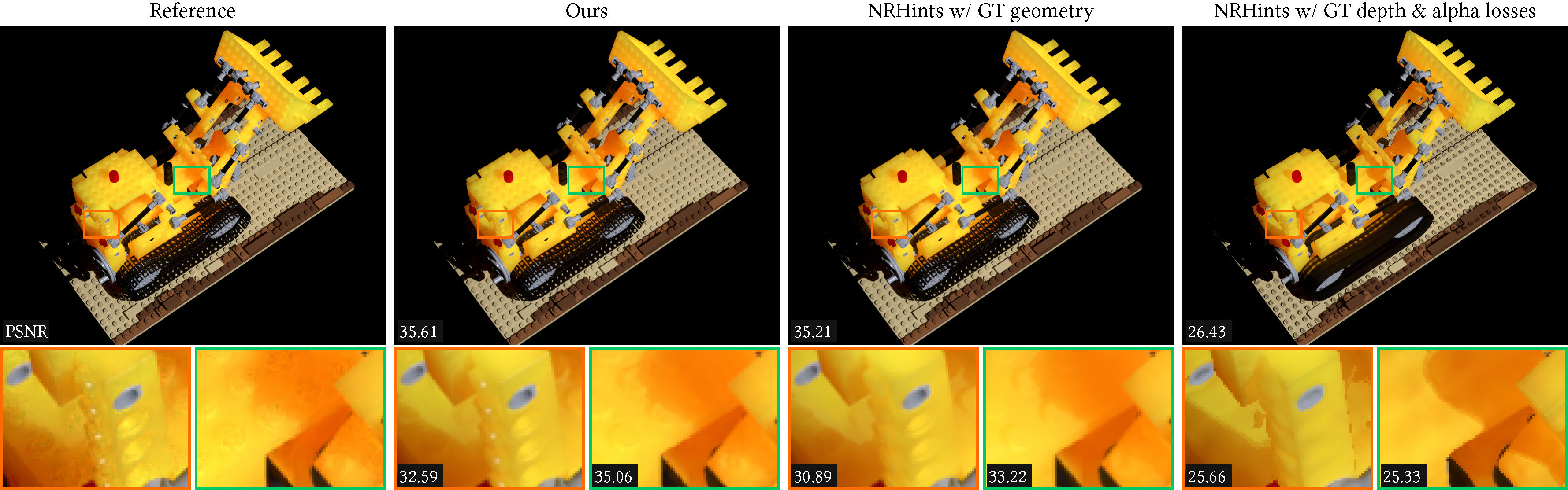}
\caption{\textbf{Comparison with two geometry-aware variants of NRHints~\cite{NRHints}.} In this comparison, we assess the performance of our method against two variants of NRHints~\cite{NRHints} that benefit from ground truth geometry. The first variant, denoted as \emph{Ours w/ NRHints neural architecture}, replaces our tri-plane and MLP neural architecture with the MLP and encoded inputs architecture of NRHints. This shows the benefit of our neural architecture, while keeping the surface-based formulation with ground truth geometry.
The second variant, referred to as \emph{NRHints w/ GT depth \& alpha losses}, modifies the NRHints pipeline by including losses to the training pipeline derived from ground-truth depth and alpha masks, which are good proxies for ground-truth geometry. Our framework consistently outperforms both variants in terms of quality, exhibiting superior inference and training speed and a reduced number of optimizable network parameters.}
\label{fig:nrhints_geometry}
\end{figure*}

%% file: chapters/compare_sai.tex
\begin{figure*}
    \centering    
    \includegraphics[width=1\textwidth]{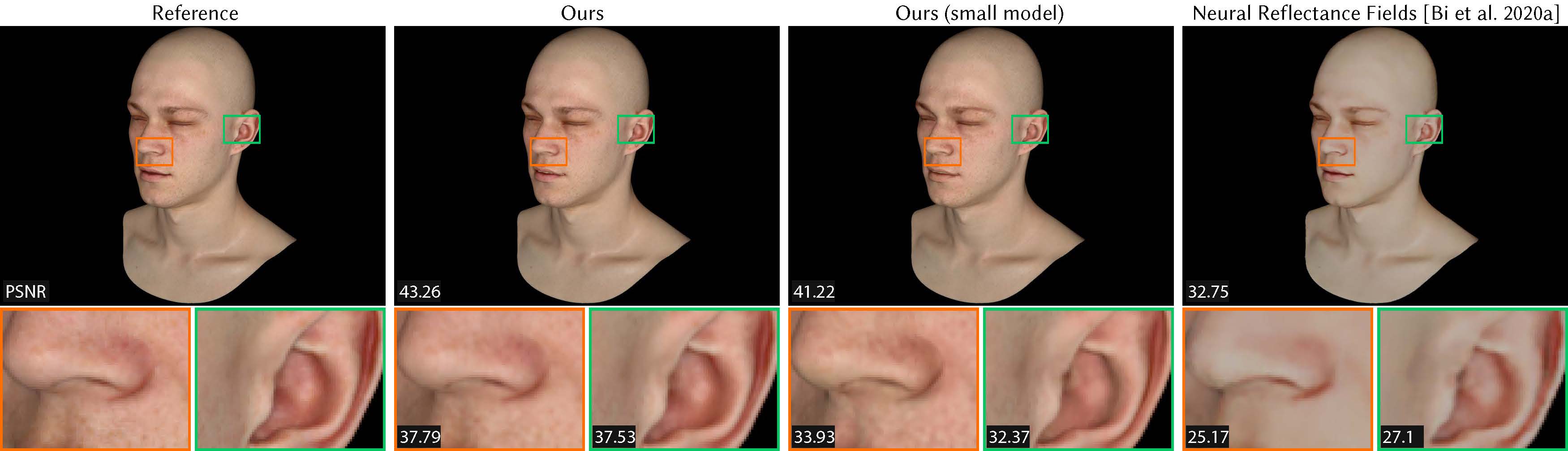}
    \includegraphics[trim={0 0 0 8},clip,width=1\textwidth]{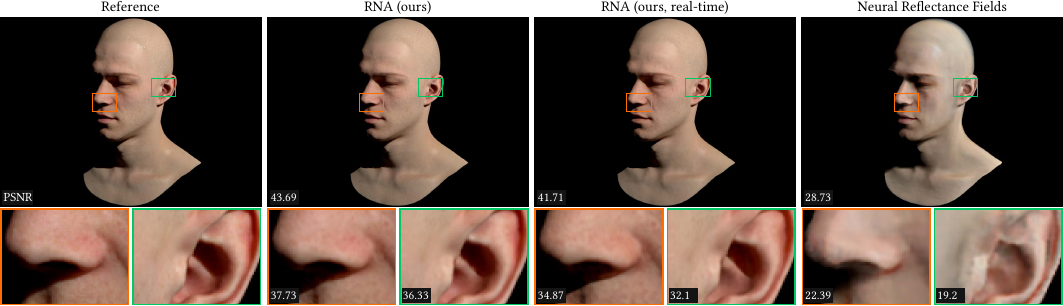}

    \includegraphics[trim={0 0 0 8},clip,width=1\textwidth]{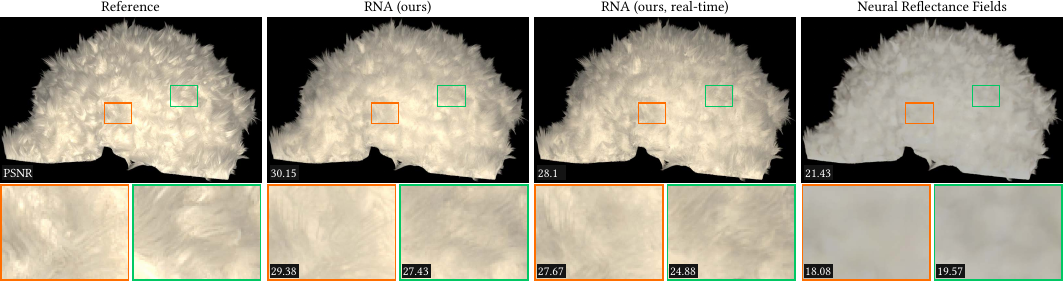}
    \includegraphics[trim={0 0 0 8},clip,width=1\textwidth]{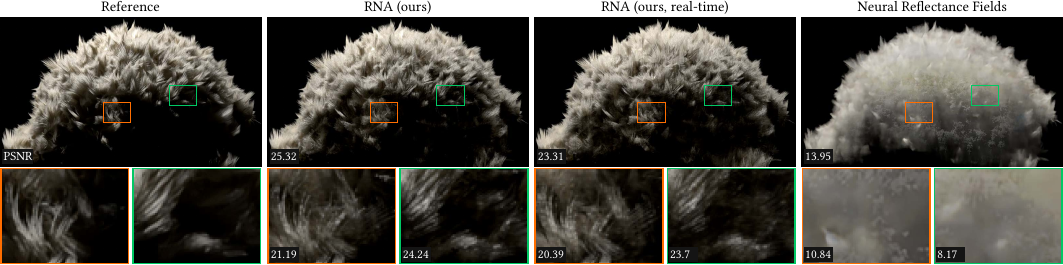}
    
    \caption{{\bf Comparison with Neural Reflectance Fields}~\cite{bi2020}, which fits views with collocated point lighting. We compare the methods with both collocated and non-collocated novel point lighting. In both cases, our method (both high-quality and real-time models) clearly outperforms \citet{bi2020}, as their method approximates the shading by estimating the parameters of a simple surface BRDF model (or Kajiya-Kay hair model for fibers). Even with collocated lighting (first and third row), \citet{bi2020} is not able to recover the same amount of detail and contrast in the shading as our method.}
    \label{fig:comparison_sai}
\end{figure*}

%% file: chapters/relighting.tex
\begin{figure*}
    \centering
    \setlength\tabcolsep{3.0pt}
    \resizebox{1.0\textwidth}{!}{
    \begin{tabular}{c|c|c|c}
         Lake &
         Office Hallway &
         City Plaza &
         Above the Clouds 
            \\
         \includegraphics[width=0.2499\textwidth]{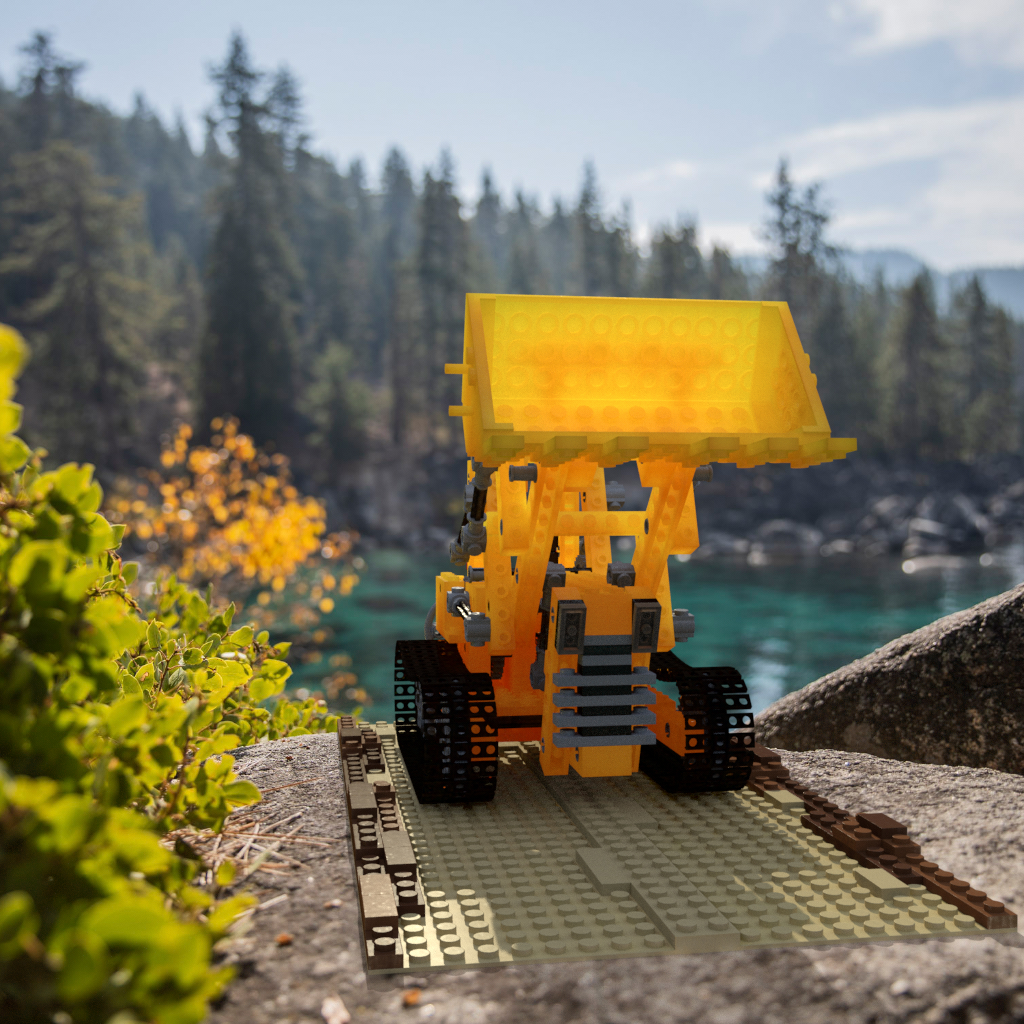} & \includegraphics[width=0.2499\textwidth]{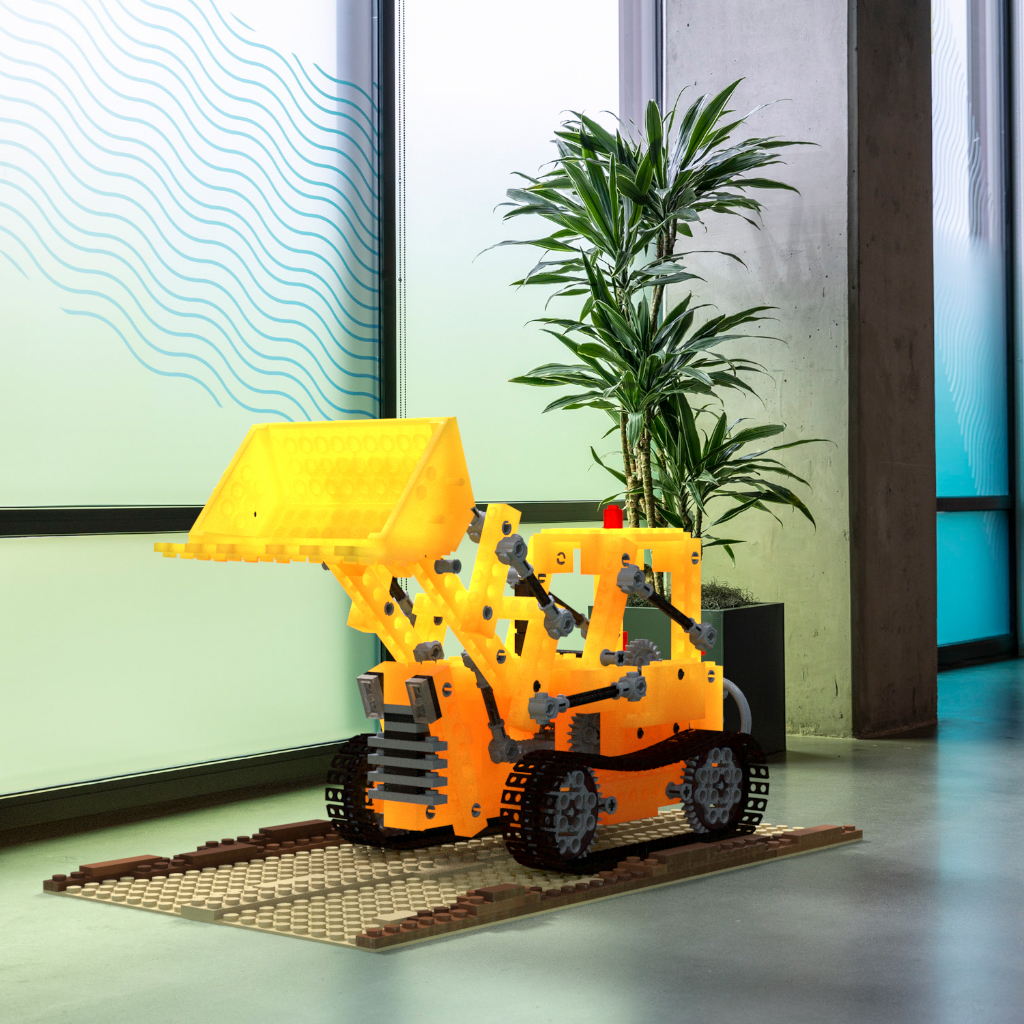} & \includegraphics[width=0.2499\textwidth]{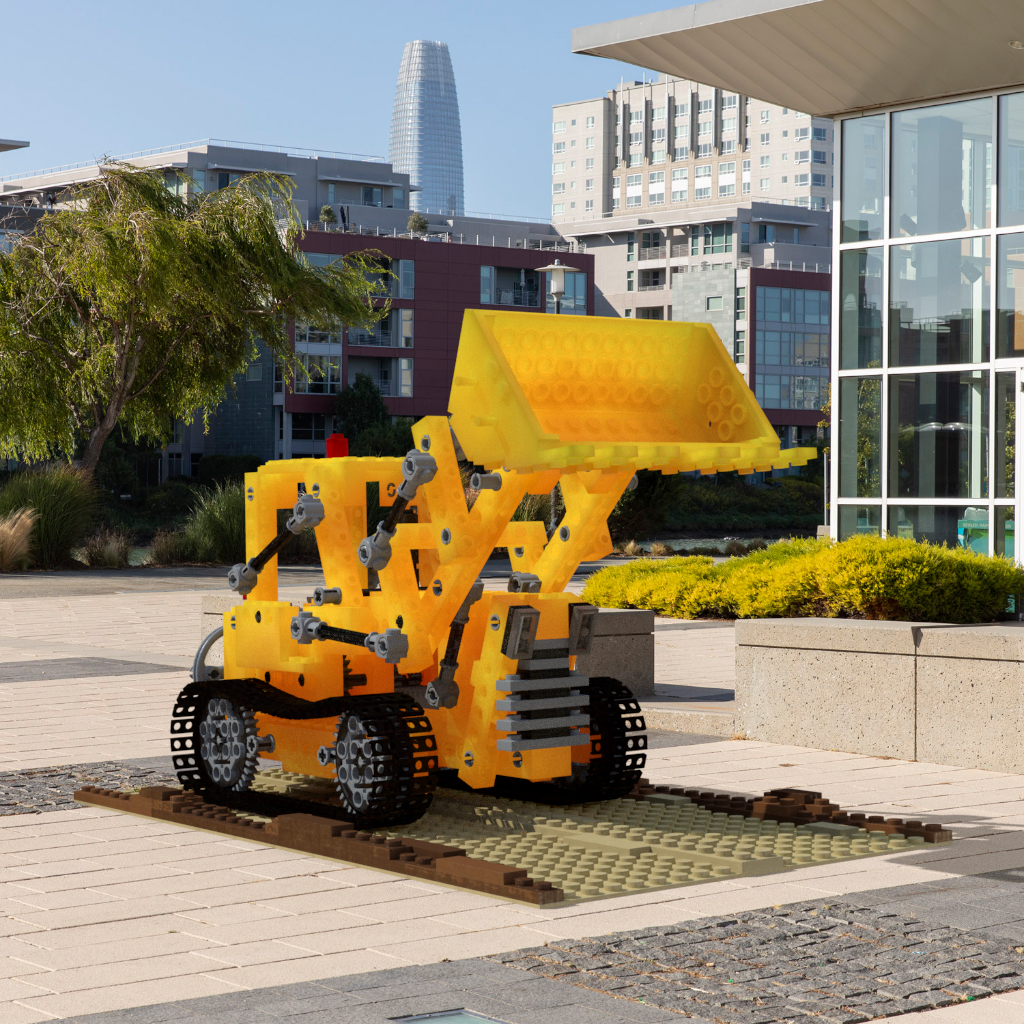} &
         \includegraphics[width=0.2499\textwidth]{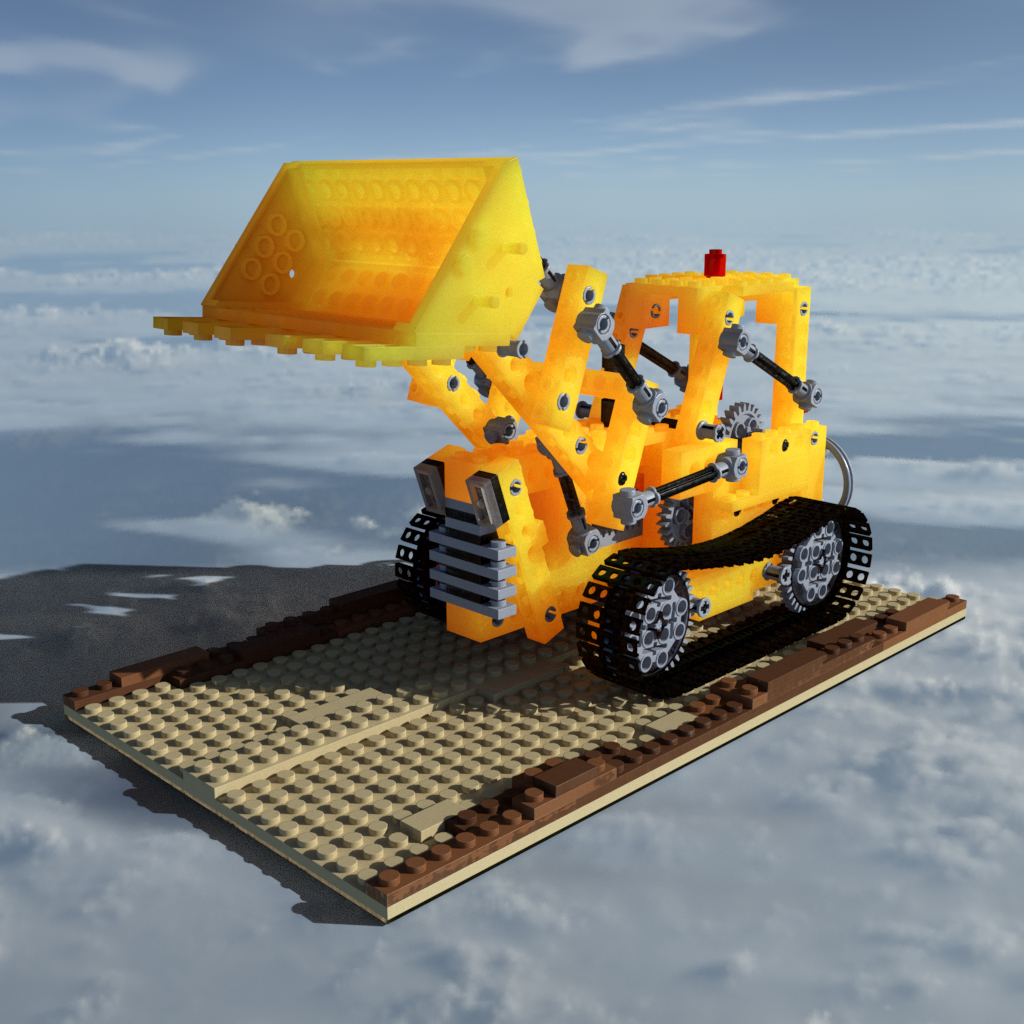} 
            \\
         \includegraphics[width=0.2499\textwidth]{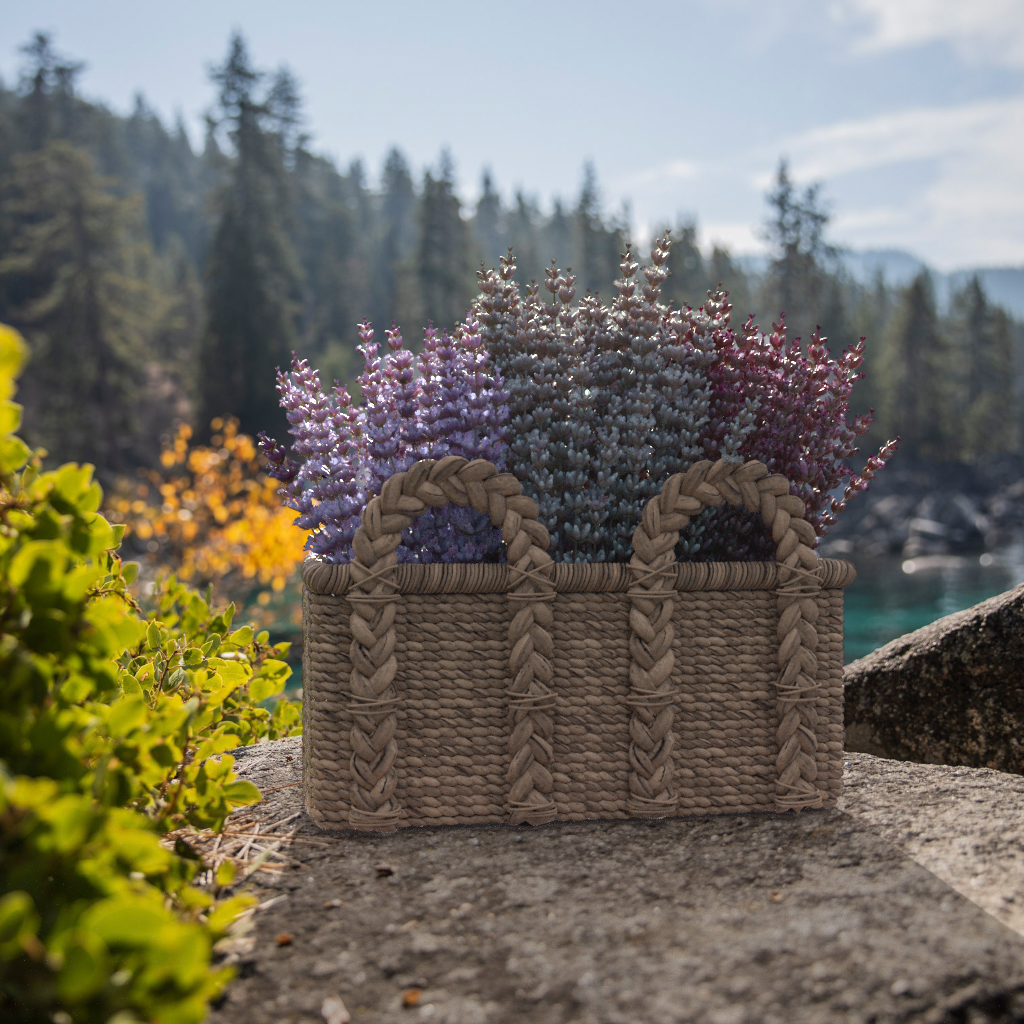} & \includegraphics[width=0.2499\textwidth]{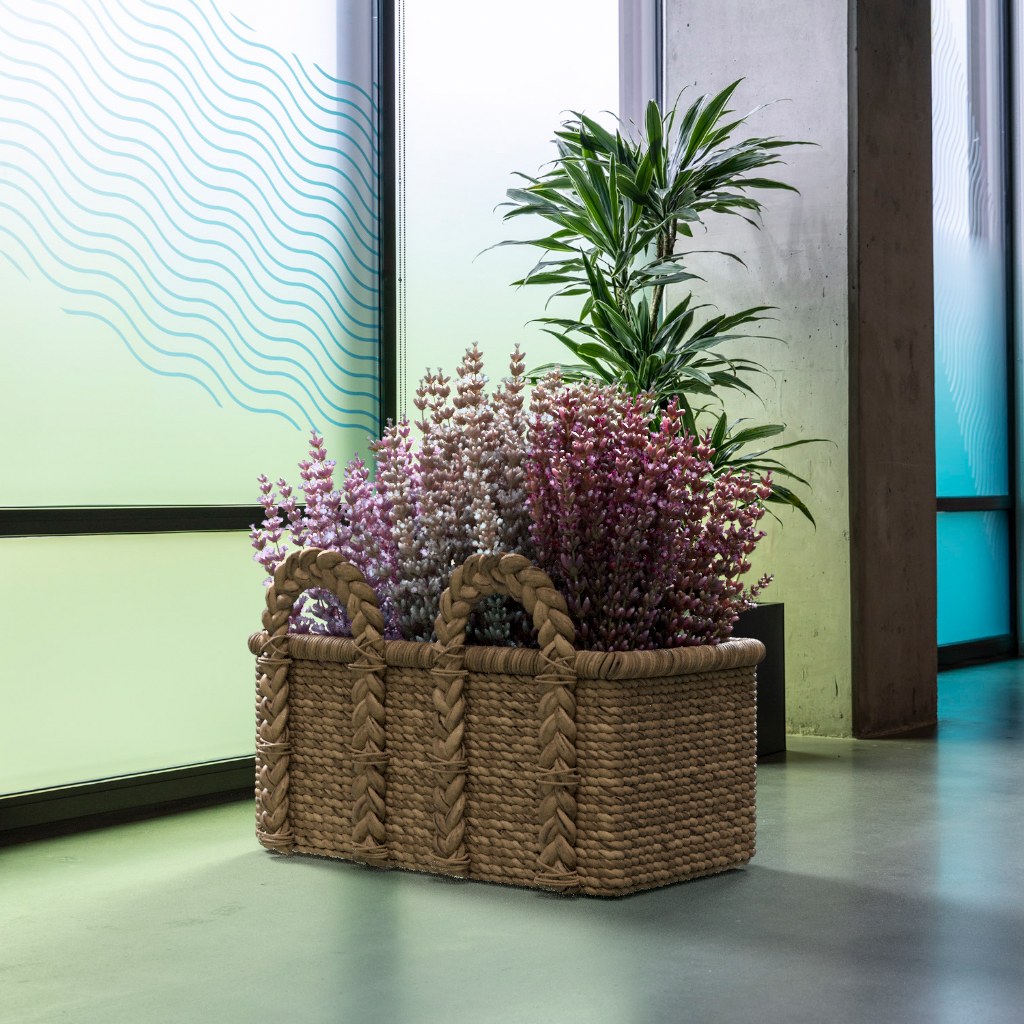} & \includegraphics[width=0.2499\textwidth]{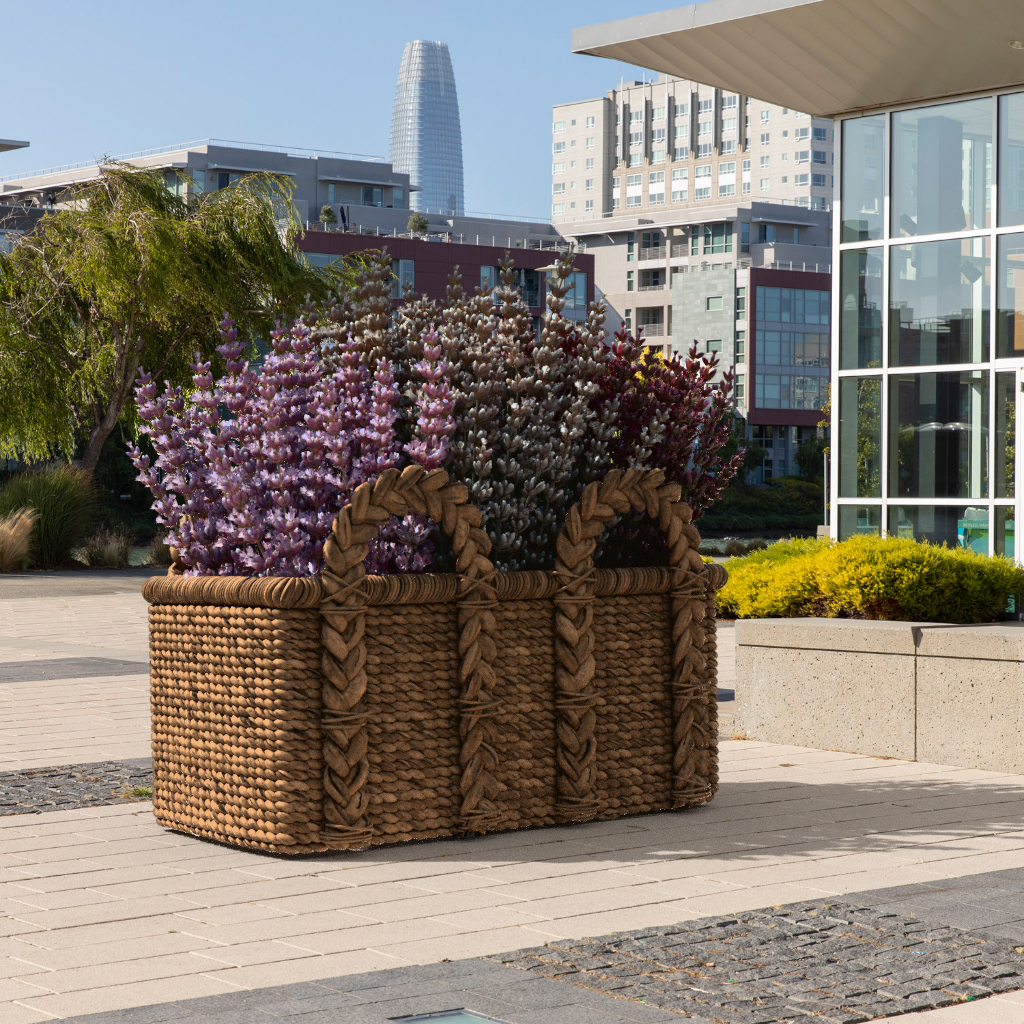} &
         \includegraphics[width=0.2499\textwidth]{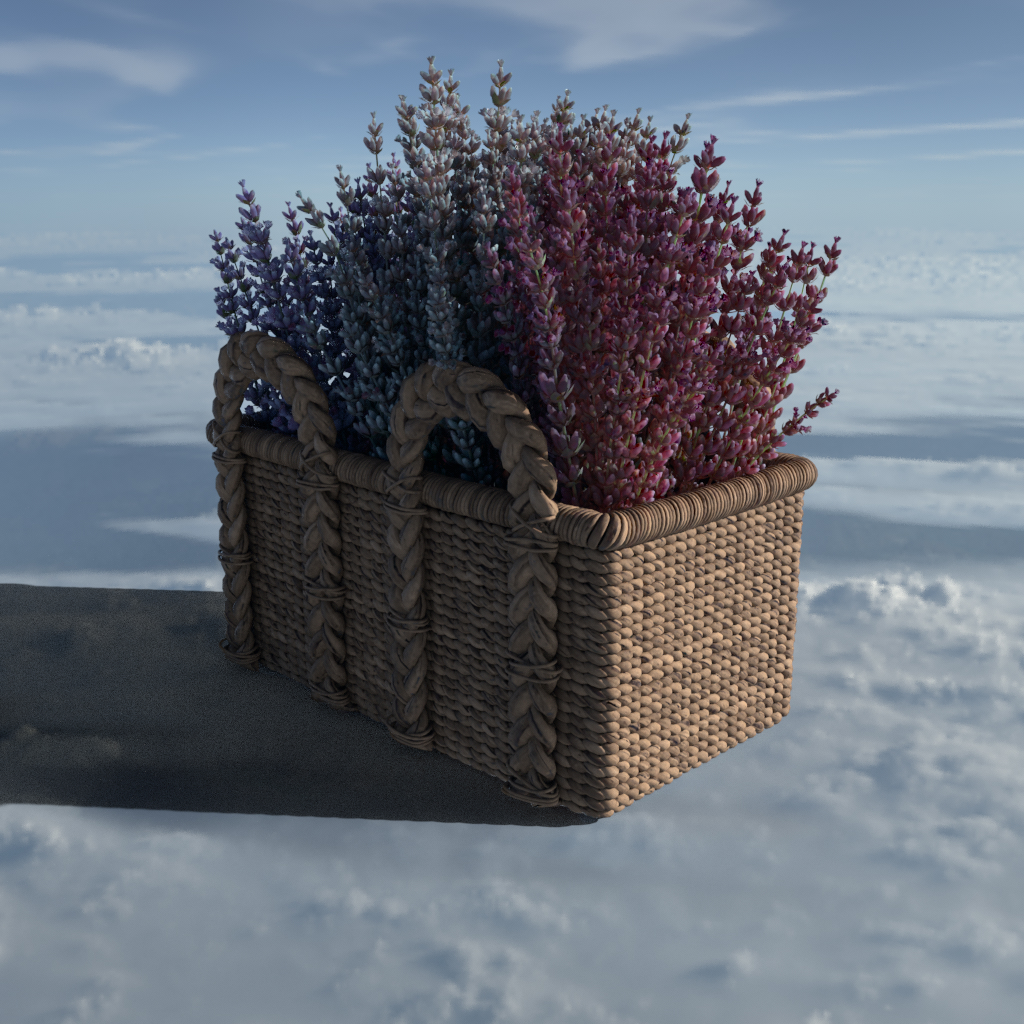} 
    \end{tabular}
    }
    \caption{{\bf IBL relighting results on surface assets.} We render our model under four different IBL environments with shadow-catching ground plane on surface assets, namely the translucent Lego and a basket of flowers, both featuring subsurface scattering. The lighting conditions vary from sharp outdoor sunlight to indoor office light, demonstrating our model's ability to faithfully react to the illuminations for relighting.}
    \label{fig:relighting}
\end{figure*}

%% file: chapters/relighting2.tex
\begin{figure*}
    \centering
    \setlength\tabcolsep{2.0pt}
    \resizebox{1.0\textwidth}{!}{
    \hspace*{-11.6pt}\begin{tabular}{ccccc}
         \raisebox{5.0em}{\rotatebox{90}{Sunset}} &
         \includegraphics[width=0.2499\textwidth, height=0.2499\textwidth]{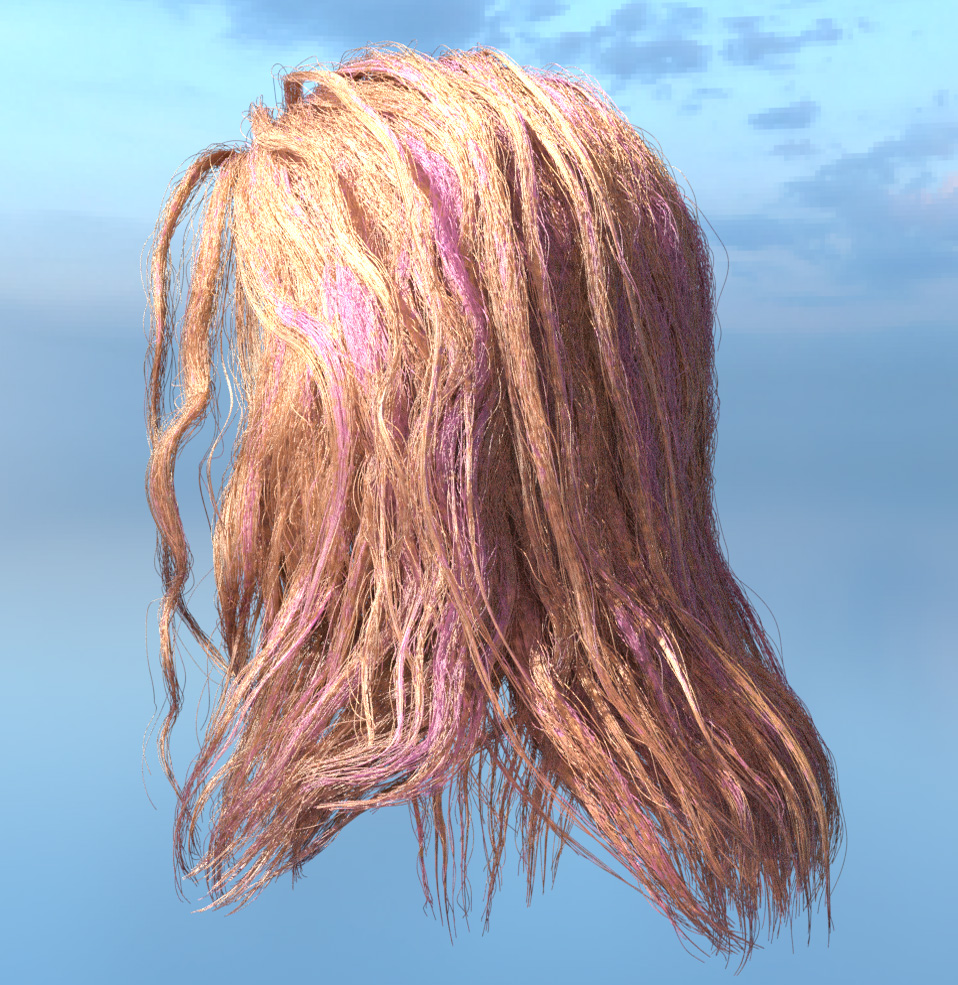} & \includegraphics[width=0.2499\textwidth, height=0.2499\textwidth]{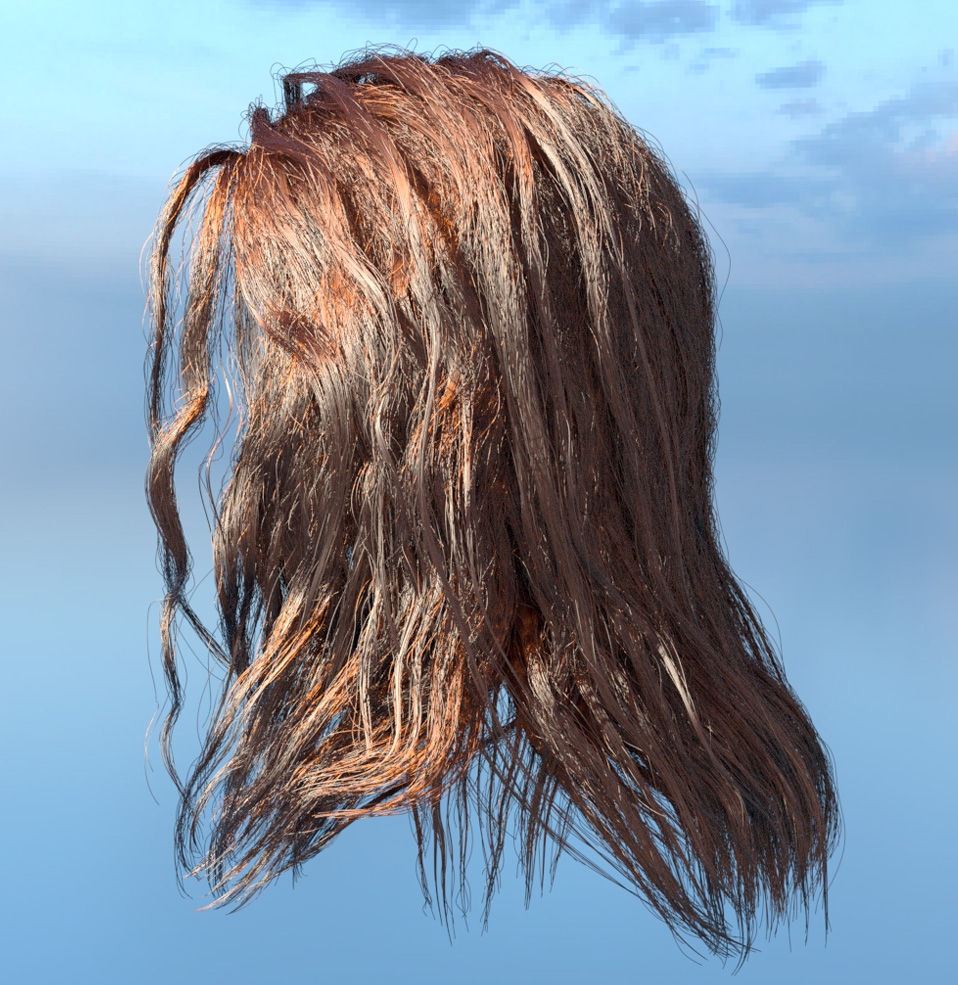} & \includegraphics[width=0.2499\textwidth, height=0.2499\textwidth]{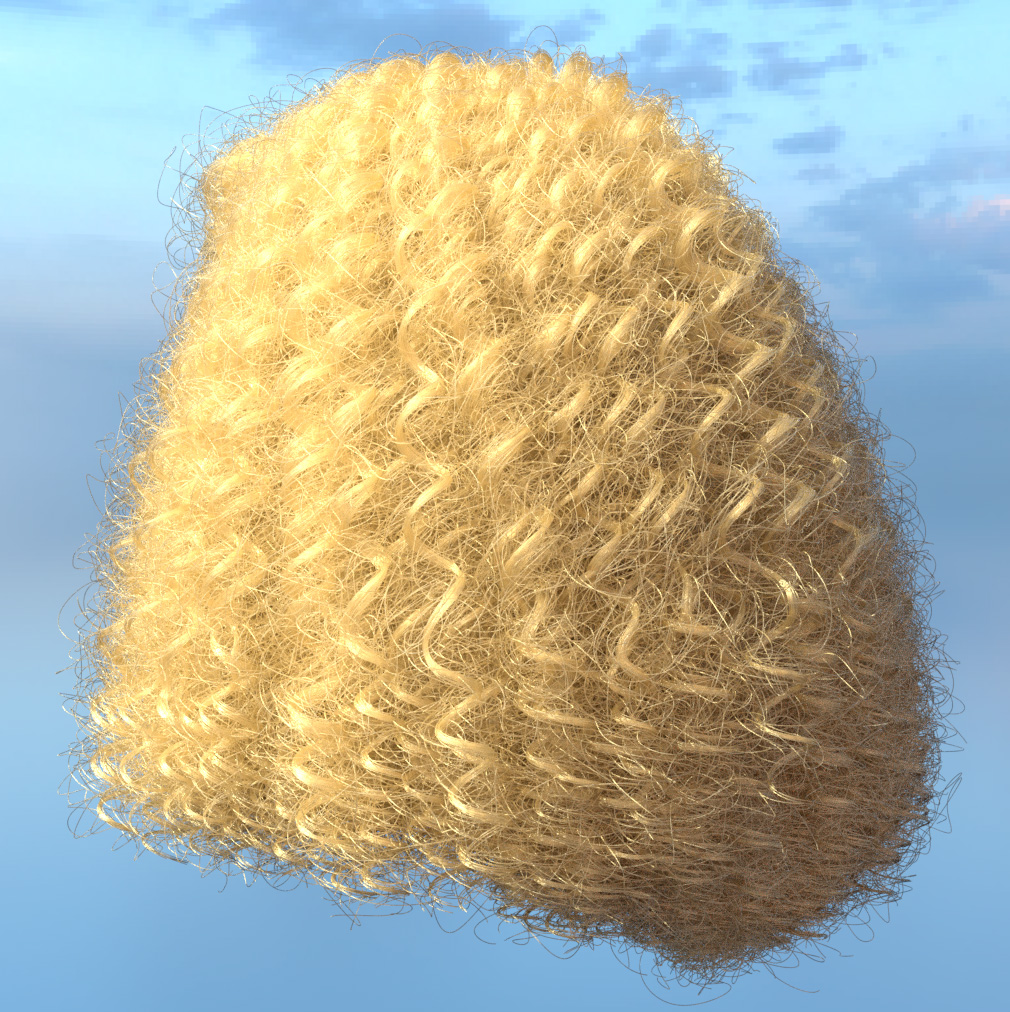} &
         \includegraphics[width=0.2499\textwidth, height=0.2499\textwidth]{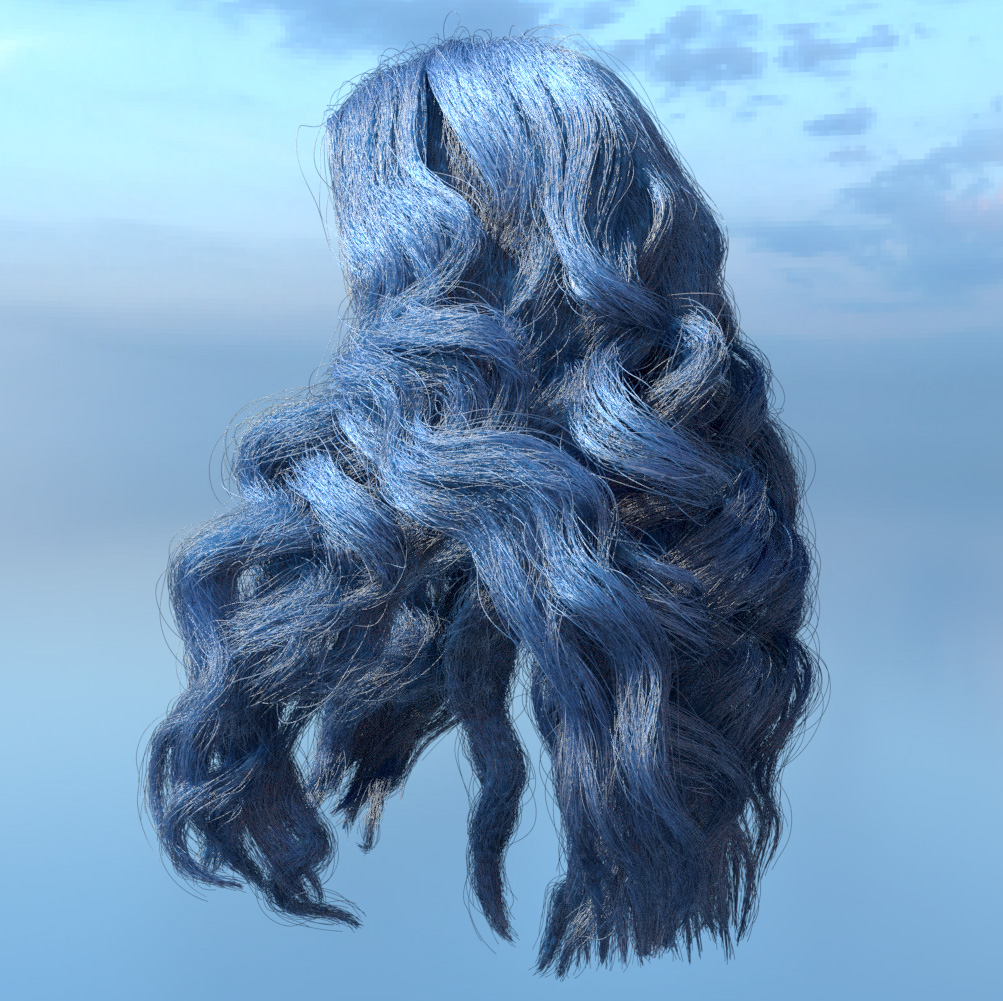} \\
         \hline 
         \\[-7.5pt]
         \raisebox{4.5em}{\rotatebox{90}{Studio Light}} &
         \includegraphics[width=0.2499\textwidth, height=0.2499\textwidth]{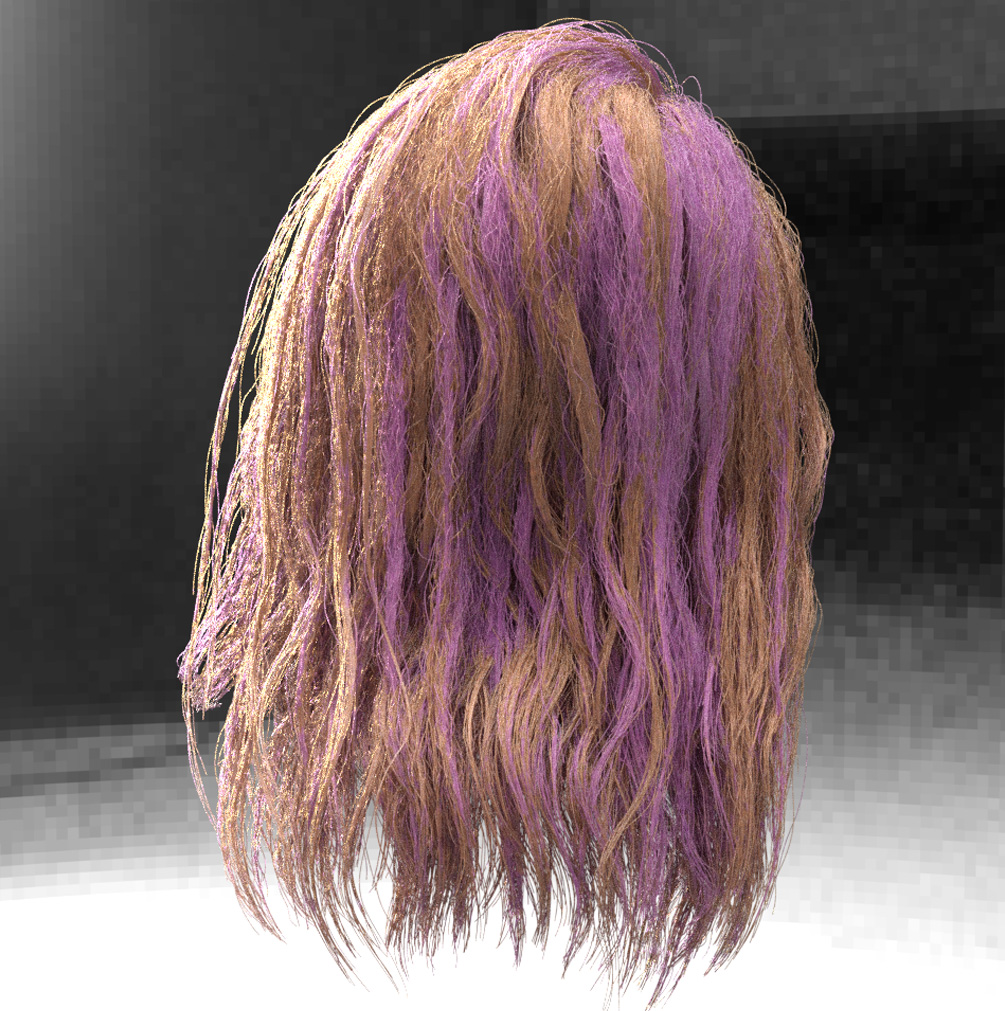} & \includegraphics[width=0.2499\textwidth, height=0.2499\textwidth]{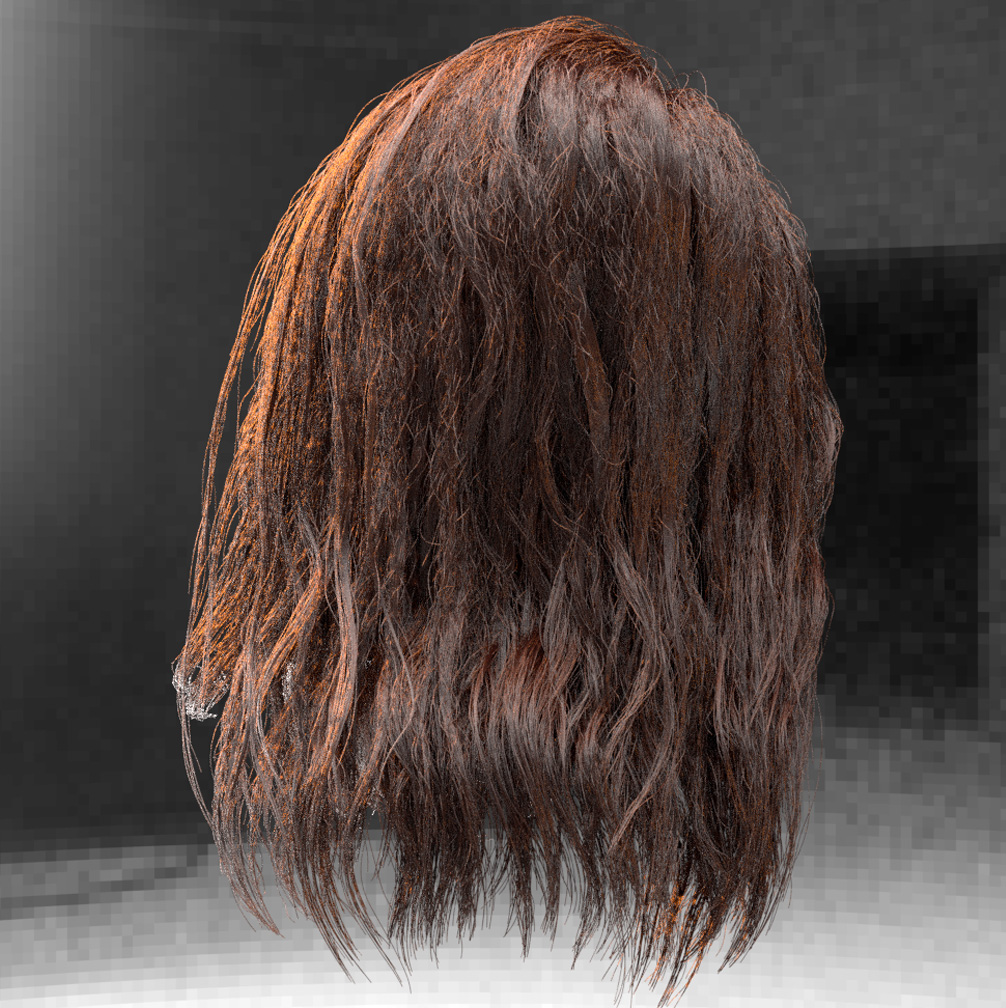} & \includegraphics[width=0.2499\textwidth, height=0.2499\textwidth]{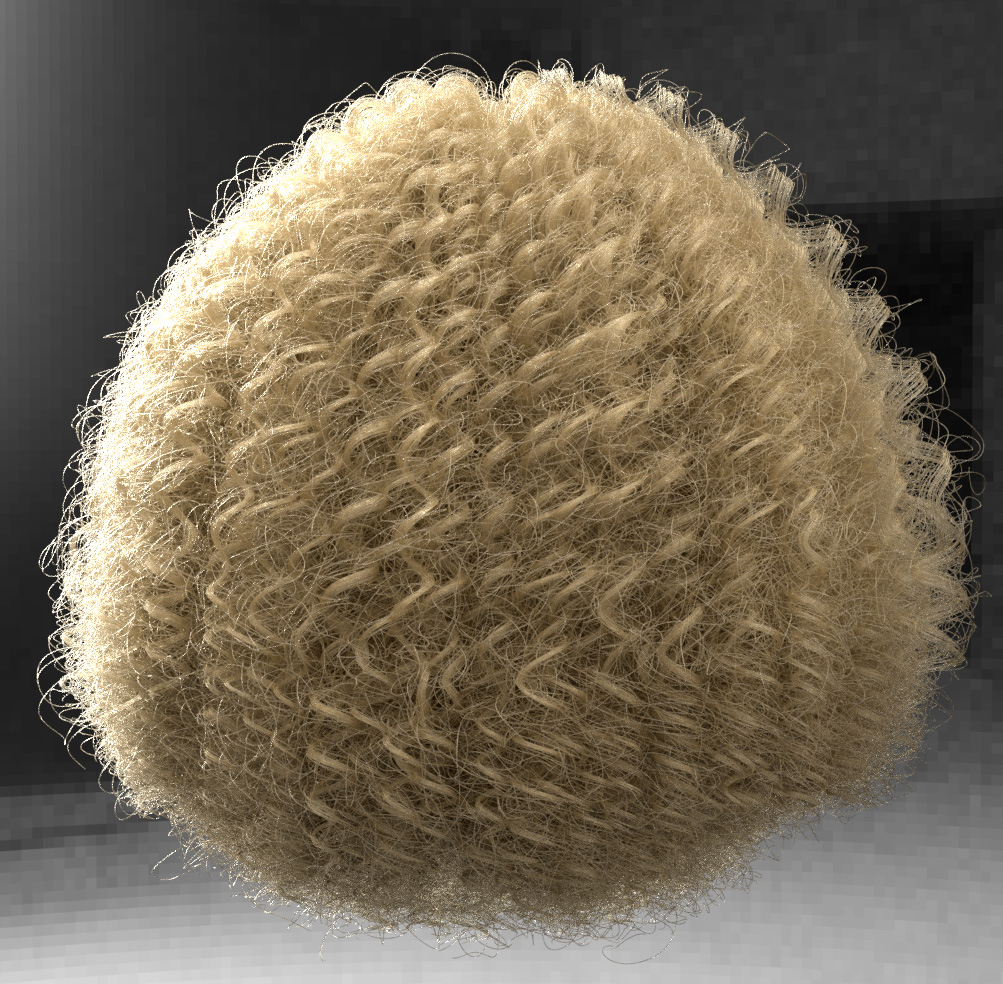} &
         \includegraphics[width=0.2499\textwidth, height=0.2499\textwidth]{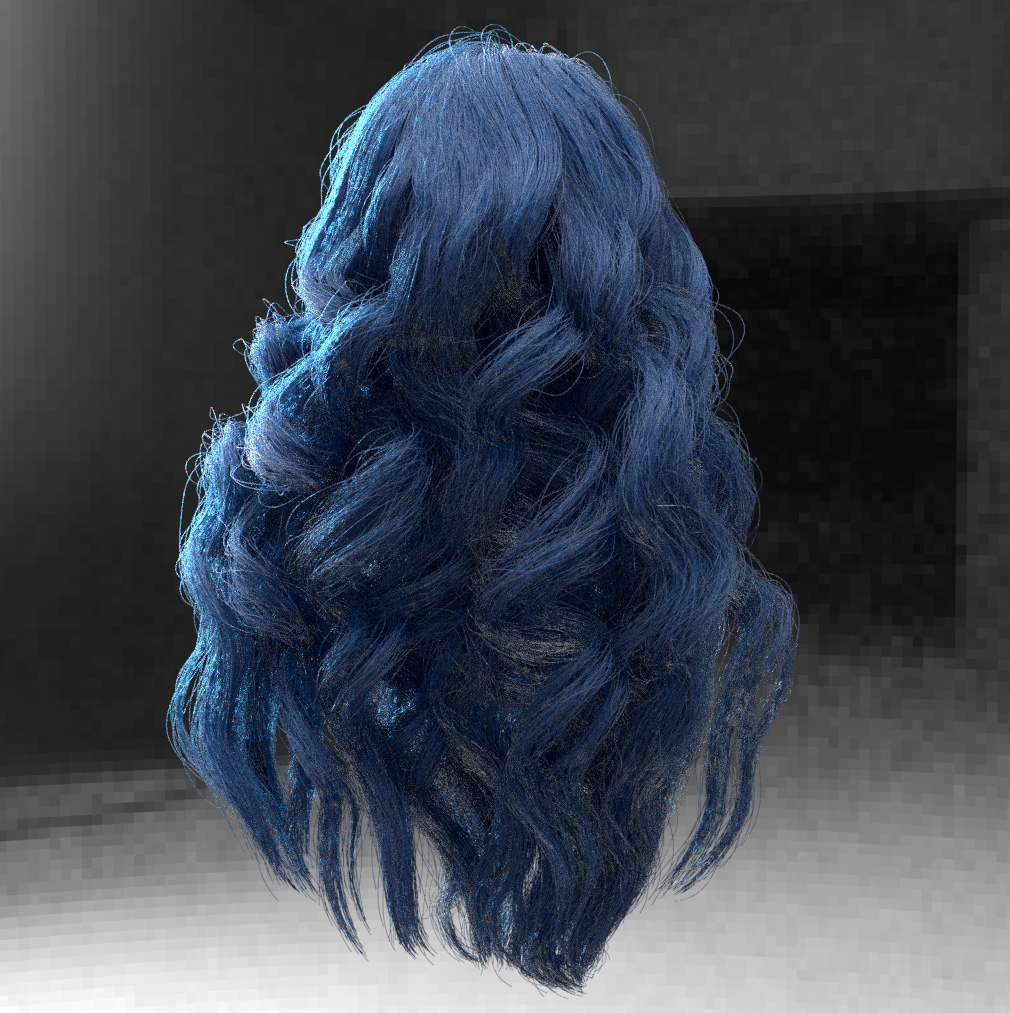} \\
    \end{tabular}
    }
    \caption{{\bf IBL relighting results on fiber assets.} Our model is rendered under precise IBL lighting environments, including sunlight and studio light, using four different hair assets with complex scattering effects. It successfully captures the texture of specular glinty highlights and the soft diffusion-like characteristics due to multiple fiber interactions, maintaining the lifelike appearance of individual hair strands and photorealism of the hair appearance, demonstrating reasonable visual accuracy while responding to various illuminations. }
    \label{fig:relighting2}
\end{figure*}

%% file: chapters/nearfield.tex
\subsection{Near-field Rendering}
\begin{figure*}\vspace{-6mm}
    \centering
    \includegraphics[width=0.8\textwidth]{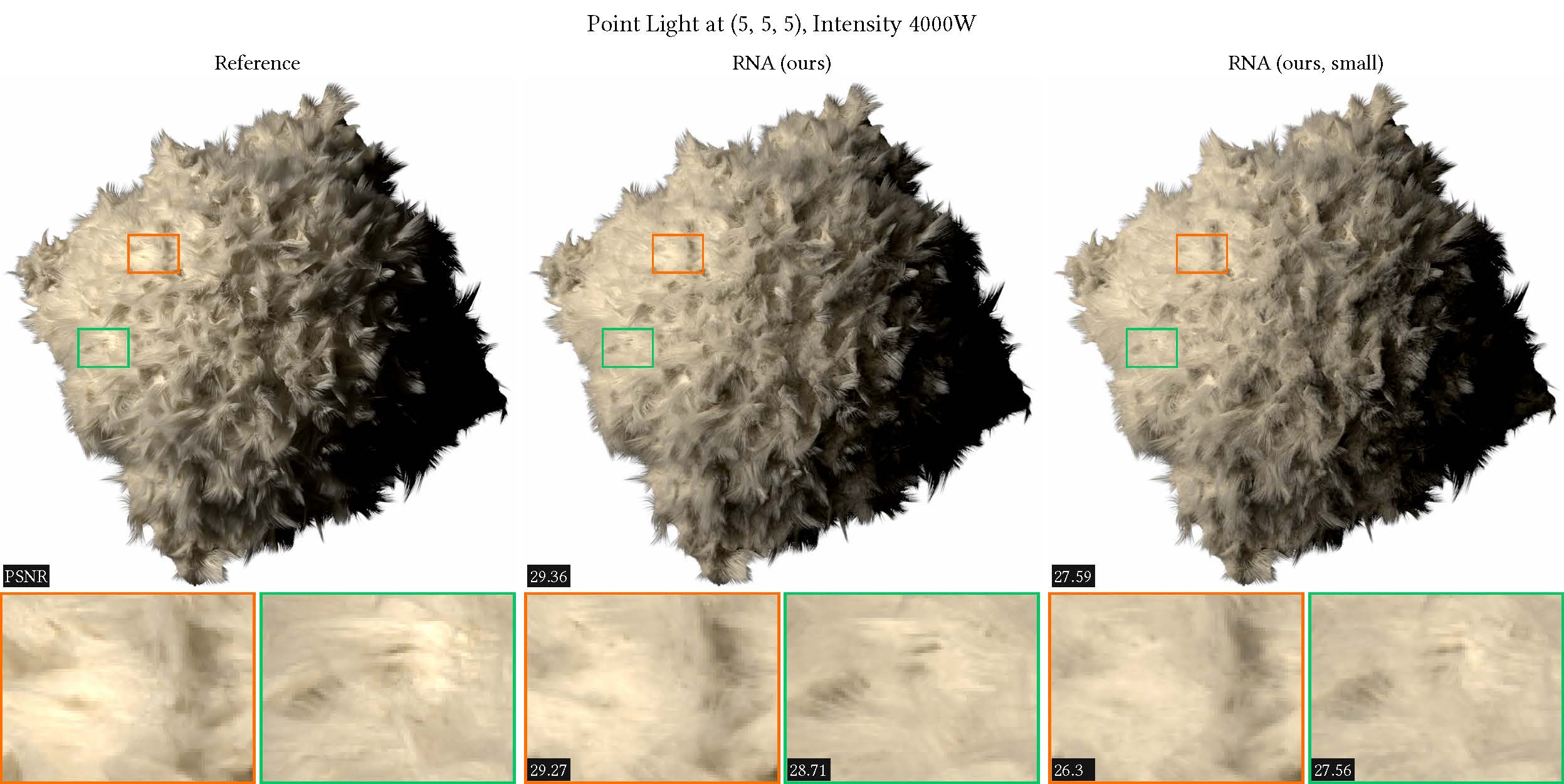}
    \includegraphics[width=0.8\textwidth]{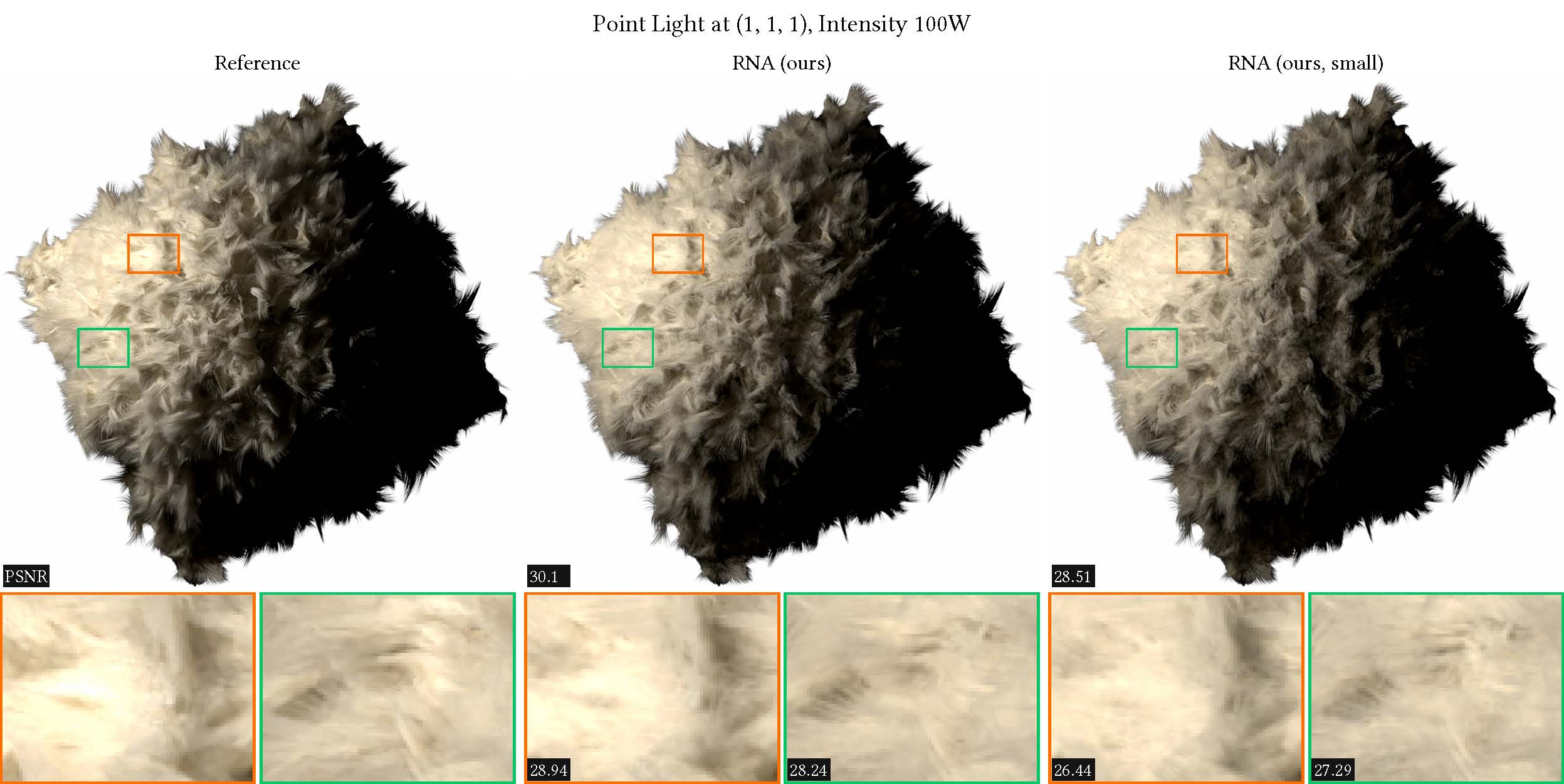}
    \includegraphics[width=0.8\textwidth]{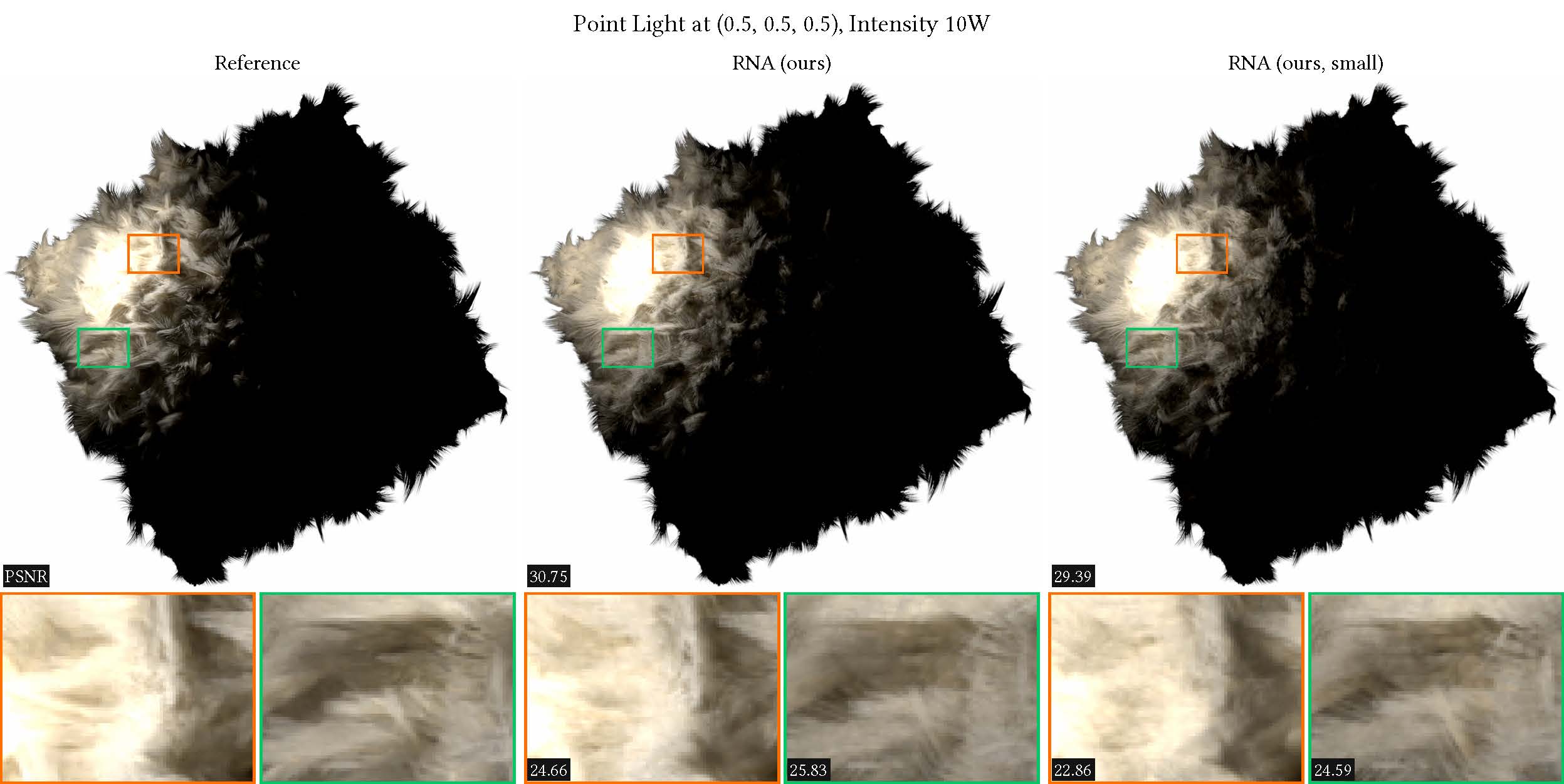}
    \caption{{\bf Near-field lighting.} We show the rendering of our white fur asset with a point light at distances that are far, close and very close to the asset to illustrate to what extent our distant lighting assumption causes error in near-field lighting.}
    \label{fig:nearfield}
\end{figure*}

As described in Section \ref{sec:model}, our model is trained under the assumption of distant lighting. In figure~\ref{fig:nearfield}, we analyze the effect of this assumption on near-field lighting with the use of a point light placed at three different distances from the neural asset. We observe that on moving the point light from a relatively far distance of $8.66$ closer to $1.73$, our model's accuracy holds up particularly well as shown by very similar PSNR values in both the full render and crops. Even when the light is moved right next to the asset at a distance of $0.86$, the overall appearance is preserved, but the close-up crops show some reduction in quality. This shows that for most practical applications, the distant light assumption does not cause our method to break down on using local lighting. We would also like to also note that all the comparisons to baselines were performed with point lights showing further evidence of near-field rendering quality.

%% file: chapters/ablations.tex
\subsection{Ablations}
\label{Ablations}

\begin{figure}
\centering
\setlength\tabcolsep{2.0pt}
    \resizebox{0.999\linewidth}{!}{
    \begin{tabular}{c|c|c}
    Ours (w/ $h$) & Ours (w/o $h$) & 
    Reference \\
    \includegraphics[width=0.33\linewidth]{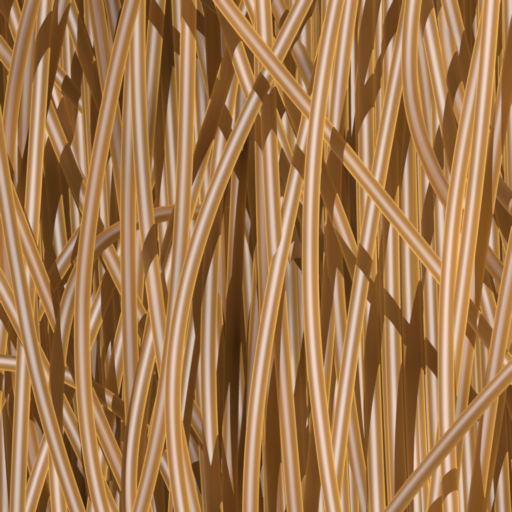} &
    \includegraphics[width=0.33\linewidth]{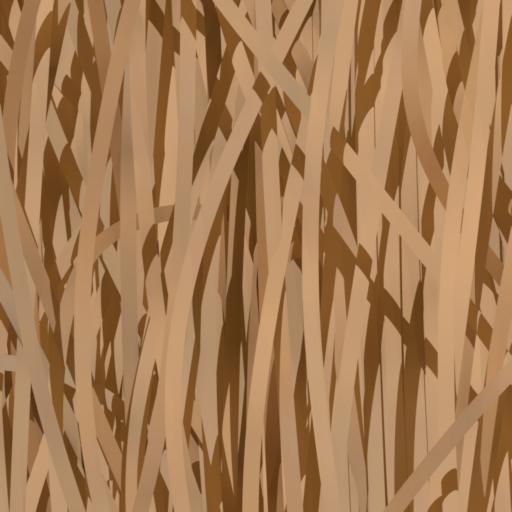} &
    \includegraphics[width=0.33\linewidth]{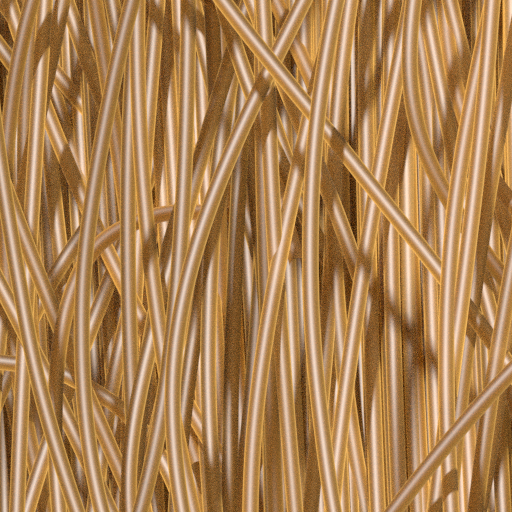} 
    \end{tabular}
    }
\caption{
    \textbf{A close-up of a small part of a blonde hair assembly.} Our approach can render fairly extreme close-ups that remain detailed and realistic, instead of blurring; however, this requires using the fiber cross-section position $h$ as one of the MLP inputs. Left: without $h$, the model cannot learn the variation. Middle: the variation across the width of a fiber is captured as well as in the reference (right).
}
\label{fig:closeup}
\end{figure}

\subsubsection{Using $h$}

In Figure~\ref{fig:closeup}, we show a close-up zoom at a small part of a hair assembly. Our model computes point-wise outgoing radiance at every fiber intersection. This lets us render fairly extreme close-ups that remain detailed and realistic, instead of blurring (which would have been the case for volumetric or mesh approximations to the fiber geometry). Note the variation in color across the width of a fiber (left), a correctly captured feature of the near-field hair BSDF model used, matching the reference (right). However, this requires using the fiber cross-section position $h$ as one of the MLP inputs. Without $h$, the model cannot learn the variation (middle).

\subsubsection{Network Architecture}

We chose MLP and triplane configurations methodically. Table~\ref{tab:network_arch} shows the various MLP architectures we tested, along with their training time and average PSNR values on the validation data for the Lego asset that has heavy scattering properties in its shading model. It is clear that increasing the network capacity improved the quality of the model, and took longer to train; the training time being a rough estimate of the model's inference performance. Of note, for the Lego asset, models with 16 and 32 hidden layer neurons were not able to capture the scattering effects in most regions and 64 was the fastest acceptable one. We chose 64 neurons as the size for the small model as it worked well with all our assets, and was also a size that generally fits on register allocations of GPU production path tracing shaders. The take-away though is that there is a quality vs. performance trade-off that can be made for any particular use case.

Table~\ref{tab:feat_grid} shows the effect of using feature grids with increasing channel counts on our high-quality model. Increasing the number of channels increases the number of inputs to the MLP and thus has an adverse effect on performance as indicated by training time. The PSNR values show that 8 channels is sufficient for good model quality, and though there is a slight benefit in going up to 16 channels, anything higher doesn't provide additional gains.

\begin{table}[b!]
\centering
\setlength\tabcolsep{2.0pt}
\caption{\textbf{Ablation on the MLP architecture.} We evaluate different configurations of the MLP, including the number of hidden layers and the number of neurons in each layer. We show the average PSNRs on the Lego asset, along with the runtime for training to give a sense of model performance. All models here use 8 channels for the feature grids. Our chosen architecture for offline model is in \textbf{bold}, and small model is in \textit{italics}.
}
    \begin{tabular}{c|c|c|c}\hline
         Hidden Layers & Neurons & Training Runtime & PSNR  \\ \hline
            4 & 16 & 34m 58s & 27.47 \\
            4 & 32 & 35m 11s & 27.36 \\
            \textit{4} & \textit{64} & \textit{31m 17s} & \textit{28.77} \\
            4 & 128 & 39m 27s & 31.02 \\
            4 & 256 & 50m 51s & 32.68 \\
            2 & 512 & 55m 51s & 33.07 \\
            3 & 512 & 1h 5m & 33.80 \\
            \textbf{4} & \textbf{512} & \textbf{1h 13m} & \textbf{34.09} \\
            5 & 512 & 1h 25m & 34.20 \\
            4 & 1024 & 2h 44m &  34.90 \\
    \end{tabular}
\label{tab:network_arch}
\end{table}

\begin{table}[tb]
    \centering
        \caption{\textbf{Ablation on Feature Grid Channels}. We evaluate use of different feature grid channels on the Lego asset with our high-quality model that has 4 hidden layers with 512 neurons each.}
    \begin{tabular}{c|c|c}\hline
        Channels & Training Runtime & PSNR \\ \hline
        4 & 1h 11m & 33.51 \\
        \textbf{8} & \textbf{1h 13m} & \textbf{34.09} \\
        16 & 1h 20m & 34.40 \\
        32 & 1h 37m & 34.33 \\
    \end{tabular}
    \label{tab:feat_grid}
\end{table}

\subsubsection{Visibility}
\begin{figure}[!tb]
    \centering
    \setlength\tabcolsep{2.0pt}
    \resizebox{0.999\linewidth}{!}{
    \begin{tabular}{c|c|c}
         Ours (w/ vis. hint) & 
         Vis. hint & 
         Ours (w/o vis. hint) \\
         \includegraphics[width=0.33\linewidth]{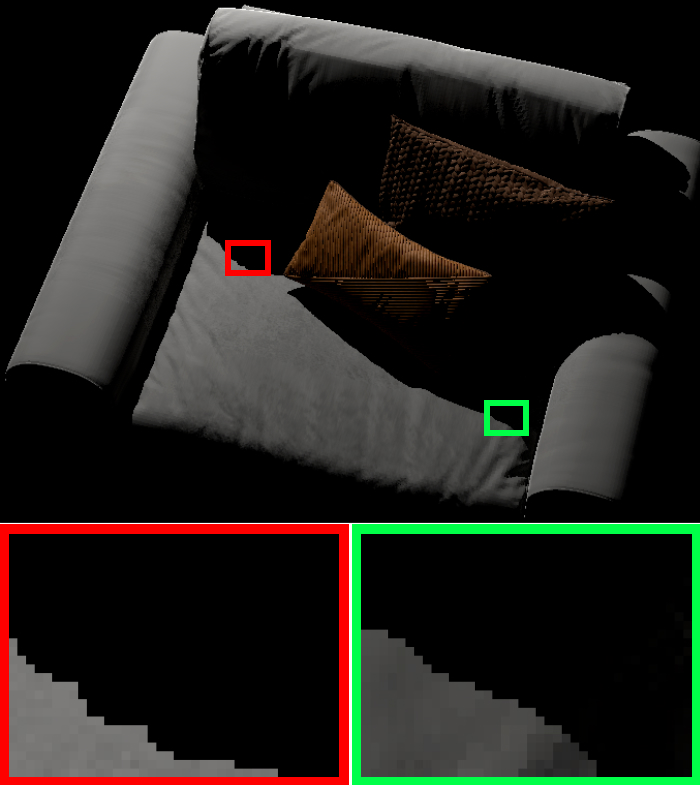} &
         \includegraphics[width=0.33\linewidth]{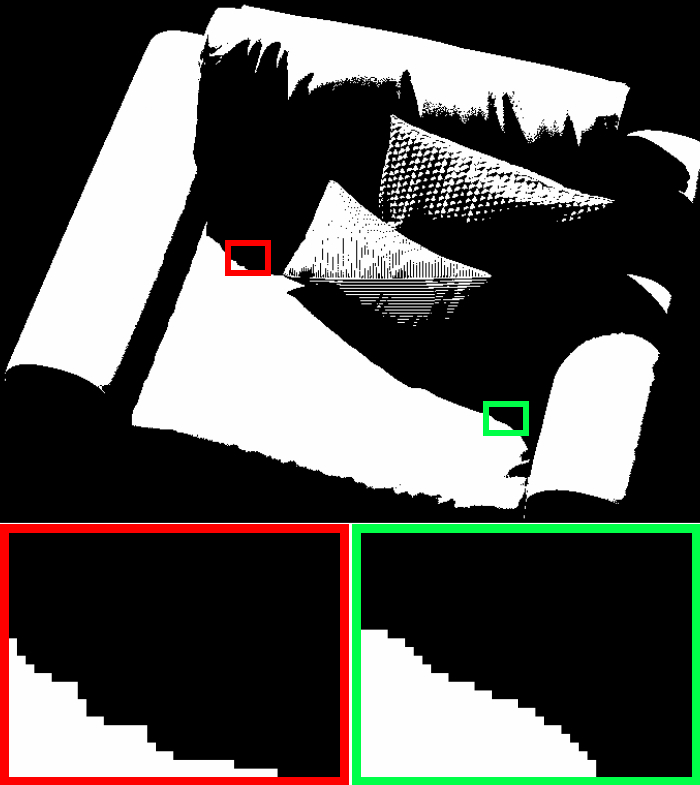} &
         \includegraphics[width=0.33\linewidth]{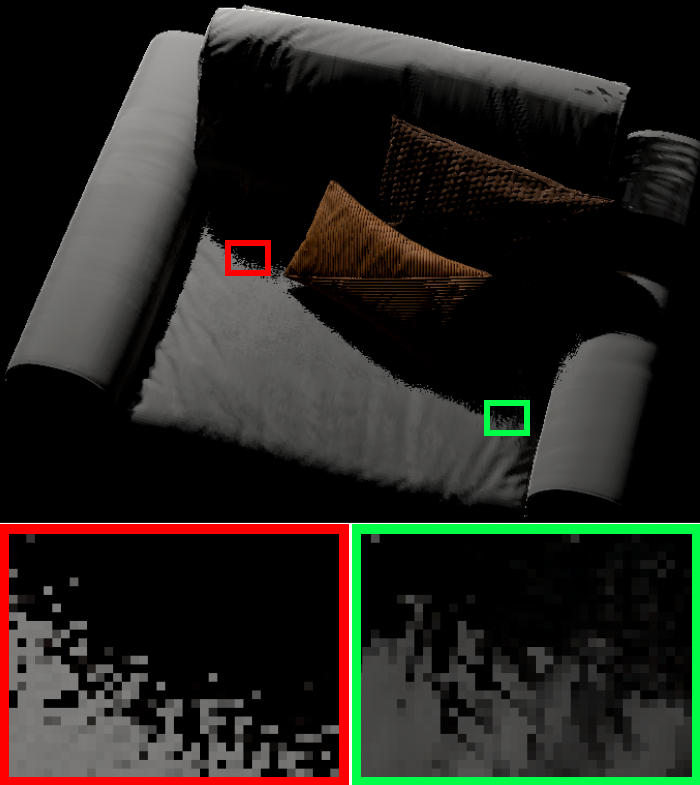} 
    \end{tabular}
    }
    \caption{{\bf Lighting visibility for self-shadowing.} We use visibility value as a hint to the MLP, which significantly improves the rendering quality around shadowing edges.}
    \label{fig:vis_hints}
\end{figure}

We compare the rendering result of our model with and without the lighting visibility hint. As shown in Figure~\ref{fig:vis_hints}, our model improves the rendering quality around shadowing edges with faithful shadows when we provide visibility hints to the MLP. This is because direct shadowing is typically high frequency signals (e.g. hard edge boundaries), which is difficult for MLP to represent well without aiding such hints. 

\subsubsection{Triplane vs. UV parameterization}
\begin{figure*}
\centering
\includegraphics[width=1.0\textwidth]{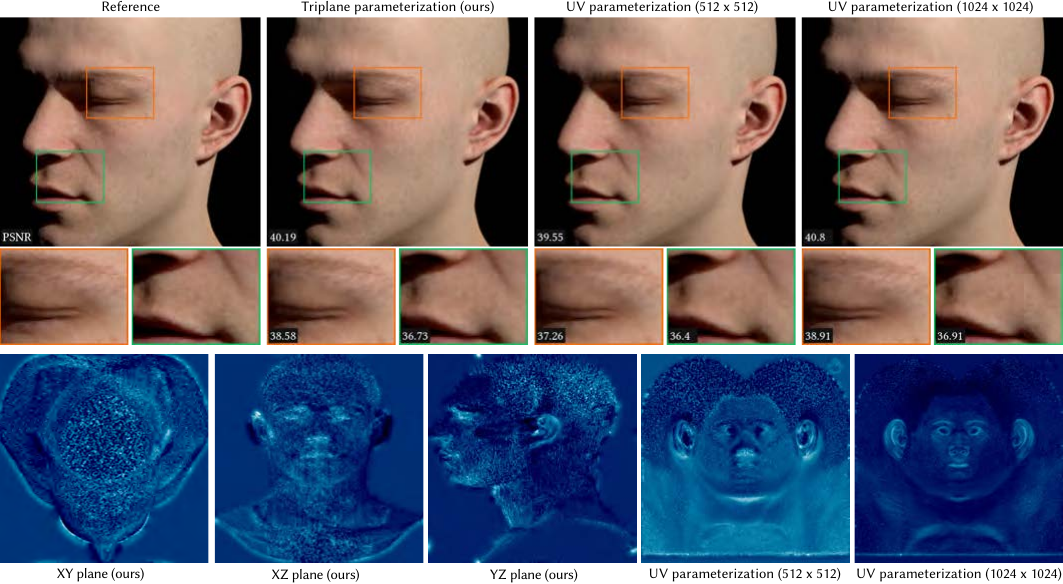}
\caption{\textbf{Ablation on the triplane vs. UV parameterization.} We examine triplane and UV parameterization on the head asset. UV parameterization at $512 \times 512$ resolution results in blurred details, becoming sharper only at $1024 \times 1024$. Conversely, the triplane representation provides satisfactory detail at $512 \times 512$. The color map is the same as in Fig.~\ref{fig:triplane_vis}, visualizing the first channel of features.} \label{fig:uv_ablation}
\end{figure*}

As shown in Figure~\ref{fig:uv_ablation}, triplane representation provides sufficient texture details at $512 \times 512$. On the other hand, UV parameterization at $512 \times 512$ resolution results in blurred details, becoming sharper only at $1024 \times 1024$. On the other hand, UV mapping is not always feasible when the mesh topology is complex (or fibers) while triplane representation does not have such limitation.

%% file: chapters/conclusion.tex
\section{Conclusion}
\label{sec:conclusion}

In this paper, we introduced a precomputed relightable neural model for representing synthetic surface-based or fiber-based 3D assets with complex materials. Our model allows full view and lighting variation; in comparison with relightable neural capture approaches, we achieve higher accuracy, though our method is specifically designed for representing digital assets, not for capture from real photographs. 
Our design enables any viewing or illumination conditions and allows for integration of our assets in full scenes, rendered in a production path tracer. Our neural model handles all shading, inter-reflections, and scattering. The benefits of our approach include increased shading performance, as well as as ability to represent complex material models and shading graphs, which do not need to be implemented in the target rendering system.

%% file: chapters/acknowledgements.tex
\begin{acks}
    We thank the authors of NRHints \cite{NRHints} for sharing some of the synthetic assets from their paper for use in comparisons, and Sai Bi for helping with Neural Reflectance Fields \cite{bi2020}. We also thank Thomas Caissard, Andrea Machizaud, Wayne Wooten and Francois Beaune for their help with implementation of RNA in the path tracing engine.
\end{acks}